\documentclass[fleqn,usenatbib]{mnras}

\usepackage{newtxtext,newtxmath}
\usepackage[T1]{fontenc}

\DeclareRobustCommand{\VAN}[3]{#2}
\let\VANthebibliography\thebibliography
\def\thebibliography{\DeclareRobustCommand{\VAN}[3]{##3}\VANthebibliography}

\usepackage{graphicx}	% Including figure files
\usepackage{amsmath}	% Advanced maths commands
\usepackage{float}
\usepackage[table, xcdraw, dvipsnames]{xcolor}
\usepackage{caption}
\usepackage{subcaption}
\usepackage{dirtytalk}
\title[Conditions for filament and star formation]{Necessary conditions for the formation of filaments and star clusters in the cold neutral medium}

\author[]{Rachel Pillsworth$^{1}$
\thanks{Contact e-mail: \href{mailto:pillswor@mcmaster.ca}{pillswor@mcmaster.ca}}, Ralph E. Pudritz$^{1, 2, 3, 4}$
\\
$^{1}$Department of Physics \& Astronomy, McMaster University, 1280 Main St West, Hamilton, ON L8S 4L8, Canada \\
$^{2}$ Origins Institute, McMaster University, Hamilton, ON L8S 4M1\\
$^{3}$ Universität Heidelberg, Zentrum für Astronomie, Institut für Theoretische Astrophysik, Albert-Ueberle-Str. 2, 69120 Heidelberg, Germany\\
$^{4}$ Max-Planck Institut für Astronomie, Königstuhl 17, D-69117 Heidelberg, Germany 
}

\date{Accepted XXX. Received YYY; in original form ZZZ}

\pubyear{2022}

\begin{document}
\label{firstpage}
\pagerange{\pageref{firstpage}--\pageref{lastpage}}
\maketitle

\begin{abstract}
Star formation takes place in filamentary molecular clouds which arise by physical processes that take place in the cold, neutral medium (CNM). We address the necessary conditions for this diffuse ($n \approx 30$ cm$^{-3}$), cold (T $\approx$ 60 K), magnetized gas undergoing shock waves and supersonic turbulence, to produce filamentary structures capable of fragmenting into cluster forming regions. Using RAMSES and a magnetized CNM environment as our initial conditions, we simulate a 0.5 kpc turbulent box to model a uniform gas with magnetic field strength of 7 $\mu G$, varying the 3D velocity dispersion via decaying turbulence. We use a surface density of $320 M_{\odot} pc^{-2}$, representative of the inner 4.0 kpc CMZ of the Milky Way and typical luminous galaxies. Filamentary molecular clouds are formed dynamically via shocks within a narrow range of velocity dispersions in the CNM of 5 - 10 km/s with a preferred value at 8 km/s. Cluster sink particles appear in filaments which exceed their critical line mass, occurring optimally for velocity dispersions of 8 km/s. Tracking the evolution of magnetic fields, we find that they lead to double the dense star forming gas than in purely hydro runs. Perpendicular orientations between magnetic field and filaments can increase the accretion rates onto filaments and hence their line masses. Because magnetic fields help support gas, MHD runs result in average temperatures an order of magnitude higher than unmagnetized counterparts. Finally, we find magnetic fields delay the onset of cluster formation by $\propto 0.4$ Myr.

\end{abstract}

\begin{keywords}
MHD -- turbulence -- methods:numerical -- stars:formation -- ISM:clouds -- ISM:structure
\end{keywords}

%%%%%%%%%%%%%%%%%%%%%%%%%%%%%%%%%%%%%%%%%%%%%%%%%%

%%%%%%%%%%%%%%%%% BODY OF PAPER %%%%%%%%%%%%%%%%%%

\section{Introduction}

Molecular clouds represent the intermediate galactic scales (100s of pc) at which large scale galactic structure and evolution gives way to more localized star formation in forming star clusters.  The multiple scale nature of this process has been explored within molecular clouds by the \textit{Herschel} space observatory, and larger galactic scale studies of molecular cloud populations by PHANGS-ALMA. Over a decade ago, \textit{Herschel} made a major breakthrough by revealing the ubiquitous filamentary networks that comprise molecular clouds in which star-forming cores form on subpc scales \citep{herschel_filaments, andre2010, henning2010, menshchikov2010}. PHANGS is a recent addition to the research,  and produced the largest catalogue (100,000) of individually resolved molecular clouds across multiple galaxies \citep{phangs_OG, sun, 2022turner, 2022brunetti}. It reveals the connection between molecular cloud properties and their galactic environments across a large array of galaxies \citep{rosolowsky_phangs}. 

Molecular clouds are believed to condense out of a phase of the interstellar gas known as the cold neutral medium (CNM) in which they are embedded \citep{klessen2016, chevance}. It is therefore essential to understand the dynamics and thermal evolution of the CNM - in particular under what conditions molecular clouds and their star clusters may form in it. For instance, \citet{sun} use the PHANGS survey to show that molecular cloud properties depend on local environmental properties such as gas surface density and star formation density. On the galactic scale, they find it is primarily galactic shear that affects the final properties of molecular clouds. These recent surveys have given a much better picture of the properties of entire populations of GMCs both within and outside the Milky Way, although they are not yet sufficient to penetrate scales below the star cluster. 

Many of the galactic observations are well supported by theoretical and computational work. As one example, the results in \citet{sun} compare well to the simulations of \citet{jeffreson}, in that both studies show that cloud properties depend on some local and some galactic environmental properties. \citet{jeffreson} showed that rotation in the galactic disk influenced the geometry and rotation of the molecular clouds which formed within it. Conversely, their work finds that gravo-turbulent and star formation properties of the molecular clouds are somewhat disconnected from the overall galactic disk dynamics, supporting an agreement with \citet{sun} that the gas and star formation densities are tied to local environment. Additionally, simulations from \citet{grisdale} showed that different global star formation prescriptions produce different local effects, especially along spiral arms in galactic disks, with turbulent star formation prescriptions forming stars in high density pockets located in the arms. These different environments also affect the final masses, radii and velocity dispersions of the GMCs that form. Finally, observations from \citet{soler2022} studied giant HI filaments towards the Galactic plane finding that the filamentary structure is closely linked to the properties of the Galactic disk, being formed as direct consequences of galactic dynamics. The interstellar medium (ISM) throughout a galaxy is also repeatedly shocked, heated, and compressed by turbulence and expanding superbubbles \citep{WatkinsKreckel2023, WatkinsBarnes2023}. 

Molecular clouds host star clusters embedded in filamentary structures \citep{KrumholzBate2014, GrudicKruijssen2021}. These are the parent structures of most of the stars (70-90 \%) that form in GMCs \citep{ladalada2003} and makes them a natural starting point for studies of star formation. On these parsec and sub-parsec scales of clusters, the observational work has also been rich and varied. Telescopes such as the Chandra X-Ray Observatory, Hubble and ALMA/SMA have allowed us to peer into Milky Way clouds and their clusters, and investigate star formation within them \citep{chandra_observestarform, hubble_observeR136, sma_observemassivecluster, he_wilson_observeYMC}.

The ionizing radiation from massive stars will also contribute to the heating in the star cluster, preventing low-mass stars from forming and effectively setting the IMF \citep{klassen}. Protostellar jets and outflows from massive stars can relieve radiation pressure around the protostar and set final masses, allowing them to have a higher final mass than without outflows \citep{krumholz_mckee, kuiper_hosokawa}. These active areas of star formation can extend out many parsecs from the protostar, affecting the environment they're embedded within \citep{beuthercore} and, therefore, the local star formation rate. Furthermore, \citet{Kuffmeier2020} used zoom-in simulations to investigate star formation and find that protostellar accretion depends on the environment in which the protostar is located, firmly linking cluster environment and protostar activity.

The feedback effects of these processes on GMC formation have also been extensively studied. For instance, recent work has determined that the tides tearing gravitationally bound clouds apart can come, in part, from massive star-forming regions within the cloud \citep{ramirez-galeano}. \citet{howard} found that the most massive clusters tend to form in high density filaments in the most massive GMCs that form, while others discuss that radiation feedback plays an essential role in the lifetime of GMCs \citep[see][and references therein]{chevance}. Other works have discussed the role of turbulence in GMC formation, finding it essential for filament formation \citep{federrathturb}, and support against gravitational collapse. Supernova explosions within clouds have been found to play a less important role in GMC destruction, as the first supernovae explosions seem to go off too late to affect their GMC, rather expanding into the local ISM \citep{smith2021}. Recently, STARFORGE \citep{starforge} has combined many of these processes into one simulation starting from GMC and investigating individual stars.

In considering regions in which most molecular clouds form in galaxies,it is well-known that our own Milky Way does not contain many massive clouds (i.e., greater than $10^6$ M$_{\odot}$). In fact, CO survey catalogues of the entire Galactic plane have found only 1064 massive clouds \citep{RiceGoodman2016}, with the outer disk's upper mass limit reaching only $10^6$ M$_{\odot}$. The same work finds that those more massive clouds are found in the inner disk of the Galaxy, closer to the dense, active Central Molecular Zone where higher line-widths suggest a more direct role from the galactic disk in the formation of clouds. Indeed, much higher surface densities, as can be found in luminous or starburst galaxies, are needed in order to form the massive star clusters that give rise to giant OB associations.  These galaxies host 1-2 orders of magnitude more molecular gas than our Milky Way and are capable of forming molecular clouds 1000 times more massive than GMCs in the Milky Way \citep{Solomon2001}. Therefore the local neighbourhood in the  Milky Way isn't ideal for simulating the processes that give rise to the most massive or numerous clouds in the galaxy. For the massive end of cluster and cloud formation, one requires a region with a much higher gas surface density than our own local galactic environment can represent. Thus, many of the previous studies we have cited likely fall short in investigating the typical massive cluster's formation.

In this paper, we investigate the sensitivity of the CNM to turbulence, gravity, and magnetic fields towards building self-gravitating giant molecular clouds for environments in which the bulk of massive molecular clouds are known to form, such as luminous starburst galaxies or the very active CMZ. In particular, we simulate a $500^3$ pc patch of the ISM in a high gas surface density environment to investigate the scaling relations dominated by ISM processes in these regions. The CNM on these scales is the starting point for the transition from diffuse, cold atomic gas to star-forming molecular gas. It is the initial state in our simulations that are designed to isolate the effects of CMN properties on structure formation. We present the argument for connecting the two scales in the following Sec.\ref{sec:bridge}. We present our models and the previous works that guided our setup in Sec.\ref{sec:methods}. A discussion of each important result from our models follows in Sec.\ref{sec:results}. Finally, we present our conclusions and the future of this problem in Sec.\ref{sec:conclusion}.

\section{Bridging scales}\label{sec:bridge}

Computationally, a plethora of different initial conditions for any scale from the 100 kpc-scale galactic disk to the pc-scale clusters have given us ample opportunity to investigate many of the intricacies of structure formation from different origins \citep[see, for example,][for simulations]{starforge, howard, brown, rieder, lane, lue, tigress}. Yet, even with all this work, there are lingering questions. For instance, is there a point we can disconnect the galactic disk and the GMC? How much influence does the disk have on star cluster formation, or the formation of individual massive stars within them? How many scales do protostellar jets, as well as feedback from massive stars, affect? We argue here that many of these questions can not be answered in isolation from the other, but rather that they require us to connect scales. 

\subsection{Filamentary Structures on Many Scales}

On molecular cloud scales, \citet{andre_paradigm} presented a comprehensive view - involving observations, theory, and simulations - on the central role played by filaments in star formation through their fragmentation into stars via gravitational instabilities. This picture shares the point made by the gravoturbulent theories of star formation \citep{klessen2016}, that supersonic turbulence creates the filamentary structure in the ISM \citep{maclow_klessen, larson}. However, it differs in emphasizing that when pushed to sufficiently high mass per unit length filaments become gravitationally unstable and fragment into star forming cores \citep{fiege_pudritz,higal, andre_paradigm}. The authors make the argument for the strengthening link between star formation and the structure of the ISM in galactic disks, motivated by this new paradigm of star formation in filaments. More recent reviews, such as \citet{chevance}, further support this view of star formation by exploring the filamentary environments we see surrounding GMCs from both observational and theoretical work. In either case, and among many other groups, there is general agreement that star formation is connected to filamentary structures in the (cold) ISM.

Observational studies strongly support the existence of filamentary structures across multiple scales. In both HI and CO, systems of filaments have been visible in the ISM, with the cold phase in particular being highly filamentary \citep[see][]{mcclure-griffiths, falgarone2001, andre2010, henning2010, motte2010}. The CNM provides a suitable initial condition for this work in a few ways. The gas in the ISM follows a cycle. Starting with the warm gas, thermal instabilities cause the gas to cool into the CNM and create giant, dense filaments \citep{hacar2022}. Out of those cold filaments, molecular gas begins to form \citep{silcc6}, which we can trace observationally using dust extinction due to dust growth being fastest and most abundant in cold gas \citep{klessen2016}. The CNM hosts most of the filamentary structures, especially those leading to and within dense molecular clouds \citep{silcc_zoom}. The molecular hydrogen forms on dust grains when it has reached a sufficiently high column density that it can shield itself from UV radiation. In our galaxy, the density required ($\sim 10^{21} cm^{-3}$) is also the density at which self gravity becomes important. Through gravitational collapse and instabilities discussed previously, molecular clouds create the bound clusters which host stars. Feedback from the stars heat the surrounding gas, completing the cycle and taking the gas from cold and molecular to its hot or warm phase once again \citep{silcc2}. Furthermore, this filamentary structure appears well before star formation starts \citep{andre_paradigm}. These filaments, independent of the observational tracer used, appear to be coherent structures, joining with each other to form dense pockets (referred to as hubs). These hubs are associated with groups of young stellar objects, or young clusters \citep{myers2009}, showing a physical connection between the cold ISM and the birth environments of massive clusters.

Magnetic fields may be playing an important role in the formation and evolution of filaments. The dense filaments show more linear structure and tend to be perpendicular to magnetic field orientation \citep{soler_planck}. On the other hand, lower density filaments are more scattered in their directions and tend to align parallel with magnetic field orientation (see also \citet{bfield_filaments}).

We can see small-scale, bound, star-forming structures form in the hubs at filament junctions, as well as along the densest filaments from gravitational instabilities. The gravitational instabilities in a fragmenting filament come about from the transition to self-gravitation, and can best be determined through filament line masses. When in the ISM though, we must also consider the physical affects of mixing and turbulence. In this case, the critical line mass includes both thermal and non-thermal contributions to the velocity dispersion:

\begin{equation}
    \sigma_{tot} = \sqrt{c_s^2 + \sigma_{turb}^2}
\end{equation}

\noindent This gives us a virial line mass, as discussed in \citet{fiege_pudritz}, defined as:

\begin{equation}
    M_{vir} = \frac{2 \sigma_{tot}^2}{G}.
\end{equation}

\noindent For magnetized filaments, (including the possibility of a helical magnetic field wrapping around a constant filament, in addition to the poloidal threading field) there is a magnetic correction to apply to the critical line mass calculation. Following from Equations 27 and 28 from \citet{fiege_pudritz}, the critical line mass becomes

\begin{equation}
    M_{mag, vir} = f_B*M_{vir} = \frac{1 + (v_A/\sigma)^2}{1 + (v_{A, \phi}/\sigma_c)^2} M_{vir}
\end{equation}

\noindent
using the poloidal Alfv\`en speed $v_A = \frac{B_{\parallel}}{\sqrt{4 \pi \rho}}$, toroidal Alfv\`en speed $v_{A,\phi} = \frac{B_{\perp}}{\sqrt{4 \pi \rho}}$, and critical velocity dispersion $\sigma_c^2 = 4\pi G \rho r^2$. Because of the explicit presence of both poloidal and toroidal field orientations in the above equation, one can see their respective effects. The poloidal magnetic increases the critical line mass, because it lends more pressure support to the gas. A toroidal field, on the other hand, squeezes the filament thereby decreasing the critical line mass. As such, it is the geometry of the magnetic field which affects line mass more so than its presence alone.

As derived, the thermal critical line mass scales in cold molecular clouds as $M_{crit} \approx 16 M_{\odot} pc^{-1} \times \bigg(\frac{T_{gas}}{10 K}\bigg)$. This gives a general minimum critical line mass, with the value changing depending on turbulent motions within denser molecular gas. In higher temperature environments such as the CNM environment, with an average temperature of 80K, we expect the average critical line mass to be $M_{crit, CNM} = 128 M_{\odot} pc^{-1}$, in atomic gas filaments - almost an order of magnitude larger than a cold molecular cloud at 10K. This suggests that larger scale structures more reminiscent of small cluster masses, can be expected to grow unstable within a CNM filament that is becoming gravitationally interesting, but not yet condensing into molecular gas. 

Supersonic turbulence in the galaxy, whatever it's source, obeys a size-line width relation $\sigma_{turb} \propto L^{1/2}$ \citep{hacar2022}. This is identical to the scaling of Burgers turbulence that arises from shocks. This scaling, when taken up to kpc scales, give line masses of the order $10^4 \rm{M_{\odot} pc^{-1}}$ indicating that massive atomic filaments can form molecular clouds via gravitational instabilities \citep{bopaper}.

While in theoretical examples the critical line mass gives us conditions for collapse, it is also supported observationally, making it a universally helpful criterion for structure formation in filaments. On the smallest scales characteristic of individual star formation, observations of \textsc{Herschel} filaments suggest that supercritical filaments are in virial equipartition and gravitationally bound. The subcritical filaments, on the other hand, are unbound and have transonic or subsonic velocity dispersions \citep{arzoumanian_2013}. In fact, subcritical filaments hold less than one third of the bound prestellar cores found in a molecular cloud, and those cores tend to be less massive and less dense than their counterparts forming in dense filaments \citep{polychroni}. Furthermore, the more massive bound cores found in filaments tend to be closer to intersections of filaments, supporting the idea of hub sections being the preferential site of massive cluster formation.

The hubs funnel the flow of gas, becoming dense and massive, and providing conditions for clustered star formation. These dense clumps will be the preferential sites for massive cluster formation, due to their connection to high rates of gas inflow from multiple filaments and their already dense environments \citep{myers2009}. Moreover, the \textit{Herschel} Gould Belt Survey found that the deeply embedded protostars (usually those which form massive stars) are found in filaments with column densities higher than $7 \times 10^{21} cm^{-2}$ \citep{andre_paradigm}. Even without limiting our view to only massive clusters and cores, we see that the vast majority ($>70 \%$) of \textit{Herschel}-identified prestellar cores are found within filaments as opposed to outside of filamentary structures, suggesting \citep[as seen in][]{menshchikov2010} that cores form along filaments via cloud fragmentation. 

In addition to gravitational instabilities giving rise to star-forming cores, the filaments act as conduits for gas flow into smaller scales. Observational works \citep[see][]{smith2012} have outlined an accretion flow along filaments onto cores, growing their mass over time. They have also suggested that a primary role of filaments is to help focus the gas towards embedded cores \citep{gomez2014}.  

However, recent simulations \citep[see][]{bopaper} have found higher accretion rates onto the filament than along them indicating more significant building up of material. This is also observed for accretion onto and flow along the filament in which the Serpens South cluster is forming \citep{kirk}, where infall rates onto filaments exceed flow rates along them by nearly a factor of 10. Regarding massive cluster formation, \citet{peretto2013} have found it possible that massive protostars accrete the majority of their mass from larger (filamentary) scales as opposed to just from their prestellar core. Dense cores also tend to share certain kinematic properties with their filaments, furthering the argument of gas flows along the structures.  It is therefore clear that filamentary structures in the ISM connect to multiple scales, from the cloud down to the single protostar, and the expected scales of instability in the filaments are approximately four times the scale length.

\subsection{The transition of the CNM to GMCs}

Given the importance of the filamentary network in the ISM, we now must ask what exactly a GMC is. Is there a distinct difference between ISM environment and GMC? \citet{chevance} argue that GMCs should not be thought of as discrete entities, but rather as an observationally defined feature in gas and dust maps. Perhaps the greatest representation of this fuzzy distinction are maps of the Orion A GMC. 

\begin{figure}
    \centering
    \includegraphics[width=0.98\linewidth]{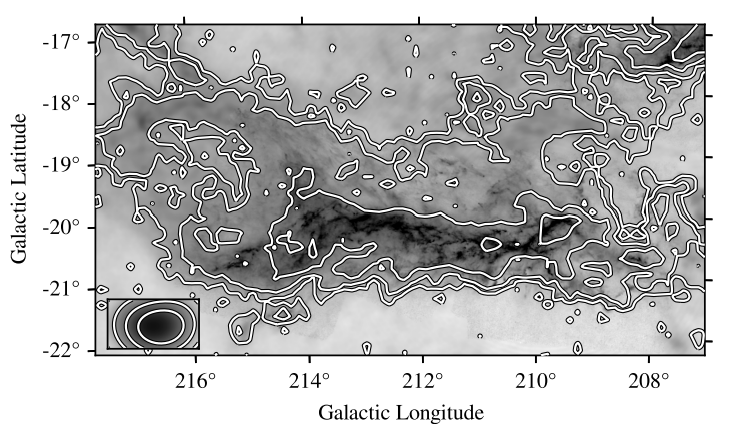}
    \caption{Image from \citet{chevance}. Orion A GMC dust emission map (grayscale) and $^{12}CO(1-0)$ emission (contour lines in position-position space.}
    \label{fig:orion}
\end{figure}

In Fig.\ref{fig:orion}, \citeauthor{chevance} show the dust emission and $^{12}CO(1-0)$ emission of the Orion A cloud together. Both tracers show what we would expect of a GMC: higher aspect ratio (>1), filamentary (or clumpy, in the case of CO contours) structure and overall connection to its environment. The largest distinction is with the resolution between the two maps. While the contour lines depict a specific contour demarcating the GMC, the dust emission shows much better the connection between GMC and environment through filamentary networks. This comparison is further complicated by the fact that, while CO emission certainly guarantees that the region is in a molecular phase of gas, it is not necessarily a tracer of molecular hydrogen in GMCs \citep{pineda2008}, making the CO emission view of a GMC an estimate of one based on resolution and limitations in the tracer itself. Thus, when looking only at gas emission, it is natural to assume a GMC as a structure which we can isolate from its environments in numerical studies. 

Dust emission, on the other hand, traces the neutral ISM, specifically the column density. Dust maps, being done with higher resolution and at sub-millimetre wavelengths where dust emission is optically thin, often allow us to resolve small-scale structure in our GMCs. With the benefit of recent GAIA surveys, dust emission is now also used to map out 3D structure of these clouds in a way that gas tracers cannot \citep{rezaei2020, Zucker_2021}. Thus, dust emission maps suggest that GMCs are inherently connected to their surrounding environment and, therefore, should not be isolated from their galactic disks to simulate their formation.

To summarize, the connection between filamentary networks and star clusters is strong. Furthermore, the GMC as a structure in \textbf{molecular} hydrogen is certainly a distinct step in the transition from neutral atomic gas to cold star-forming gas. In galactic scale simulations of cluster formation, the cold phase of the ISM gas is crucial in setting realistic cluster mass functions (CMF) on par with the Milky Way \citep{martapaper}. It is clear that dynamical processes in the CNM are important for the formation and final properties of massive star clusters and vice versa. This motivates our choice of conditions in the cold ISM as the initial condition of our work.

\section{Numerical Methods}\label{sec:methods}

Our simulations use \textsc{Ramses}, which is a magneto-hydrodynamics code \citep{ramses} popularly used for cosmological simulations \citep[for an example of star formation research, see][]{ramsescosmo1, decataldo2017, bopaper, BrucyHennebelle2021}. On the smaller ISM scale and below, the code has become increasingly popular to use due to its capability for high resolution and efficient run times, arising from its implementation of an adaptive time step. For example \citet{BellomiEcoGal, han, ntormousi, BrucyHennebelle2020, BrucyHennebelle2023} all use \textsc{Ramses} at parsec scales to investigate star formation with high resolution. Given the increasing use of the code on smaller scales, and the link this paper has to our larger galactic simulations (also done in \textsc{Ramses}), we choose to use it for our ISM scale work, as it sits between the aforementioned scales already tested. In order to achieve its high resolutions, \textsc{Ramses} uses adaptive mesh refinement, an important mechanism that we implement in our work in order to resolve our filamentary structure.

\subsection{Physical Mechanisms}

When investigating the CNM and molecular cloud formation, turbulence plays an important role in creating the structure. As we have already discussed, it is an essential addition to the consideration of critical line masses, influencing the gravitational collapse of filaments and thus pushing the subsequent structures that form to higher masses. Turbulent mixing from supernovae explosions specifically is most crucial in the cold and warm phases of the ISM \citep{tigress}, where we see velocity dispersions of 5-12 $km s^{-1}$ in simulation domains of size $0.5 \times 0.5 \times 2$ kpc. Dispersions of this magnitude are often associated with turbulence, as the typical turbulent rms velocity is $v_{rms} \approx 5 \frac{km}{s}$, which becomes the dominant speed compared to typical sound speeds of $c_s \approx 0.2-1.0 \frac{km}{s}$. Large-scale turbulent motions have also been found to have significant effects on the structure of the ISM at kpc scales, with the turbulent power spectra of the filaments and molecular clouds containing signatures from an imposed larger-scale power spectrum \citep{colman_ECOGal}.

Generally, magnetic fields are difficult to measure, especially in GMCs, due to the fact that most GMCs sit in or within only a few degrees of the Galactic mid-plane which makes line-of-sight magnetic field observations next to impossible with current technology \citep{pattle2022}. Yet, despite this fact, we have been able to uncover a substantial amount of information regarding the effects of magnetic fields from theoretical simulations as well as our relatively limited observations. Magnetic fields link gas across multiple physical scales, being most dynamically important around the pc-scale of a molecular cloud. On scales of 10 pc, molecular clouds have highly ordered magnetic fields, with the orientation of their internal structure closely aligning with the orientation of the fields \citep{pillai2020, pattle2022, tahani2022}. These highly ordered fields are indications that the field is strong enough to resist distortions due to turbulence.  They also provide support against compressive shocks, as well as general pressure support against collapse, thus delaying the onset of molecular cloud formation by more than 20 Myr \citep{silcc5}. On the other hand, though it will take longer to form molecular clouds, fragmentation in the ISM will happen sooner with strong magnetic fields due to the action of global Parker instability modes, and support long filaments that extend in either radial or azimuthal directions in the disk, as opposed to hydro cases which see predominantly ring-like structures in the filaments \citep{kortgen2019}.

Cluster formation will also be affected by radiation effects and galactic shear. Radiative feedback and stellar winds from massive stars will control the star formation rate through their influence on local gas properties, thus forcing clusters that form after to be lower mass\citep{silcc6}. Galactic shear has a loose correlation with local gas properties around molecular clouds \citep{sun}, while also having a moderately strong effect on the overall rotation of a cloud and, therefore, the angular momentum available to transfer from cloud to cluster \citep{jeffreson}. 

While we recognize the importance of both of these effects, we neglect them in this work for simplicity, as we focus on short times in the early formation and evolution of molecular clouds such that neither shear nor radiation will have significant effects on the dynamics of the gas. Our goal here is to focus on the sensitivity of molecular cloud and cluster formation to dynamical processes that shape the CNM - magnetic fields, turbulence, and gravity. The level of turbulence in the galaxy may be regulated by star formation feedback to a value that is able to offset global gravitational collapse \citep{KimOstriker2007, Kim_Ostriker2015}. The question we address here is what do different levels of CNM turbulence imply for filament and molecular cloud formation? 

In a separate study of molecular clouds in a galactic disk, presented in \citet{bopaper}, we include full galactic environmental effects, including galactic shear, and investigate the role of galactic shear on the clouds. In future work from that paper, we will further resolve the cluster scale and compare to the present work, including the effects of radiation.

\subsection{Simulations}

\subsubsection{Initial Conditions}
We simulate the CNM via a turbulent box setup for two general cases; a magnetic and non-magnetic case. We start with a 0.5kpc $\times$ 0.5kpc $\times$ 0.5kpc box, to match the size of the PHANGS-ALMA hexagonal kpc-scale observations \citep{sun} as well as being on par with similar theoretical works \citep[cf.][]{BellomiEcoGal, colman_ECOGal, tigress}. This size is chosen such that it can contain entire cloud complexes, with GMCs on the scale of 70 pc, as well as the properties and physical mechanisms acting in the local environment, such as those discussed above. However, we do note that this will not account for the gravitational potential in a galaxy due to its stars and dark matter, instead only allowing a central collapse by assuming no initial distribution in the potential. The origin is set in the center of the box, such that half of the z-dimension of our box can be imagined as containing 250 pc both above and below the plane, similar to pressure scale heights of CNM gas \citep{kortgen2019, soler2022}.

We choose a domain size and surface density to align with the lower end of Figure 2a from \citet{WilsonElmegreen2019}, such that a domain of half height 250 pc places our CNM gas at the less luminous end, more similar to that which would be present in a starburst galaxy. Scatter in these numbers show scale heights of up to 316 pc, so our choice of 250 pc places us within a very reasonable assumption of scale height. This idea is further supported in observational works which find scale heights of 270 pc minimum for low SFE and full vertical disk thicknesses greater than 0.5 kpc \citep{FisherBolatto2019, BournaudElmegreen2009}. Additionally, previous works discussed in \citet{FisherBolatto2019} argue that other methods of kinetic energy injection, such as supersonic turbulence, would increase these scale heights even more. So our domain size is perfectly placed within reasonable values for our gas surface density. We discuss our choice of gas surface density further on.

While a larger box size would be possible with a more explicit disk setup in the density profile, and simulations of collimated flows are popular for this very reason \citep[see, for example, the SILCC projects:][]{silcc2, silcc5, silcc6, silcc_zoom}, they require additional manual setup in the initial conditions to the velocity and angular momentum. This is something we wished to avoid in our project, as we wanted to isolate some of the most basic physical processes that regulate the CNM and under which conditions it gives rise to cluster forming filaments. Due to both the possible vertical extents of gas and the simplicity we aimed for in our initial conditions, we keep the full extent of our z-direction 0.5 kpc, making it symmetric with the rest of the box.  

Observations of pc scale star-forming filaments in molecular clouds have shown typical widths of 0.1 pc, scaling with gas temperature \citep{arzoumanian, 2011pineda, 2010pineda}. Given these results, we allow our box to refine up to a highest resolution of 0.48 parsec per cell, which is achieved consistently throughout our dense filaments, so that we highly resolve our clouds but do not resolve individual star-forming filaments and our sink particles can act to represent protoclusters. Our simulations employ periodic boundary conditions. This allows for some effects of galactic gas flow through the simulation volume, without additional computational expense. We do not include shear across our box. Thus, local angular momentum arises purely from the local angular momentum that comes about from turbulent mixing and shocks moving through the medium.

Since we do not include shear or any effects of galactic rotation in this box, we use a mixture of solenoidal and compressive turbulence to simulate general motions and mixing in a patch of the ISM and allow for filament formation within our box. The turbulence is initialized with a 1/3 compressive fraction. We allow our turbulence only to decay, and keep driving forces turned off such that the turbulence is never evolved past initialization, as motivated by previous work such as \citet{howard}. Finally, the power spectrum follows a Burger's power-law shape as supported by previous studies of supersonic turbulence in the ISM and GMCs \citep{federrathturb, algorithmPDF}, given by the energy spectrum $E(k)$ of the turbulence, where k is the wavenumber:

\begin{equation}
    E(k) \propto k^{-2}
\end{equation}

\noindent wherein the velocity dispersion is $\sigma_{turb}^2(k) = \int E(k) dk \propto k^{-1}$.

In setting up the turbulence in a simulation, RAMSES has effectively two control parameters that need to be set for the strength of the initial turbulence field: the source term of the kinetic energy (called the "RMS forcing parameter"), and the decay time scale. By selecting various decay timescales of our turbulence, the initial velocity dispersion measured in our simulation can be fixed.

Given the use of decaying turbulence, one must be careful with the decay timescale and the timescale of structure formation in the simulation. In order to achieve the 3D velocity dispersions we find necessary in this work, the decay timescales must be set sufficiently long enough to maintain the desired dispersion for structure formation. We then run our simulations for a maximum of 10\% of their decay time in order to ensure turbulence has not significantly decayed throughout formation, as well as to remain sufficiently early enough in protocluster formation that our assumptions surrounding the lack of radiative transfer can remain valid. As such, decay timescales are set to be 100, 65 and 50 Myr which create dispersions of 5 km/s, 8 km/s and 10 km/s in our models, respectively.

Figure \ref{fig:turbspectra} in Appendix C shows the turbulent energy spectra for times of 1.90 and 3.79 Myr in our disp8MHD simulation. Early times show the turbulent cascade has not reached Burger's scaling for the majority of the scales and oscillates around it at small scales. The spectrum at 3.79 Myr shows the model has reached Burger's scaling at large and small scales, while the middle of the spectrum trends closer to Kolmogorov scaling. This transition to Kolmogorov scaling is expected for later times as magnetic energy increases \citep{Schleicher2013}. The differences between early and late time do not correspond to a significant decay fraction in the energy spectra as there are no appreciable differences in upper energy magnitudes nor any deviation away from Burger's scaling at the small scales with time. In reality, the simulations run up to the point their oldest sink particle reaches 0.5 Myr in age, in order to avoid feedback and radiation effects that are here neglected. Finally, it is this decay time scaling which affects the velocity dispersions in our runs. We stress that velocity dispersion is not a parameter in the setup of our simulation explicitly, and the values we are studying are direct consequences of the strength of the turbulence initialized in the simulations.

\begin{table*}
\centering
\resizebox{1.0\linewidth}{!}{%
\begin{tabular}{|l|l|l|l|l|l|l|l|l|l|l|l|}
\hline
\textbf{Run} & \textbf{Density($H/cm^3$)} & \textbf{Box Size (kpc)} & \textbf{Initial Temperature (K)} & \textbf{$\Delta x_{start}$ (pc)} & \textbf{$\Delta x_{final}$ (pc)} & \textbf{$\sigma_{turb, initial}$ (km/s)} & \textbf{$\sigma_{turb, sink}$ (km/s)} & \textbf{B field ($\mu G$)} & \textbf{$\lambda_{J, thermal}$ (pc)} & \textbf{$\lambda_{J, total}$ (pc)} \\[7pt] \hline
\textit{\textbf{disp5}}              & 30.0                       & (0.5,0.5,0.5)           & 58.8                             & 1.95                             & 0.48                           & 2.72                                     & 5.51                                & 0.0         & 17.44  & 48.23               \\[4pt]
\textit{\textbf{disp8}}              & 30.0                       & (0.5,0.5,0.5)           & 58.8                             & 1.95                             & 0.48                           & 6.96                                     & 6.99                               & 0.0        & 17.44   & 123.3                \\[4pt]
\textit{\textbf{disp10}}             & 30.0                       & (0.5,0.5,0.5)           & 58.8                             & 1.95                             & 0.48                           & 10.5                                     & 11.5                                & 0.0        & 17.44   & 186.2                \\[4pt]
\textit{\textbf{disp5MHD}}           & 30.0                       & (0.5,0.5,0.5)           & 58.8                             & 1.95                             & 0.48                           & 2.72                                     & 3.2                               & 7.0        & 12.16   & 48.23                \\[4pt]
\textit{\textbf{disp8MHD}}           & 30.0                       & (0.5,0.5,0.5)           & 58.8                             & 1.95                             & 0.48                           & 6.96                                     & 7.0                                & 7.0        & 12.16   & 123.3                \\[4pt]
\textit{\textbf{disp10MHD}}          & 30.0                       & (0.5,0.5,0.5)           & 58.8                             & 1.95                             & 0.48                           & 10.5                                     & 10.26                                & 7.0       & 12.16    & 186.2                    \\[4pt] \hline
\end{tabular}%
}
\caption{Parameters for all models computed. The first three simulations represent our fiducial models with no magnetic fields and maximum resolution of 0.48 pc. We set our magnetic field runs to have a field strength of 7 $\mu G$, in accordance with average values of ISM magnetic field strengths. $\sigma_{turb,initial}$ and $\sigma_{turb, sink}$ represent the velocity dispersion in the simulations on initialization and once sink particles have formed, respectively. $\lambda_{J, thermal}$ and $\lambda_{J, total}$ give the initial Jeans lengths of each simulation calculated using the average temperature and the velocity dispersion, respectively.}
\label{tab:models}
\end{table*}

Table \ref{tab:models} lists the initial conditions we use in each of our models, and we outline here the decisions behind these parameters. In order to mimic the magnetized CNM we initialize a constant magnetic field of magnitude $7 \mu G$ in only the y direction, allowing it to evolve with the gas over time. We choose this setup because large scale ordered magnetic fields in galaxies are toroidal in the plane, for which a small patch will look approximately constant and oriented in one direction. The gas is initially isothermal and set at a density of 30 $cm^{-3}$ and a temperature of 58 K, comparable to median values for the CNM, as outlined in Table 1 of \citet{klessen2016}. We then evolve the chemistry as described in the following subsection. These values give surface densities of $320 M_{\odot} pc^{-2}$, which are not dissimilar to surface densities within the first few hundred parsec of the Milky Way disk, otherwise known as the central molecular zone (CMZ). Gas surface densities in this portion of the disk can be on average $10^2 - 10^3$ M$_{\odot}$ pc$^{-2}$ for the atomic gas, especially in its early evolution \citep[see Figures 5 and 9 of][]{TressSormani2020}. Additionally, this chosen surface density reflects the more active starburst and luminous galaxies, in which one expects higher gas densities, on the order of $10^2$-$10^4$ M$_{\odot}$ pc$^{-2}$, due to the much more star-forming regions these galaxies represent \citep{WilsonBemis2023, WilsonElmegreen2019}. 

\subsubsection{Heating \& Cooling}

Through the chemistry code \textsc{Grackle} \citep{grackle}, we include non-equilibrium chemistry with a 9-species network to form both atomic and molecular hydrogen. Grackle includes chemical heating and cooling, as well as metal line cooling, radiative cooling and $H_2$ formation on dust grains (which introduces dust cooling and dust-gas heat transfer). Through the addition of molecular hydrogen formation, Grackle also initializes a default ISRF strength of $2.72 \times 10^{-3}~erg~s^{-1}~cm^{-2}$, but we do not include radiative transfer from \textsc{ramses} nor do we include a UV background. Due to the lack of UV background and radiation in the simulation, explicit self-shielding of $H_2$ is ignored. 

A constant photoelectric heating rate is used such that at a hydrogen number density of n=1 cm$^{-3}$, the heating rate is $8.5 \times 10^{-26}$ erg cm$^{-3}$ s$^{-1}$. Because we do not include any radiative transfer due to computational limitations, there is no photodissociation. Our molecular hydrogen abundance therefore represents an upper limit, as destruction of the molecule is not accounted for in the abundances.

In our simulations we argue this is reasonable, as the photodissociation rate for gas at temperatures below 10000 K is sufficiently low as to be reasonably ignored in the vast majority of our gas \citep{CoppolaDiomede2011}. For gas below densities of 10$^4$ cm$^{-3}$, the probability of dissociation via UV is only 15\% and, even in those higher density areas, a sufficiently high column density can contribute to an order of magnitude drop in the dissociation rate (such column densities translate to $\sim$ 10$^{21}$ cm$^{-2}$ for the diffuse ISM) due to self-shielding. In many cases the dissociation rate of H$_2$ is not so high as to make it completely unavoidable in cases where one is not studying in detail the chemistry of the gas \citep{klessen2016}.

Furthermore, given the inverse relationship between density and photodissociation rate, combined with the timescale of destruction being on the order of 1 Myr or higher for dense, optically thick gas \citep{AbgrallLeBourlot1992}, we determine that the inclusion of photodissociation would not alter the dense gas enough to significantly affect our results. The overall timescale of formation of H$_2$, in order to have enough present that its destruction would be necessary, is on the order of 10$^9$ yr, with only an order of magnitude decrease in supersonically turbulent environments \citep{klessen2016}. This is comparable to the crossing and dynamical timescales of our simulations, both of which we do not model for any large fraction of their scale.

Overall, self-shielding of H$_2$ plays a large role in CNM clouds with no nearby massive stars, such as we are modelling here, and \textsc{Grackle} does perform estimations of self-shielding tied to H$_2$ formation on dust grains, even without radiative transfer or UV sources. We conclude that our molecular hydrogen content acts sufficiently as an upper limit to the true abundance one would have. To account for this being an upper limit, we henceforth discuss molecular hydrogen through column density cuts of 10$^{21}$ cm$^{-2}$ and above. In the case of volume density discussions, we have used the column density cut to determine an average volume density of 100 cm$^{-3}$ and higher for the `molecular hydrogen'. Finally, we are tracking the formation of molecular hydrogen in these simulations so we follow its rapid formation by starting with an entirely atomic gas that converts to molecular quickly.

\subsubsection{Clumpfinding \& Sink Particles}

At any time during the simulation run time, the gas can form both clumps and sink particles. We define our clumps with a minimum density of 100$cm^{-3}$ and a minimum mass of 100$M_{\odot}$, such that they represent molecular gas clumps and GMCs (with enough mass gained over time). These are formed through the 3D clumpfinder algorithm in \textsc{Ramses}, outlined in \citet{phew}. As a brief overview, this algorithm works by identifying density peaks in a data set using a peak-based approach. Peaks with significant height compared to their valleys, defined as topological relevance, are identified and labelled, while nearby peaks with lower relevance are merged into the same clump as the larger, given a certain saddle ratio between them. This algorithm allows us to identify and label larger, extended structures that cannot be accurately represented by a particle, and keep track of their properties. In our simulations, this translates well to tracking the clouds in the complex we form from our CNM, allowing for a matching between the clusters that form and which clouds they form within. 

The clusters form via sink particles, initialized at very low masses in order to allow the sink's mass to be a product of accretion using a density threshold scheme, outlined in \citet{ramses_sinks}. Starting with very low masses ensures that we follow the formation of the protoclusters as early as possible, such that sink particle ages correspond very closely to the age of the protocluster and we can accurately limit our runtimes to timescales well before star formation and feedback would take place. Sink particles are a useful catch-all particle that can easily represent any point-like density peak, given the appropriate sub-grid physics is applied, so we have ample freedom in our code to allow them to represent clusters. In general, the agreed upon rules for sink formation are those outlined in \citet{federrathsink}. These rules outline conditions for density thresholds, refinement, boundedness checks, gravitational potential minima, stability, and location. \textsc{Ramses} adopts the same general conditions of sink particles, but simplifies the rules. Sink particles are formed from the clumps found in clumpfinder, which asserts they form within already dense structures, and we set our sink formation density threshold to 5000$cm^{-3}$ to ensure we are only picking cluster candidates out of the star-forming gas density peaks within a clump. 

Once sink candidates are identified, they undergo 3 checks. First is the virial check, in which the code checks if the gas that would be contained in the sink is able to gravitationally collapse. More precisely, the virial check analyzes the gravitational field at a possible site for sink formation, ensuring it is universally compressive and strong enough to overcome internal pressure support. Second, the collapse check verifies that the gradient of the velocity is negative across all principle axes, ensuring that not only is the gas collapsing, but it is contracting. The final check is the proximity check, which ensures that gas which is being accreted by a sink particle cannot form its own sink particle. Contrary to the outline of \citet{federrathsink}, \textsc{Ramses} does not carry out separate tests for boundedness, Jeans instability or gravitational potential minima. We refer the reader to the original methods paper for sink particles in \textsc{Ramses} for more detailed explanations \citep{ramses_sinks}. We also note that discarding these tests allows for a more general implementation of the existing tests. For example, while virial checks give valuable information on gas collapse, the authors point out it is possible for gas to exist in virial equilibrium but not be collapsing. They then argue that their virial check approaches both collapse and boundedness by being implemented more generally.

We split our simulation runs into two groups: those with and without magnetic fields. Within these two groups, we further split it into three runs with different turbulent velocity dispersions. Velocity dispersions are set via the turbulent auto-correlation time \citep{federrathturb} and the rms acceleration \citep{schmidt2009}, defined as $f_{RMS} = 3v^2 / L$. These represent the parameters through which one can set an initial turbulent velocity. The auto-correlation time sets the amount of time needed for one wave or shock to cross half the length of the box, such that it is half of the crossing time. The rms acceleration, on the other hand, sets the amplitude of the turbulence by setting the acceleration of the material. Due to their dependence on the rms velocity, the two must be set to determine the velocity field within the simulation at startup. We set these to correspond to speeds approximately double the desired dispersions, found through trial and error for all other parameters of our box already set, such that we would reach the desired dispersions at $10\%$ of the global crossing time, or 1 local crossing time for a 100 pc patch. Actual dispersions achieved at initialization ($\sigma_{turb, initial}$) and at the onset of sink formation ($\sigma_{turb,sink}$) are given in Table \ref{tab:models}. In the following section we compare results from the different dispersion values as well as between hydro and magnetic field runs. 

\section{Results}\label{sec:results}

One of the most important physical attributes of any phase of the ISM is the velocity dispersion that it supports. As is discussed in \citet{klessen2016}, there is a theoretical range of velocity dispersions (both in 1 dimension and 3) which can be present in the ISM, specifically depending on the phase of the gas. From Table 4 of \citet{klessen2016}, the velocity dispersion of the gas decreases as we transition from atomic to molecular hydrogen. Assuming a scale height of 0.5 kpc in the solar neighbourhood, the HI gas can display velocity dispersions of $\sim 12 km~s^{-1}$, whereas molecular gas may have velocity dispersions $\sim 5 km~s^{-1}$. Taking the average, assuming completely equal portions of HI and molecular hydrogen gives a 3D velocity dispersion of $8.5 km~s^{-1}$.  What is the origin of these values?  This question motivated our first set of numerical experiments.

\subsection{Self consistent turbulent amplitudes in the CNM}

One approach to understanding turbulent amplitudes is to assume that the ISM achieves a hydrostatic balance between turbulence gas pressure resulting from stellar feedback, and the weight of the gas \citep{Kim_Ostriker2015}. Too little turbulence and gravitational collapse of the medium ensues - while too much will heat and disperse it. The velocity dispersion predicted by this hydrostatic balance condition is 
\begin{equation}
       \label{eqn:balance}
      \sigma_z^2 = \frac{\pi}{2} \frac{G \Sigma^2}{\rho}
\end{equation}

\noindent where $\sigma_z$ is the vertical component of the velocity dispersion, $\Sigma$ is the average column density of the gas, and $\rho$ its average volume density.

We first chose initial velocity dispersions near the predicted values of \citet{klessen2016}. We ran a large number of simulations for a box without periodic boundary conditions. For this setup high velocity dispersion strongly shock heats the gas and disperses gas out of the box, while too low a velocity dispersion leads to general collapse. After some trial and error, we find minimum 3D velocity dispersions of $\sim 5 km~s^{-1}$, and a maximum of $\sim 10 \rm{km~s^{-1}}$, in order to create a filamentary ISM, similar to \citet{JoungMacLow2009}. 

With these experiments completed, we then used these initial conditions in periodic box simulations, which we show in all subsequent figures. Additionally, to consider the formation of dense filaments and clusters, we further limit that range between 8 and 10 $km~s^{-1}$, due to the fact that our 5 km/s run experiences a very slow global collapse. This makes structure transient, as the turbulent shocks are not strong enough to maintain sharp filamentary structure. The formation of filaments and subsequent clusters is therefore happening on much longer timescales. We find that the 5 km/s simulation fails before reaching its burst of sink formation due to overcooling of the completely empty edges as the entire simulation globally collapses. For these reasons, we do not take the 5 km/s models into account for any of our conclusions surrounding sink particles. We still include said models in the discussion for completeness.

Our numerical results are also reflected in observational work such as HISA observations of GMF38.1-32.4a \citep{WangBihr2020}. These cite velocity dispersions in HI of $\sim 4-6 km~s^{-1}$, with some regions reaching  in $^{13}$~CO as high as 10$km~s^{-1}$. Averaging between HI and $^{13}$~CO gives dispersion ranges of 6-12$km~s^{-1}$.

\begin{figure*}
    \centering
    \includegraphics[width=0.9\linewidth]{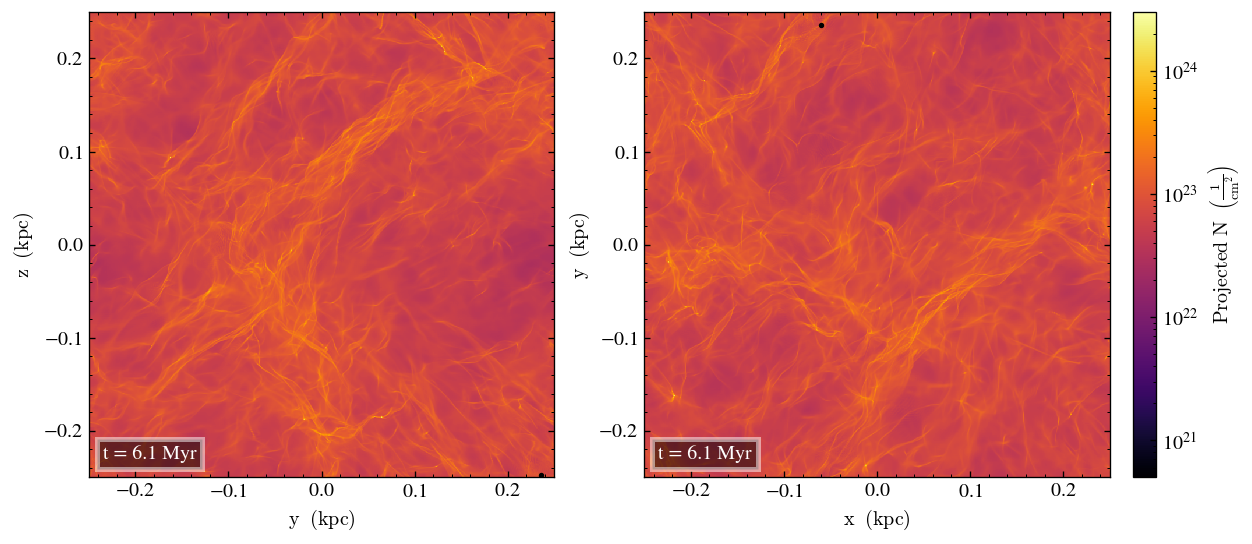}
    \includegraphics[width=0.9\linewidth]{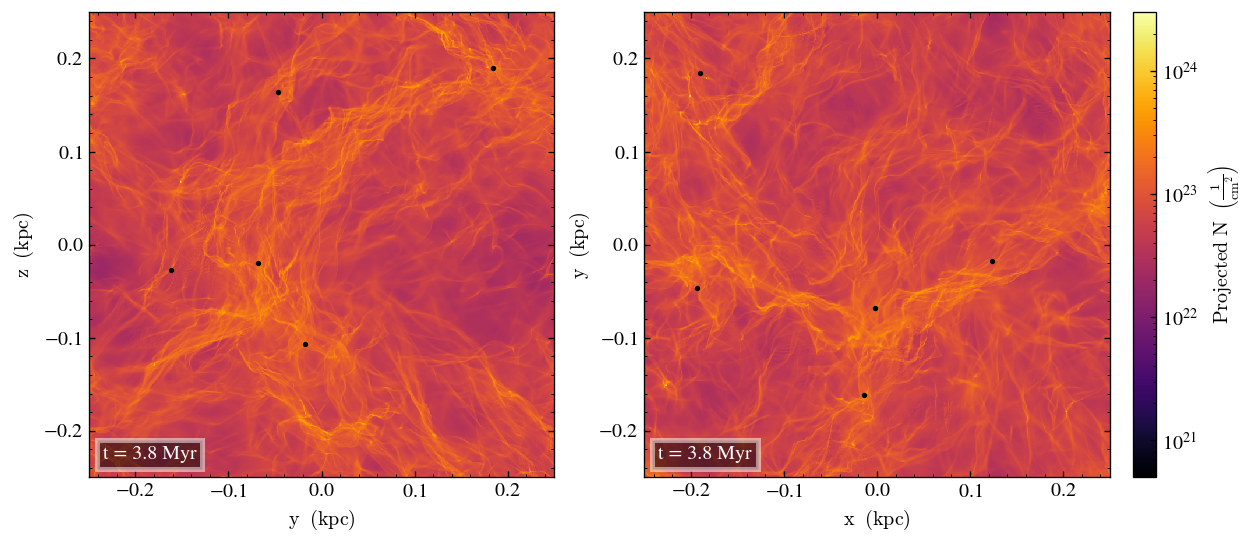}
    \includegraphics[width=0.9\linewidth]{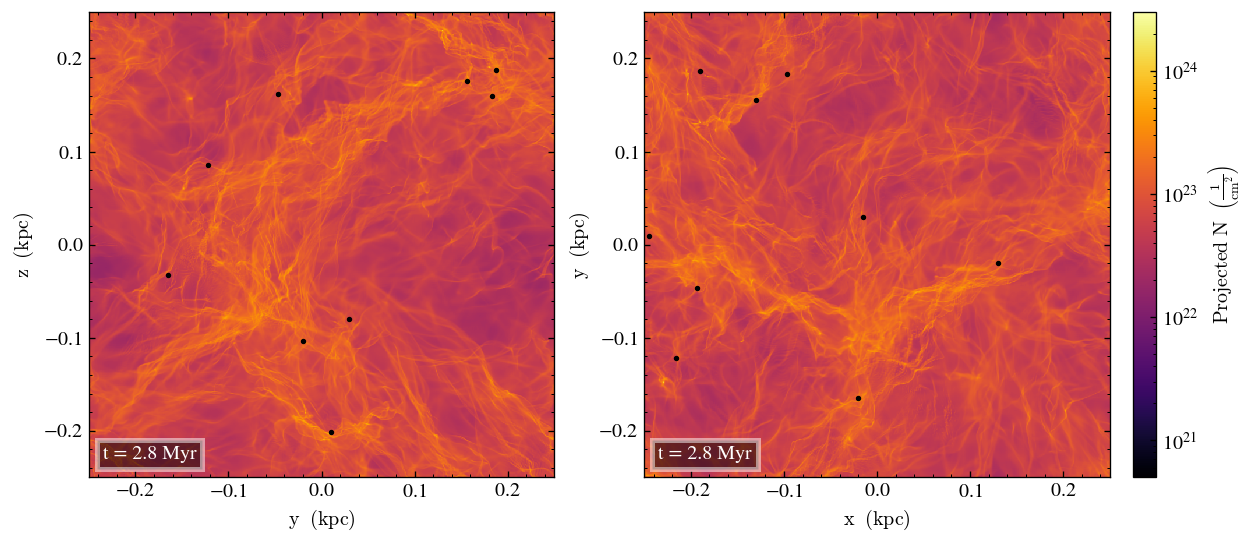}
    \caption{Column density projections of our hydro runs once the first sink particle has formed. \textit{Top:} Model for our 5 km/s dispersion case at 6.1 Myr. Projections in both the x and z plane are shown, with the center of our box centred at (0,0). \textit{Middle:} Column density projection of the 8 km/s dispersion case at a time of 3.7 Myr. \textit{Bottom:} Projections for our 10 km/s dispersion case at 2.8 Myr. Sink particles are shown as black filled circles.}
    \label{fig:Dispersions}
\end{figure*}

Figure \ref{fig:Dispersions} shows the column density plots in both the x and z plane for each of our hydrodynamic runs at the beginning of cluster formation. While the plots of the 8 km/s and 10 km/s runs give very similar structure, the properties of the structure are different. With a dispersion of 8 km/s, the structure is sharper, due to the self-gravity of the filaments being strong enough to pull them together, yet not so strong as to dominate completely and collapse all the gas. In our 10 km/s run, the structure becomes more diffuse looking. In our 8 km/s models, the ratio of kinetic to gravitational energy is 0.01, indicating turbulent motions play a dynamic role in the formation of the structure. On the other hand, as this global ratio reaches 0.03 for our 10 km/s models, we see turbulence overcome gravity on scales of $\sim 100 pc$, creating more transient structures. This higher proportion of turbulent energy to gravitational potential energy contributes too much mixing throughout the box, creating structure that cannot be pulled together by gravity as much as in the 8 km/s case. A dispersion of 5 km/s, on the other hand, displays the opposite. In this case, the gravity is far stronger than the turbulence can be, not allowing it to create nearly as much structure. The gravitational potential dominates the gas, causing a global collapse towards the centre of the box, still creating structure, though with far shallower density fluctuations than our more turbulent setups. Based on structure alone, it is clear that a dispersion of 8 km/s can create a well balanced and realistic ISM slice. 

\begin{figure*}
    \centering
    \includegraphics[width=1.0\linewidth]{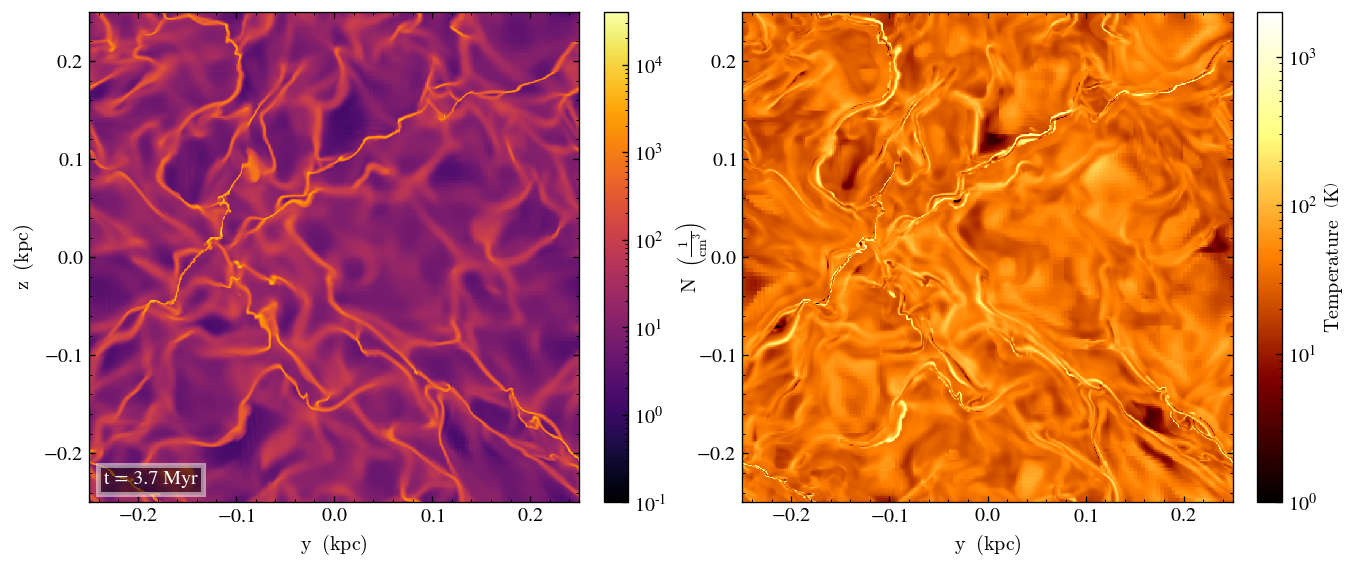}
    \caption{Slices through the center of the box of our disp8 model with density (left) and temperature (right) plotted at 3.7 Myr, just before sink formation begins. Sharp filamentary features are seen in the density, mirrored in the temperature slice, where hot gas outlines the edges of filaments. This slice has an average density of $\sim 10^2 cm^{-3}$ and an average temperature of $\sim 10 K$, lending an average sound speed of $0.3 km s^{-1}$ throughout the simulation.}
    \label{fig:densitytemperature}
\end{figure*}

If one compares each of these cases to Figure \ref{fig:orion}, it can also be seen that the 8 km/s case creates structure much more akin to observations of Orion A, specifically the structure visible in the 250$\mu m$ dust emission. Figure \ref{fig:densitytemperature} shows densities and temperatures in a slice of the 8 km/s model, in which we can see the densest filaments rest in temperatures $\sim 10 K$. We see higher temperatures ($\leq 600 K$) at the boundaries of the filaments, from shock waves creating higher temperature shock fronts. Our filament temperatures compare well with observations of the Orion A GMC, which has temperature ranges of $\sim 10-30 K$ \citep{lombardi_herschel}, while the hot, diffuse gas surrounding the filaments would likely be too diffuse to be observed. We note some cold gas of temperatures $\sim 1 K$, tending to appear in low density pockets of the gas. Given the placement of these cold pockets, we conclude that their presence is due to shock waves emptying the area of gas, pushing it into the hot boundaries of the filaments. The average temperature of the entire simulation is $10 K$. The presence of a large range of temperatures (spanning $1-1\times 10^3 K$), emphasizes the importance of cooling processes that give rise to a multiphase ISM in simulations of filamentary structure, most especially at the current scale. Additionally, in Fig. \ref{fig:Dispersions}, we notice a large structure just below the center on the 8 km/s density projections. The size and components of gas present in this structure are similar to Orion A, which is 90 pc in length and has approximately 45\% of its gas sitting at high column densities \citep{3dshape_orion_gaia}. We further discuss comparisons between our models and observations of Orion A in the following Sec. \ref{moleculargas}. 

Finally, we note that using our quoted values for $\Sigma$ and $\rho$ in Equation \ref{eqn:balance}, we find $\sigma_z \simeq 30  \rm{km \: s^{-1}}$. This is at least a factor of 3 higher than our own simulations suggest, yet matches well with extragalactic observations such as \citet{WilsonElmegreen2019} who find dispersions of 30-80 km/s for unresolved galaxies. Given $T \propto \sigma_{tot}^2$, a velocity dispersion of 30 km/s gives temperatures of more than $10^4$ K, representing highly shocked gas that will quickly destroy the CNM. In fact, our own simulations do see some of the shocked gas create high temperatures over 1000 K in Fig. \ref{fig:densitytemperature}.

However, we note that all the simulations of \citet{Kim_Ostriker2015, KimOstriker2013, KimKim2011} also have velocity dispersions (see Figure 2 of their 2015 paper) of 5-10 km/s, in excellent agreement with our own values. They use setups typical of the local ISM with gas surface densities of $\approx 1 - 10~\rm{M}_{\odot}/\rm{pc}^2$ and an effective scale height of $\sim 60$ pc on average \citep{Kim_Ostriker2015}. However, they also report that these velocity dispersions seem to be somewhat independent of the surface density used; increasing $\Sigma$ by a factor of 10 \citep[see Figure 11 in][]{KimOstriker2013} made no difference to their resulting velocity dispersion. Our simulations, sitting a factor of 10 higher in column densities than theirs also agree with this finding. Thus, Equation \ref{eqn:balance} appears to be only a rough guide to the link between velocity dispersion and gas column density with large deviations possible. Taken together, all of these results point towards a more general idea that velocity dispersions in the ISM are a consequence of the gas phases themselves.

\subsection{Molecular gas and cloud morphology} \label{moleculargas}

\begin{figure*}
    \centering
    \includegraphics[width=1.0\linewidth]{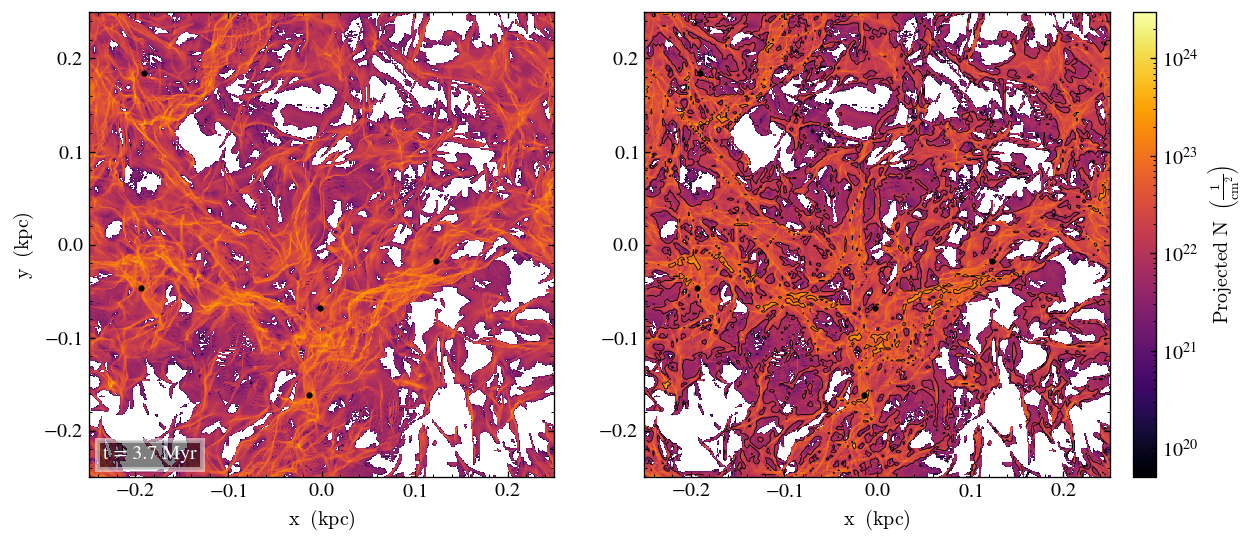}
    \caption{Side-by-side comparison of the molecular gas structure for the disp8 model. \textit{Left:} Z-plane projection of gas, cut to contain only that above $10^{-21} g\cdot cm^{-3}$. Maximum resolution is 0.48 pc, chosen to approach typical filament widths of 0.1 pc. \textit{Right:} Same, with added contours to simulate observations of molecular gas structures. Contours are drawn from $10^{20} cm^{-3}$ to $10^{23} cm^{-3}$ to mimic resolution range of \citet{lombardi_herschel}. Sink particles are denoted by filled black circles.}
    \label{fig:moleculargas_cut}
\end{figure*}

In our models we define our molecular gas at densities of 100 $cm^{-3}$ or higher in order to best compare to dust emission maps of molecular clouds, which contain no chemical information about the molecular gas. Using our 8 km/s models, we can consider the masses of gas at densities of 100, 1000 and 10000 $cm^{-3}$ to determine amount of molecular and star forming gas in our model. Table \ref{tab:gascuts} shows these masses. At the time of formation of the first sink particle, we see $\sim 50\%$ of our gas mass lies at densities of 100 $cm^{-3}$ or higher, showing that a significant fraction of our gas is tied up in dense structures that will eventually form stars. Our highest density cut, representing gas which has surpassed star-forming densities and, therefore, the threshold for our sink particle formation, contains only $0.2\%$ of our total gas mass, showing a significant jump from molecular gas densities to star-forming densities. From this, we can see a considerable amount of gas in the CNM will become molecular gas, as we can also see in Fig. \ref{fig:moleculargas_cut}, but very little of that will continue to climb in density to eventually form stars.

\begin{table}
\centering
\begin{tabular}{c|c|c|}
\cline{2-3}
\textbf{}                                                      & \textbf{Hydro}   & \textbf{Magnetic Fields, B=7$\mu$G} \\ \hline
\multicolumn{1}{|c|}{\textbf{$M_{T}  (M_{\odot})$}}             & $9.3\times 10^7$ & $9.3\times 10^7$         \\ \hline
\multicolumn{1}{|c|}{\textbf{$M_{100}  (M_{\odot})$}}           & $4.9\times 10^7$                 & $4.7\times 10^7$         \\ \hline
\multicolumn{1}{|c|}{\textit{\textbf{$M_{1k}  (M_{\odot})$}}}  &  $1.2\times 10^7$                & $1.4\times 10^7$         \\ \hline
\multicolumn{1}{|c|}{\textit{\textbf{$M_{10k+sink}  (M_{\odot})$}}} & $1.9\times 10^5$                 & $4.7\times 10^5$         \\ \hline
\end{tabular}
\caption{Comparison of gas masses between the hydro and magnetic field cases for our disp8 model. From top to bottom we give total gas mass, and total masses for gas above densities of 100, 1000, and 10000 $cm^{-3}$, respectively. For density cuts of 10000 $cm^{-3}$ we include total mass in sink particles as well.}
\label{tab:gascuts}
\end{table}

We also note a large complex of molecular clouds just below the center of the box, extending roughly 100 pc in length and a maximum of 30 pc in breadth. We find this size is comparable to that of the Orion A GMC in Fig. \ref{fig:orion}. The similarities between the two also extend into the contours tracing molecular gas. In both figures we see the majority of the molecular cloud complex outlined by one contour, with the resolution of the contours unable to depict any of the fine filamentary structure creating the complex. Additionally, this is where we see many of the sink particles form, indicating it is the primary area of star-forming gas. We emphasize that while there are similarities with the Orion A cloud properties, the environment of our simulated cloud is very different. Our point is just that the formation of GMCs with recognizable properties is natural in a turbulent CNM setting.

\begin{figure*}
    \centering
    \includegraphics[width=1.0\linewidth]{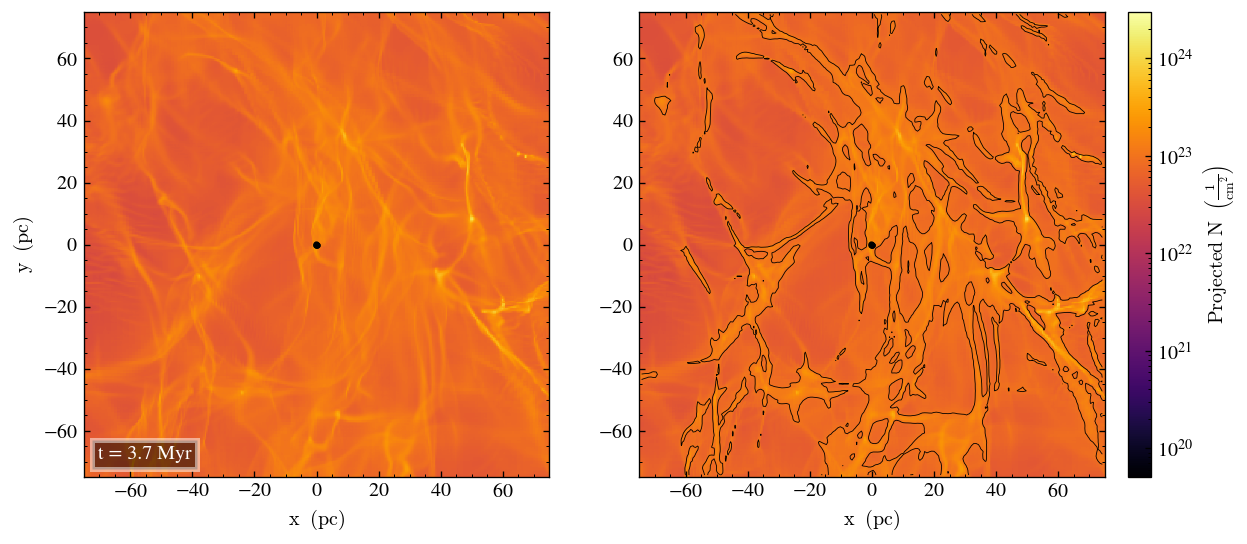}
    \caption{Zoom-in of contoured region in Fig. \ref{fig:moleculargas_cut}. Density projection and contours are the same, though plotting has been recentered on the center of the location and the domain plotted is 150 pc. Sink particles are represented by filled black circles.}
    \label{fig:moleculargas_zoom}
\end{figure*}

We provide a zoomed in look at the molecular cloud complex noted above in Fig. \ref{fig:moleculargas_zoom}. While the contours break up into smaller clumps, they still closely trace the filamentary structure and outline a cloud sized complex of molecular gas. For the purposes of comparison, the Orion A GMC complex is $\sim 7.57 \times 10^4 M_{\odot}$ \citep{lombardi_2mass-orion}, whereas the mass of the molecular gas contained in this cut out is $\sim 5.78 \times 10^4 M_{\odot}$. Our entire zoom region is larger (150 pc) compared to the size of Orion A, which is $\sim 90$ pc \citep{3dshape_orion_gaia}. Additionally, we find we have a lower percentage of high density gas (25\%, compared to 45\% in Orion A according to \citet{3dshape_orion_gaia}), which can also be attributed to the young age of our simulation. Our average velocity in the area is $13.9$ km/s, higher than the assumed 10 km/s from \citet{wilson1970}, though not entirely outside an acceptable range. The velocity dispersion in this region is $7.5$ km/s, more than double the estimates of $\sim 2.5$ km/s for Orion A\citep{orion_kinematics}. 

Overall, this molecular cloud candidate can be compared to observations of molecular clouds, though it does not fully represent the Orion A cloud complex. We attribute the differences to the short timescales we run our simulations for, and intend to investigate the evolution of this area at later times in future work. We find our simulations produce clouds similar to observed molecular cloud structure and properties though we hasten to add that a more detailed comparison requires a statistical sample of simulated clouds, which is beyond our computational capability in this paper.

\begin{figure}
    \centering
    \includegraphics[width=1.0\linewidth]{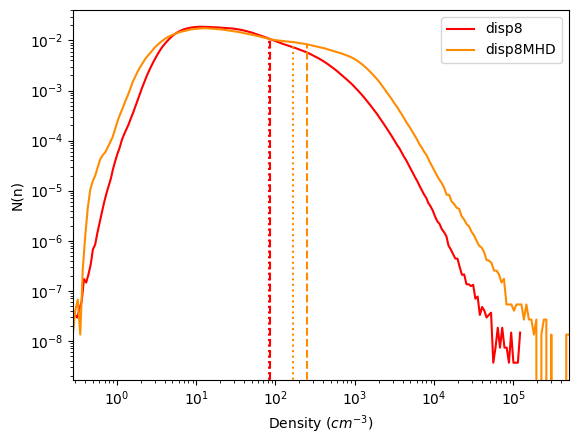}
    \includegraphics[width=1.0\linewidth]{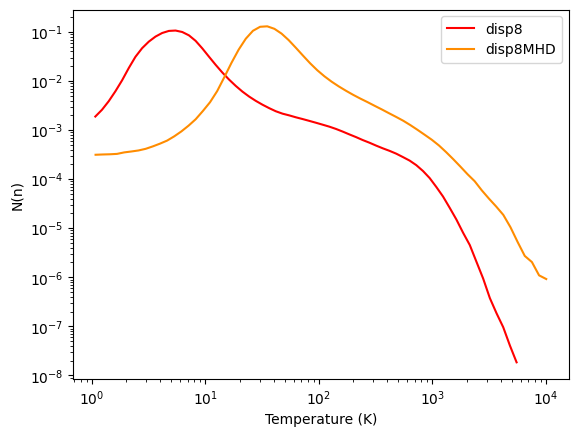}
    \caption{Density and temperature PDFs for the 8km/s models. Dotted vertical lines on the density PDF mark the average density of all gas, whereas dashed lines represent the average of cold gas in the simulation. The hydro model shows a cooler temperature peak by an order of magnitude compared to the MHD case, significant of the higher compression of the gas in these models.}
    \label{fig:densityPDF}
\end{figure}

Fig. \ref{fig:densityPDF} shows the density and temperature PDFs of our 8 km/s dispersion model. We use dotted vertical lines to show the average density of all the gas. We also do a temperature cut of all the cold gas based on where our temperature PDF flattens out. The dashed vertical lines in our density PDF mark the average density of the colder gas. Temperature PDFs show an order of magnitude difference in the peak temperatures of our models, corresponding with the results discussed for Fig.\ref{fig:temprho} in Section \ref{sec:hydromag} below. Discussion of the temperature differences follows there.

While the density PDFs between the two models are quite similar, we notice a second peak beginning to form in our magnetic model. As this also corresponds to a shift in our average density, we can relate this to the cold gas, such that we are producing a double log-normal, with each peak corresponding to a region in the temperature PDF (roughly split here between cold and warm gas). We note that our peaks are quite subtle as we are limiting ourselves to very early times in the gas, whereas the work of \citet{JoungMacLow2009} show much more distinct peaks. However, we still see similar general trends between our work and theirs, expecting that later stages of our simulation would evolve quite similarly. 

In the high density range of the PDF we notice a deviation from a normal shape, indicating a power-law tail produced via self-gravity as explored in works such as \citet{kainulainen}. The density PDF of our MHD model is also wider than our hydro model, possibly due to the correlation between gas density and temperature, as higher temperatures would be found in more diffuse gas, and the high-density shocked gas.

Finally, the bottom panel of Fig. \ref{fig:densityPDF} shows that the temperature of the peak in the density PDF in the hydro simulation is lower by nearly an order of magnitude than is the case for the MHD run (a few degrees K in the former versus 30 K in the latter case). This is an important consequence of the fact that magnetic fields prevent the high gas compression that can occur in purely hydrodynamic shocks. Lower densities imply lower cooling rates so more of the gas remains warmer than in the pure hydro simulation. This is further emphasized in our disp5 models shown in Fig. \ref{fig:disp5pdfs}, which show that the exclusion of magnetic fields allows the gas to overcool due to the global collapse of the model. This brings a non-negligible amount of gas below 3 K and effectively stops our run.

\subsection{Hydro vs. magnetic fields}\label{sec:hydromag}

\begin{figure*}
    \centering
    \includegraphics[width=0.75\linewidth]{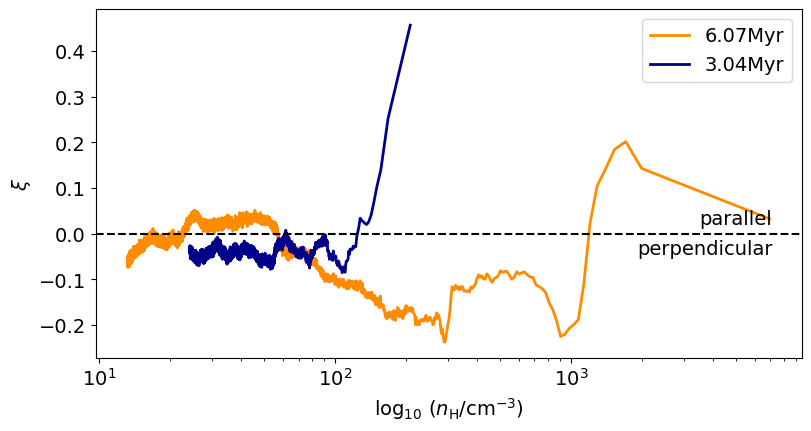}
    \includegraphics[width=0.75\linewidth]{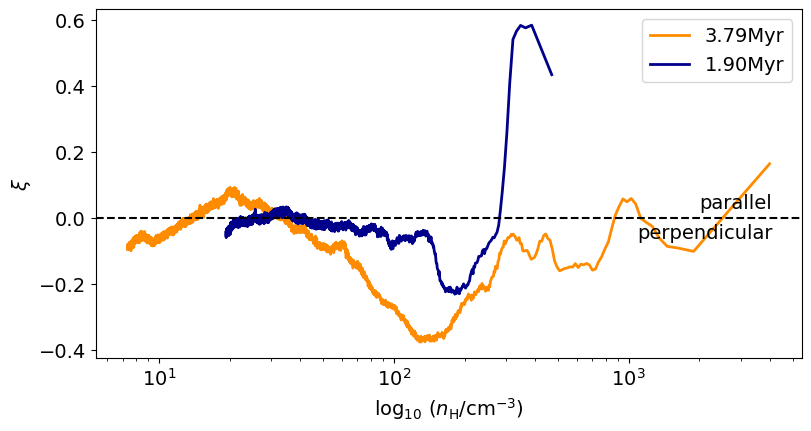}
    \includegraphics[width=0.75\linewidth]{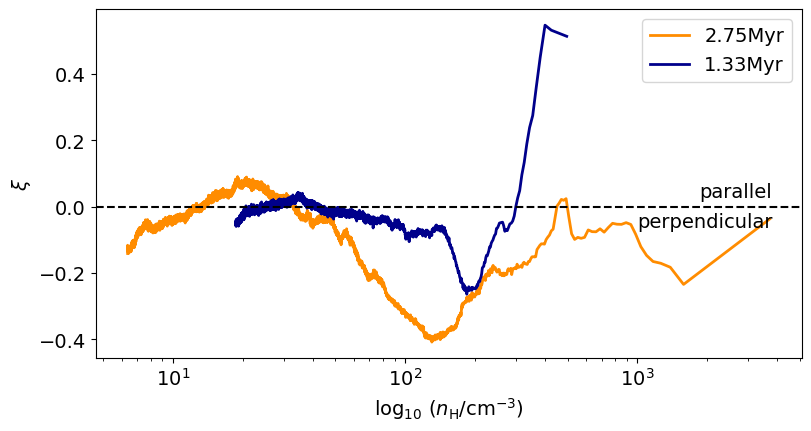}
    \caption{Magnetic field orientations relative to filaments shown through $\xi$ - the concavity parameter of the histogram of relative orientations from \citet{SolerHennebelle2013} for each of our MHD models. From top to bottom: 5 km/s dispersion, 8 km/s then 10 km/s. A $\xi$ value greater than 0 indicates parallel alignment of magnetic field with the filament axes, whereas less than 0 indicates a perpendicular alignment. Evaluated for times of the onset of sink formation (orange lines) and halfway between simulation start and sink formation time (blue lines).}
    \label{fig:magneticfield_lineconv}
\end{figure*}

To begin our analysis of the effects of magnetic fields on structure formation in the CNM, we first look at its effects on the structure created in different scenarios of turbulent strength. In figure \ref{fig:magneticfield_lineconv}, we show the orientation of the magnetic field to the filament. We use Magnetar \citep{SolerHennebelle2013}, which analyzes the relative orientation of the magnetic field and the filament using density gradients to determine the orientation of the filament. The cosine of the angle between filament and magnetic field will be normally distributed for a gas with parallel magnetic field alignment. A flat distribution signifies random orientations with no net alignment. In Fig.\ref{fig:magneticfield_lineconv}, we plot the $\xi$ parameter, which describes the concavity of the histogram of $cos(\phi)$. When this value is positive, the magnetic field is aligned parallel to a filament while for negative values, the magnetic field is aligned perpendicular. For each of our MHD simulations, we plot this parameter against the density for two times: one at which sink particle formation has just begun, and one halfway between the start of the simulation and the onset of sink formation. In this way, we are able to analyze not only the orientation of the magnetic field, but the evolution of the orientation as well.

We begin by noting that the early time frame has a smaller density range as the gas in our box is still evolving at this point. It is still highly turbulent and has not yet formed any stable filaments, and therefore has not yet reached high densities. Even with this smaller density range though, it is still clear that there are significant differences in the orientation of the field. In all three early time cases, magnetic fields are mostly parallel to filamentary structure, or showing no real alignment, as signified by the line hovering around zero in low densities. However, both the 8 and 10 km/s cases show small peaks into perpendicular alignment at molecular gas densities, with magnitudes of 0.2. These peaks of perpendicular alignment could be the beginning of molecular filament formation in the simulations, indicating a link between the slightly larger range of densities in these two compared to the 5 km/s case and their perpendicular field alignments. 

At the point of sink formation, we see a more evolved gas, as evidenced by the larger range of densities. The magnetic field orientations have changed over this time scale, showing higher magnitude concavity, indicating a stronger alignment parallel to the filaments. For all three models, the magnetic field begins to have a perpendicular alignment at densities of $10^2~\rm{cm}^{-3}$, corresponding to our molecular gas filaments. Beyond this point, the models' similarities end. 

Dispersions of 5 km/s show a weaker perpendicular alignment to filaments. At densities of $10^3~\rm{cm}^{-3}$, the alignment of the magnetic field switches to parallel. In the highest densities, the concavity approaches zero, possibly showing a random alignment in the innermost regions of a filament and around clusters. Dispersions of 8 and 10 km/s show stronger perpendicular alignment. Once perpendicularly aligned, the 8 km/s model gradually climbs back to random and even weakly parallel alignments, becoming parallel at densities of approximately $3000~\rm{cm}^{-3}$. While the change from perpendicular to parallel alignment is smoother in the 8 km/s model, the alignment still definitively changes at the innermost high density portions of filaments. Our 10 km/s model does not display any parallel alignment in the gas though it is trending towards parallel alignment at high densities. Instead, the gas remains at least weakly perpendicularly aligned throughout the entire density range, indicating a higher density needed in filament centres in order to sweep magnetic fields into parallel alignment.

As we reach the highest density filaments, we see that fields parallel to the filament occur at the time that sink particles appear.   This is the signal that gravity now is stronger than magnetic field tension of the perpendicular field. The gravity driven flow towards the clusters (i.e., sink particles) bends the field into a parallel configuration \citep{KlassenPudritz2017}.  This alignment aids in funnelling material into dense clumps along the filaments. We see evidence of this phenomenon in all three of our models, but notably the density at which this occurs is not constant. Instead, the density necessary to pull magnetic fields into a parallel alignment is connected to the amount of turbulent support in the gas, where high velocity dispersions reach higher densities in the centres of filaments before the magnetic field aligns parallel with gas flows within the filament.

\begin{figure*}
    \centering
    \includegraphics[width=1.0\linewidth]{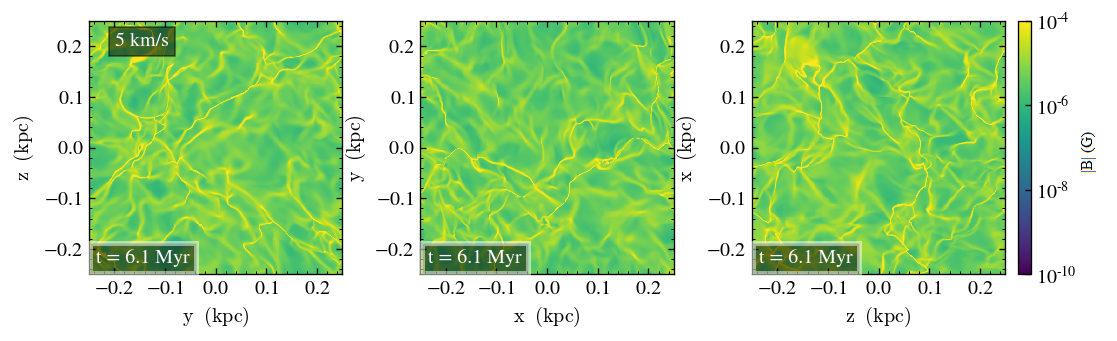}
    \includegraphics[width=1.0\linewidth]{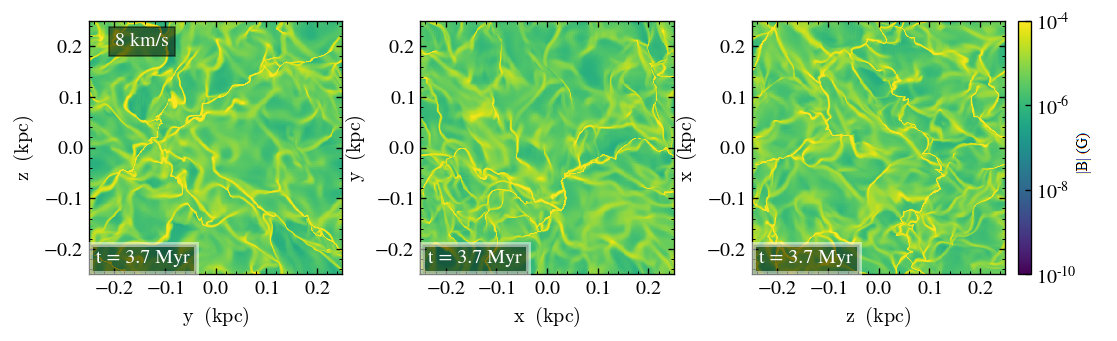}
    \includegraphics[width=1.0\linewidth]{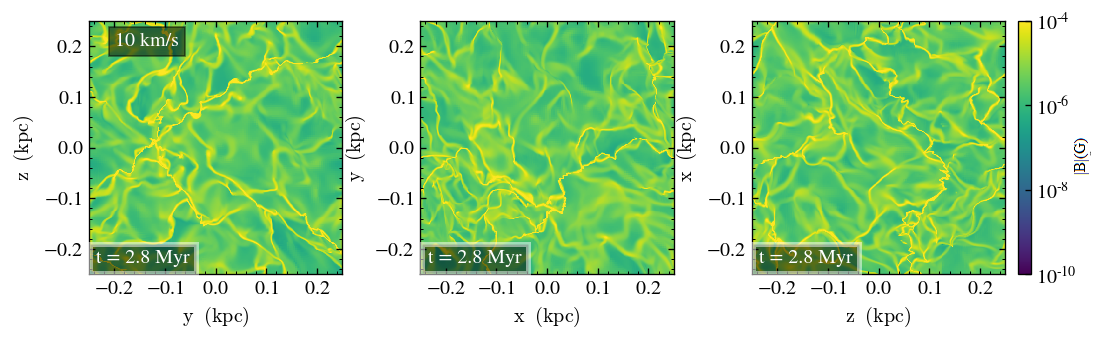}
    \caption{Magnetic field magnitude maps for each of our MHD models. From left to right: x, z and y plane slices through the center. Sink particles are represented by white filled circles.}
    \label{fig:magneticfied_magnitude}
\end{figure*}

Figure \ref{fig:magneticfied_magnitude} presents the structure of the magnetic field via its magnitude through a slice of the simulation box. These maps reveal where our highest magnetic field strengths lay. We can see strong magnetic field lines tracing the high-density filaments, displaying a link between field strength and density in our gas. However, though all three cases show the same link between gas density and field strength, we see differences in the filaments traced in each model. 

In our least turbulent model, the 5 km/s dispersion case, the filaments of strong magnetic field are thin, sharply defined and quite long. Many of the filaments trail the length of the entire simulation domain, and shorter filaments seem to quickly diffuse out into lower field strength (and lower density) gas. We see some bending in the field around the area we see our highest density structure forming, but the majority of the structure is smooth. 

In our 8 km/s dispersion model, we see more intense field variations. While our filaments can still span the length of the domain, there are fewer short filaments overall, and those that are present do not diffuse in the background field as quickly. Additionally, we see that though our filaments are wider, we have more structure variation throughout. Notably, we can distinguish the location of our highest density structure from the complication of the field lines around the area. As an example, in our z-plane view we know the structure to exist just below the center point of the box, spanning across the x direction. Likewise, in the slide of our magnetic field, we see a very dense filament traced by highly magnetic gas breaking off into smaller filaments and loops just below the center of the domain. We can as well see many filaments turn around completely causing sharp u-bends and overlapping with other filaments. 

Finally, in our highest turbulence case, the 10 km/s dispersion model shows many of the same features. Magnetic field strengths increase with gas density, effectively showing strong magnetic fields in dense filament centres. Filamentary areas where we see cluster formation in our gas also contain the strongest field strengths along the direction of the filaments. The difference between this and the 8 km/s model largely comes in through the filament widths. We can see our more dynamic case lends wider and more diffuse looking filaments than our 8 km/s case, not unlike our discussion of the densities, where we find the 10 km/s models to be more diffusive due to the stronger turbulence.

Magnetic fields also delay structure formation \citep{banerjee_clumps}. In our simulations, we can see some ties between magnetic field strength, turbulent velocity dispersion and the onset of structure formation. The time at which the first sink particle forms is also linked to structure formation timescales. Typical structure formation timescales of a turbulent medium correspond to the crossing time of the gas, the time in which it takes for a wave to cross the length of the domain. On the scales of structure necessary for cluster formation, we consider the crossing time across the typical size of a cloud (100 pc), corresponding to $10\%$ of the initial auto-correlation time of the turbulence. However, we find that structure formation begins rapidly in our simulations, and the formation of sink particles as clusters can equally represent the fragmentation of structures. When comparing these two, we find that each of our simulations form filaments well before they reach $10\%$ of their crossing time, as can be seen Table \ref{tab:timescales}.

\begin{table}
\centering
\begin{tabular}{|l|l|l|}
\hline
                   & \textit{$t_{cross, 10\%}$ (Myr)} & \textit{$t_{sink}$ (Myr)} \\ \hline
\textbf{disp5}     & 9.7                              & 5.76                          \\ \hline
\textbf{disp5MHD}  & 9.7                              & 6.12                      \\ \hline
\textbf{disp8}     & 6.1                              & 3.21                      \\ \hline
\textbf{disp8MHD}  & 6.1                              & 3.71                      \\ \hline
\textbf{disp10}    & 4.8                              & $\geq$ 2.7                          \\ \hline
\textbf{disp10MHD} & 4.8                              & 2.77                      \\ \hline
\end{tabular}
\caption{Comparison between time of first sink formation ($t_{sink}$) and $10\%$ of the global crossing time ($t_{cross, 10\%}$) for each of the six simulations we run. Note the disp10 model, which does not form any sink particles before resource limitations force us to end the run. }
\label{tab:timescales}
\end{table}

We see some distinct differences in timescales with the inclusion of magnetic fields. While all of the models form sink particles well before reaching one crossing time, the time of sink particle formation can vary. In the 5 km/s and 8 km/s models, we see sink particles form roughly 0.4 and 0.5 Myr sooner, respectively, in the absence of magnetic fields. As such, there is an obvious delay in cluster formation with the inclusion of magnetic fields demonstrating the important dynamic role that fields can play in the ISM. As magnetic fields support the gas and slow the formation of structure, so too will it delay the onset of star formation. 

However, as an outlier, we note that our 10 km/s model does not display this same behaviour. In fact, while the hydro and magnetic cases are run for the same amount of time, we see no sink formation start in our hydro model whereas the inclusion of magnetic fields begins sink particle formation just before the end of the run.

\begin{figure}
    \centering
    \includegraphics[width=0.98\linewidth]{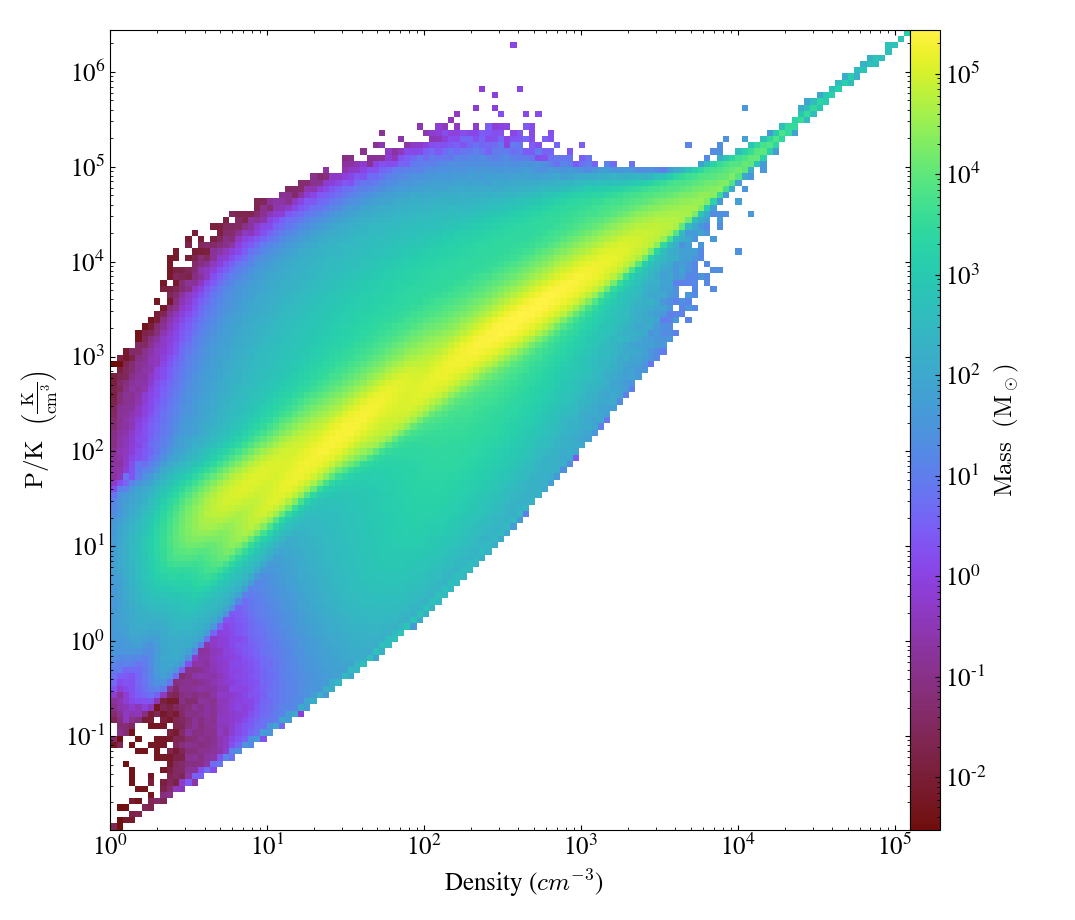}
    \includegraphics[width=0.98\linewidth]{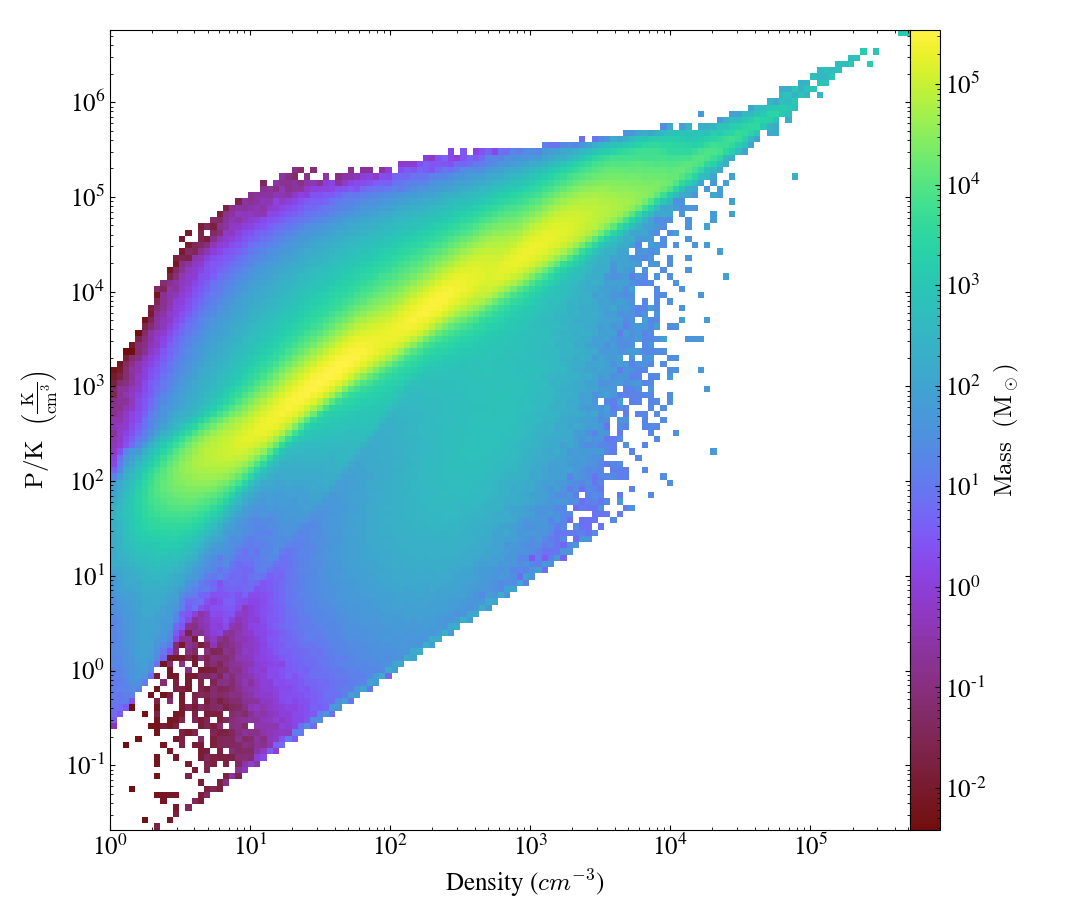}
    \caption{Mass-weighted pressure-density phase plot of the disp8 (top) and disp8MHD (bottom) model, for densities above $1 cm^{-3}$. The majority of the mass is found in median pressures, molecular to star-forming gas.}
    \label{fig:pressurephase}
\end{figure}

\begin{figure}
    \centering
    \includegraphics[width=0.98\linewidth]{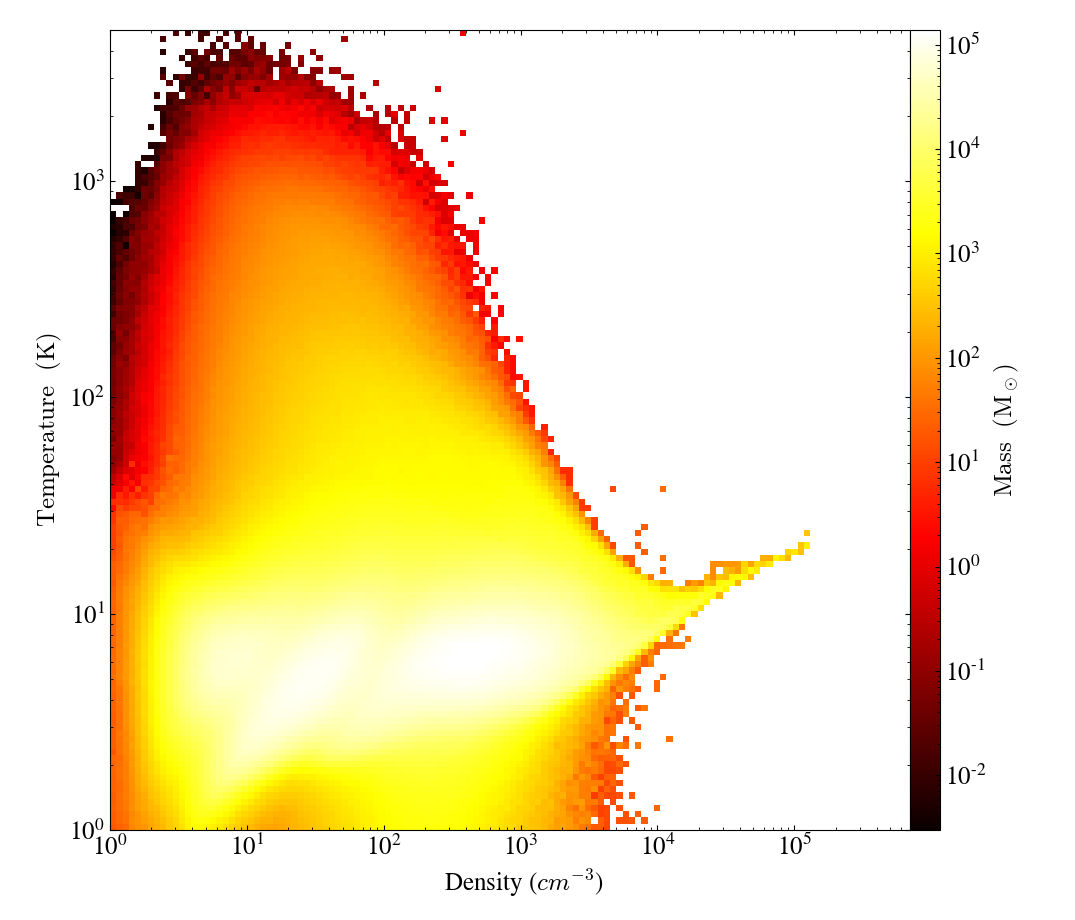}
    \includegraphics[width=0.98\linewidth]{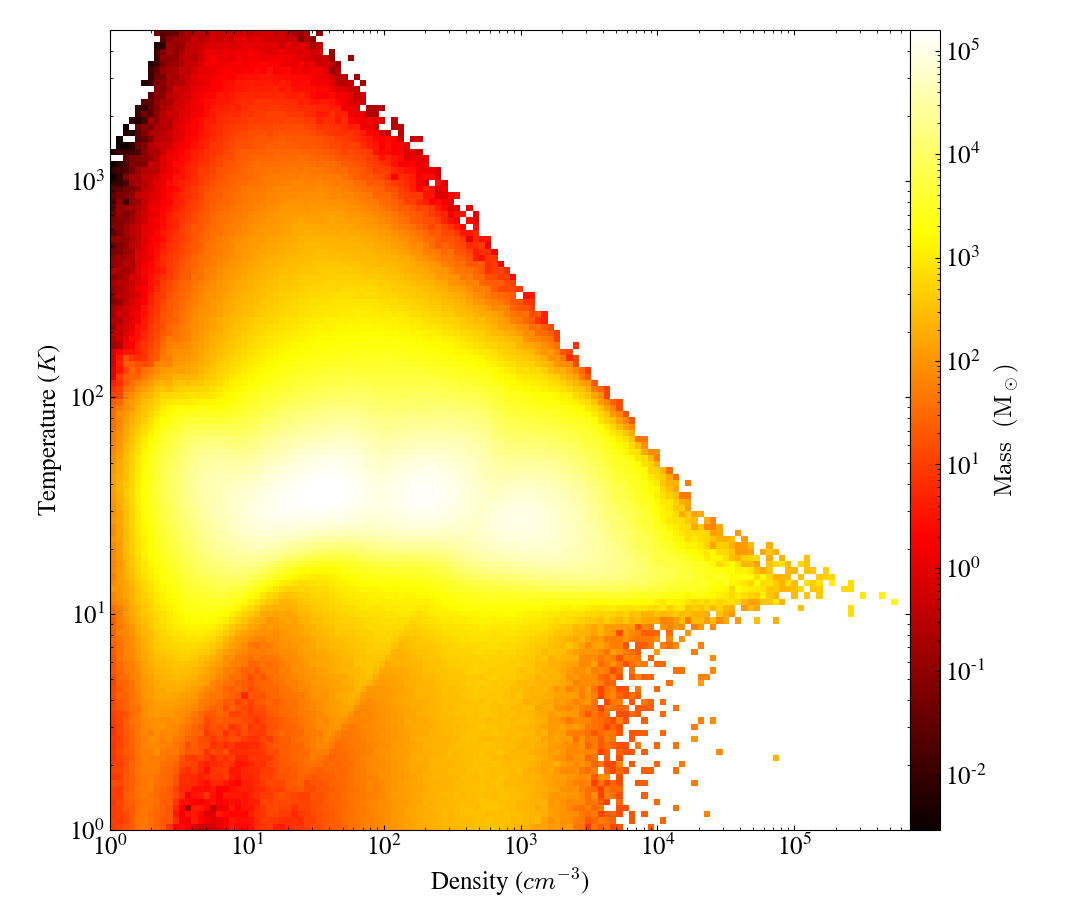}
    \caption{Mass-weighted temperature-density relation for disp8 (top) and disp8MHD (bottom) models, for a temperature range between $1-10^3 K$ and densities of $1-5\times 10^5 K$.}
    \label{fig:temprho}
\end{figure}

The gas densities present in our simulation also vary between hydro and magnetic cases. While we have discussed the breakdown of the gas in our hydro models in Section \ref{moleculargas} with Table \ref{tab:gascuts}, we also note the breakdown when including magnetic fields. While we have the same total gas mass, we can see the percentage of gas in each density cut compared to the total drops with each higher cutoff. In our disp8MHD model, $50\%$ of the gas is molecular densities of 100 $cm^{-3}$ or higher, when we increase to 1000 $cm^{-3}$ for our cutoff, this drops to $15\%$ of the total gas mass, showing that most of the gas is remaining at molecular gas densities or lower. When we consider the star-forming gas, with density cutoffs of 10,000 $cm^{-3}$ or higher, only $0.49\%$ of the gas in the box reaches this level. While our magnetic field cases can create an abundance of molecular gas, very little of that gas reaches star-forming densities, leaving few clusters to form. Specifically comparing to the hydro case, we see that the fractions vary slightly between the two. While cuts for densities of $100 cm^{-3}$ and $1000 cm^{-3}$ contain similar gas masses, we note a large difference in the star-forming gas cuts. 

We have, therefore, discovered an important distinction between these two cases. Even though the two models have similar amounts of molecular gas, our hydro case has roughly half the amount of gas at star-forming densities, indicating a lower number of possible sink particles that can form. This lower fraction of high density gas would suggest that the inclusion of magnetic fields does not significantly affect the cold gas, but can contribute to a higher star formation rate as the magnetic fields allow more gas to funnel to higher densities. Indeed, we also see this in the number of sink particles each model forms, where our hydro cases formed only a couple cluster sinks but our MHD models were able to form anywhere from 5-10 cluster sinks in less than 0.5 Myr.

\subsection{Phase diagrams}

Comparing the mass-weighted pressure-density relations in Figure \ref{fig:pressurephase}, we note the changes in slope of the relation for the highest masses. In particular, we notice the hydro case contains a majority of its mass in molecular gas at pressures of $\sim 10^3 K cm^{-3}$, up to a maximum of $\sim 10^5 K cm^{-3}$ with a minimum of $\sim 10^1 K cm^{-3}$. In the MHD models, on the other hand, the gas sits in higher pressures by about an order of magnitude, with most sitting at pressures of $\sim 10^4 K cm^{-3}$. Thus, even considering what pressure at which the majority of the gas measures, the magnetic fields contribute to overall higher pressure, a result which has also been found in previous works such as \citet{fiege_pudritz}.

In Figure \ref{fig:temprho}, we plot the temperature-density phase diagram of our two 8 km/s dispersion models. Both have near constant slopes, indicating an isothermal gas in both, though the average temperature is different by an order of magnitude between the two. While the majority of the gas in our hydro model sits around $10^1 K$ cooler than our initial temperature, the MHD models are supported at the initial temperature of our gas just by the inclusion of magnetic fields. This net cooling present in the hydro models is likely due to higher compression of the gas. Without magnetic fields, the gas experiences less support against shock waves and compression, ultimately allowing for more compressed gas than if magnetic fields are present. This compressed gas cools faster, and the average temperature of our simulations will decrease as the amount of highly compressed gas increases. 

We see evidence of the increased compression in our hydro models in Figures \ref{fig:pressurephase} and \ref{fig:temprho}. Note that high densities have their pressure and temperature spreads significantly thinned, and we see fairly sharp relation with little to no spread in both temperature and pressure for our hydro models. This indicates the compression of gas, as high densities are pushed to a stronger pressure or temperature relation. On the other hand, our MHD models do not display such a sharp relation, especially so in our temperatures, where high densities follow our isothermal line and are free to have larger spread in temperature. The overall temperature trends, in tandem with our results from Table \ref{tab:gascuts}, indicates that the presence of magnetic fields could be necessary to allow for the creation of hot, gas as well as setting the star formation rate in the ISM.

These temperature spreads also indicate the presence of multiple phases. While the majority of our gas exists in a CNM phase, we can also see a spread towards warmer temperatures at similarly low densities, indicating a warm, neutral phase of hydrogen present. This multiphase ISM structure - specifically the heating to warm and even hot temperatures - comes about purely from shock waves. However, we do note that an ionized medium phase is missing because we lack supernova feedback in our simulations. In order to push to these dense, hot regions of the temperature-density space it is likely necessary to add those ISM heating mechanisms we know to be important at kpc scales like these (i.e. photoelectric heating and UV radiation, as discussed in \citet{draine, klessen2016}).

\subsection{When do clusters form?}\label{clustertime}

\subsubsection{Timescales of formation}
\begin{figure}
    \centering
    \includegraphics[width=0.45\textwidth]{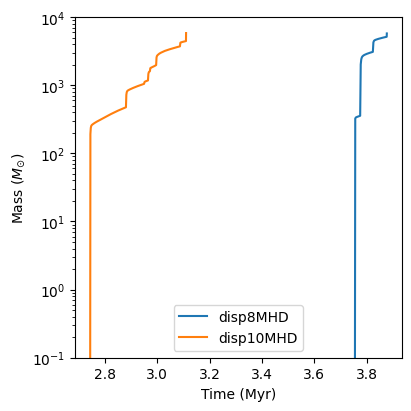}
    \caption{Mass growth of sink particles over time shown for MHD models disp8MHD and disp10MHD.}
    \label{fig:sink_times}
\end{figure}

One can also consider the timescale of cluster formation, as well as how many clusters form. In Figure \ref{fig:sink_times}, we show the mass function of the sink particles that form across our disp8MHD and disp10MHD models. Disp5MHD is excluded as it only forms one sink particle during the simulation. Mass growth of clusters is rapid early on, quickly gaining more than 7 orders of magnitude of mass in less than 0.2 Myr. The mass growth starts to exhibit a turnover, slowing down significantly after the initial burst of growth, consistent with results from \citet{howard}. Both the 8 and 10 km/s dispersions display similar mass growth, indicating this pattern is not linked to the dynamics of the gas.

We additionally compare the formation time for sink particles across all our simulations. From Table \ref{tab:timescales}, we can see that sink formation starts well before one crossing time has passed on a cloud scale and all of our runs begin forming young cluster sinks at least 2 Myr before one local crossing time has passed. 

\begin{table}
\centering
\begin{tabular}{l|l|l|}
\cline{2-3}
                                                  & \textbf{Lowest Mass ($M_{\odot}$)} & \textbf{Highest Mass ($M_{\odot}$)} \\ \hline
\multicolumn{1}{|l|}{\textit{\textbf{disp5MHD}}}  & $4.92 \times 10^2$                 & ---                                    \\ \hline
\multicolumn{1}{|l|}{\textit{\textbf{disp8MHD}}}  & $1.32 \times 10^2$                 & $2.94 \times 10^3$                  \\ \hline
\multicolumn{1}{|l|}{\textit{\textbf{disp10MHD}}} & $2.28 \times 10^1$                 & $1.36 \times 10^3$                  \\ \hline
\end{tabular}
\caption{Lowest and highest mass sink particles for each of our MHD models. In the case of our 5 km/s dispersion, we note only one sink particle forms and therefore no upper bound is available for masses.}
\label{tab:sinkmasses}
\end{table}

Both the 8 and 10 km/s simulations show rapid sink formation before reaching 4 Myr. In our non-magnetic case, the first sink to form for the 8 and 10 km/s cases form at times of 3.33 and $\geq 2.7$ Myr, respectively. The two have similar bursts of star formation, displaying a rapid onset of formation immediately. The magnetic cases display remarkably similar behaviour. While onset of sink formation occurs later, 3.71 and 2.77 Myr for the 8 and 10 km/s cases respectively, the two cases still display early bursts in star formation, forming many sinks within 0.1 Myr. Both these time frames are consistent with findings from \citet{huili_paper1}. 

For our 5 km/s simulations, sink formation occurs significantly later and many fewer sink particles form. In our non-magnetic case, sink formation begins at 5.76 Myr, and our magnetic case begins sink formation at 6.12 Myr. While this is significantly later than the 8 and 10 km/s models, these times are still well ahead of the local crossing time, unanimously indicating cluster formation begins well before one crossing time. Additionally, both 5 km/s cases produce very few sink particles, indicating we have not reached the burst of formation for these models yet. Computational limitations prevent us from running these models longer.

In the case of disp5MHD, we see only one sink form at 6.12 Myr, while the simulation is cut off at 6.2 Myr. Similarly in the hydro case, our disp5 model forms 1 cluster sink before the simulation is ended. With sink particle formation beginning so late in the run, and the formation rate being significantly slower than either the 8 or 10 km/s cases, the cloud is close to the end of its lifetime before forming any massive clusters and may never form more than a few. Further evolution of the cloud is needed to investigate the later cluster formation for a burst as is seen in our 8 \& 10 km/s simulations. Additionally, though the gas is self-gravitating, the density fluctuations formed via shock waves in the medium are not strong enough to create the extra compression needed for supercritical filaments and their associated star clusters. For these reasons, we conclude that a velocity dispersion of 5 km/s is too low to create the extensive filamentary structure observed in the CNM. We conclude that 8 km/s is ideal for both cluster formation onset and structure formation.

\subsubsection{Cluster accretion}
\begin{figure}
    \centering
    \includegraphics[width=1.0\linewidth]{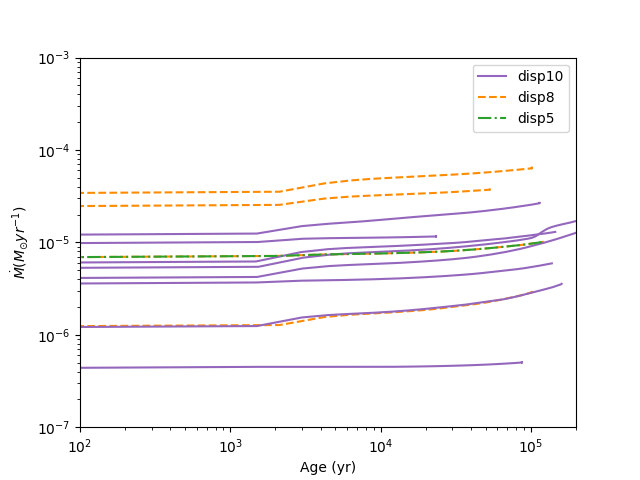}
    \caption{Accretion rate vs. age of the sink particle for our three MHD cases: disp5MHD, disp8MHD and disp10MHD. Age represents how long the cluster sink has been around, not the physical time in the simulation.}
    \label{fig:sink_accretion}
\end{figure}

Another important factor that governs the time scale for cluster formation is the accretion rate onto the region. In Fig.\ref{fig:sink_accretion}, the accretion rates of each individual sink particle in our MHD simulations are plotted against the age of the sink particle, for better sink-to-sink comparison. We can see that when sinks are just forming, their accretion rates can range from $10^{-6} - 10^{-3} M_{\odot}/yr$, depending on the strength of turbulence and the location of the sink particle. For example, our highest number of sink particles formed occurs in our disp10 simulation, where turbulence is very strong and the filamentary structure forms quickly. The sink particles in this case (represented by the purple lines) show a spread in accretion rates, and each sink's rate stays steady, peaking upwards at later ages, when the sink is more massive and more active. While the average accretion is low, we note that these cluster sinks are still very young and we expect their accretion rates to increase rapidly as time goes on. 

Other published simulations of cluster formation show accretion rates of at least $10^{-4} M_{\odot}/yr$ and up to $10^2 M_{\odot}/yr$ \citep[see, for example,][]{martapaper,satin,howard}. Furthermore, the average mass of a cloud in our simulations is $1.3 \times 10^5 M_{\odot}$, comparable to the $7.5\times10^4 M_{\odot}$ of Orion A, but lower than average cloud masses from galactic simulations of GMC formation \citep{grisdale}. Given that our clouds are lower mass, the scaling relation between cloud and cluster mass discovered in \citet{howard} shows that cluster masses would also be in the low mass end of cluster mass ranges, particularly in the time frame for which we run our simulations. 

We also note that over longer periods of time we would expect a varying accretion rate, and the smoothness of the rates presented here is due to the relatively short timescale on which we are looking at these clusters. An analysis of these clusters at later times will follow in future work. No mergers take place between sink particles to grow their mass, though the early phases of a cluster's growth will be dominated by mergers \citep{starforge1}. Instead, we see mass growth only through initial formation and gas accretion as the timescales discussed here are still shorter for the clusters than discussed in the paper referenced. This result is in agreement with simulations of massive isolated GMCs wherein only the most massive clusters undergo significant mergers with smaller clusters \citep{howard}. On average, our sink particles reach masses of $10^3 M_{\odot}$ before our simulations end - i.e. an Orion Nebula Cluster type of mass scale. Our 8 km/s dispersion model is responsible for the most massive cluster sink, reaching a mass of $2.94 \times 10^3 M_{\odot}$, whereas its lowest mass cluster is roughly $10^2 M_{\odot}$. Lowest and highest masses of each of our MHD models can be seen in Table \ref{tab:sinkmasses}.

\subsection{Angular Momentum and Filament Stability} \label{sec:linemass}

In this subsection, we analyze two other key factors governing structure formation; angular  momentum and gravitational stability.  On the first issue, we note that the local specific angular momentum distribution inside our simulations is a consequence of both the initial conditions - the power injected in solenoidal modes - as well as the subsequent interaction of oblique shocks. 

\begin{figure}
    \centering
    \includegraphics[width=1.0\linewidth]{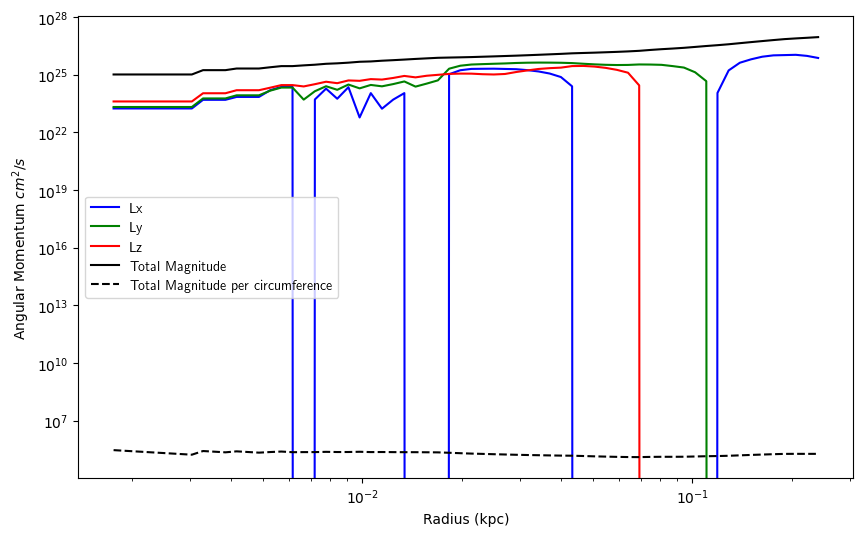}
    \caption{Specific angular momentum profile of gas throughout the box, moving radially outwards from the center, for our disp8MHD model. Both the individual components and total magnitude of the gas's specific angular momentum are plotted.}
    \label{fig:angularmomentum}
\end{figure}

In Fig. \ref{fig:angularmomentum}, we plot each individual component of the specific angular momentum, as well as the total magnitude, as a function of radius from the center of the box. Most notably, one can notice the shallow, positive slope of the total magnitude, indicating a lack of net rotation in the box. We further support this conclusion with the flux of specific angular momentum, shown by the dashed line. As the flux per radius is constant and low, we conclude that global rotation is not present in the simulation. Instead, any local rotational motion is essentially random across the box, brought on by the interaction of shock waves moving through the domain \citep{Jappsen2004}. The directional components of the specific angular momentum show relatively even levels with each throughout most of the box, indicating they are balanced and no dominant axis for rotation stands out. However, the outer edge does show a dominance from the x component, which may indicate some rotation along the outer edge of the box, but is not dynamically important as we do not see any other evidence of the possibility of rotation. 

We conclude that turbulence does not create any net rotation in our simulation domain, which is a good check on the veracity of the RAMSES code since no net initial rotation was put into our simulation box.  Turbulence does produce local fluctuations in angular momentum which will carry over into the formation of protostellar disks on the smallest scales (not probed in our simulations).  We also note that the effects of galactic shear on our box could be important, and this we investigate in \citet{bopaper}.

\begin{figure*}
    \centering
    \includegraphics[width=1.0\linewidth]{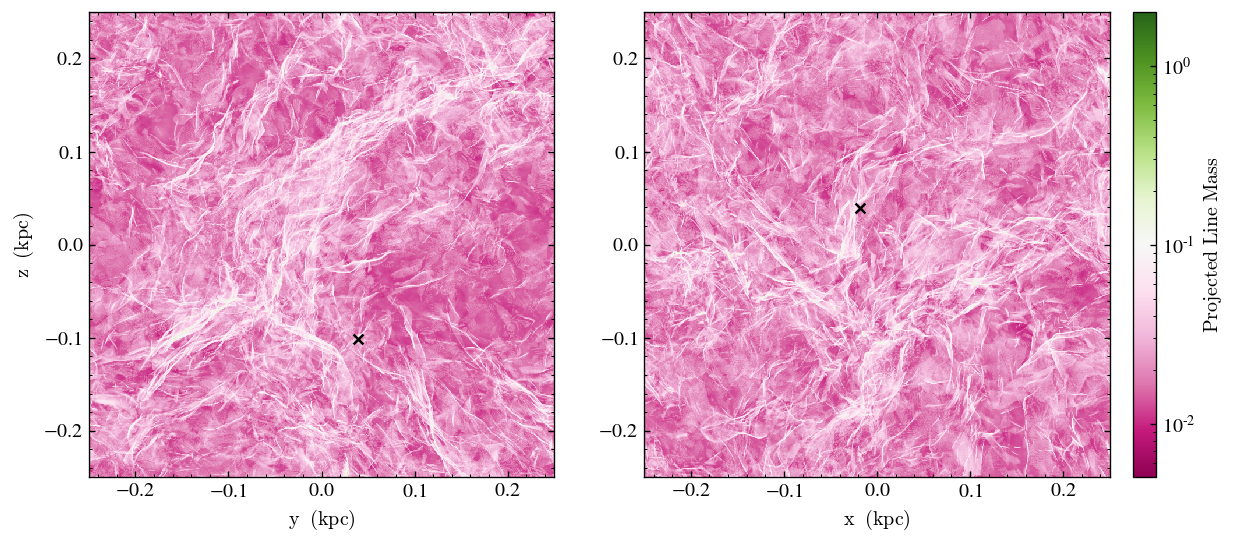}
    \includegraphics[width=1.0\linewidth]{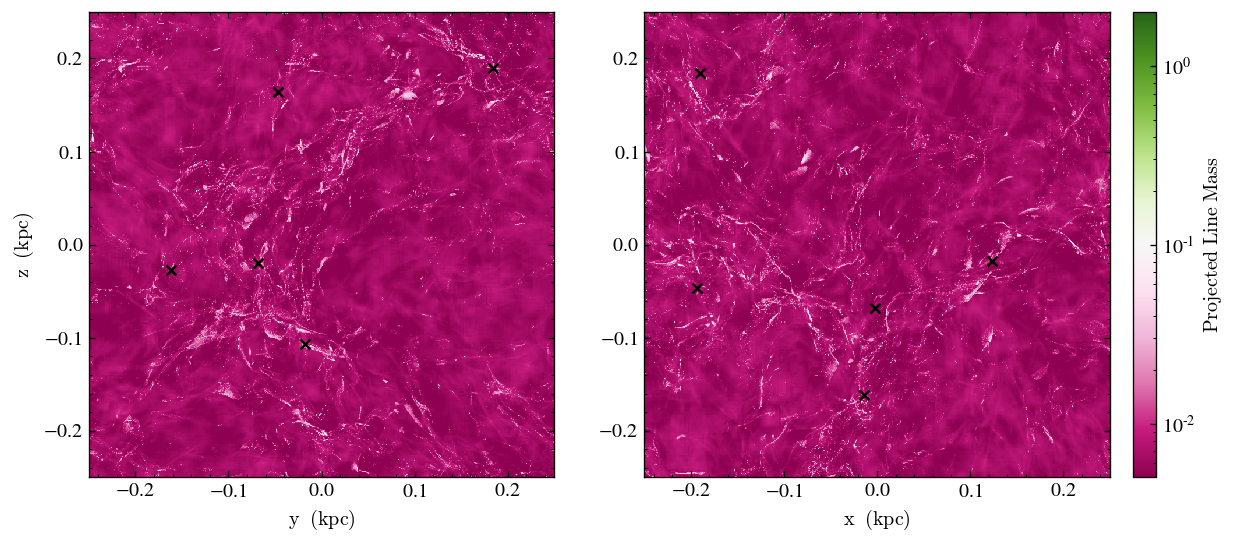}
    \caption{Line mass ratio maps for projections of \textit{top:} our disp8 model and \textit{bottom:} our disp8MHD model. Black X's show the positions of sink particles. Projections are taken for the x and z plane, as they showcase the most structure. Critical line mass is calculated as $M_{crit} = 2c_s^2/G$, thereby indicating only thermal stability. Line mass is taken as cell mass/cell width and sampled for each cell, therefore providing a lower bound.}
    \label{fig:line_mass}
\end{figure*}

We consider now the stability of the filamentary structure created in our 8 km/s dispersion models. Figure \ref{fig:line_mass} depicts projections of the line mass ratio of the gas in our models. We note that our simulations have considerable turbulent amplitudes on the supra molecular scales, and these are much larger than found within molecular clouds (recall the turbulent scaling relations; $\sigma_{turb} \propto L^{1/2}$). \citet{andre2010} considered nearly thermal filaments in low mass molecular clouds. At the same time, the CNM has much higher temperatures than the cold molecular clouds so that thermal speeds and critical line masses in CNM filaments are higher. In order to make comparisons then, we use only the thermal motions for our critical line mass as they do. We do not include any turbulent motion or magnetic field corrections. As such, our ratios in Figure \ref{fig:line_mass} represent lower bounds on the true line mass ratios of the gas.

In both the hydrodynamical and MHD models, we see largely subcritical thermal line masses for gas throughout the domain, though the MHD model contains gas that is more subcritical than our hydro model. Without magnetic fields, the gas is allowed to approach much higher line mass ratios, the majority of the gas sitting around a line mass ratio of 0.1, corresponding to transcritical line masses. One can also notice that the structures in line mass projections trace the structures we see in Fig. \ref{fig:Dispersions}. Our densest filaments also contain the gas closest to supercritical ratios. Specifically the region between z=-0.1 and z=0.0 in the top left plot of Fig. \ref{fig:line_mass} contains gas approaching the critical ratio, as indicated by the faint green colour in the filamentary structure. Empty pockets of gas which do not make up the filamentary structure are shown in dark pink, with very low ratios. In our hydrodynamical models, there are very few of these areas, indicating most of this gas is still close to collapse and that the structure is either overall more unstable or younger and therefore has had less time to settle.

The bottom row of Fig. \ref{fig:line_mass} shows the same projections for our disp8MHD model, to compare the effects from magnetic fields. Here we see far less gas approaching critical values, with most of the domain having significantly subcritical line masses. The structure in these projections is therefore more evolved than our hydrodynamical models, with only the densest areas of the filaments being near fragmentation, and most of the gas having already fragmented into clumps. Likewise, we see fewer filaments of transcritical line mass in these projections. Instead, we note only the densest filaments of the projections are close to approaching the critical value. Furthermore, we see no elongated structures approaching supercritical values in green. Instead, we see clumpier structure overall, and very small areas approaching supercritical values and starting to fragment. We therefore conclude that the magnetic fields have an overall stabilizing effect on the gas when only considering thermal stability. In future work we intend to analyze the impact of accounting for magnetic field affects in the critical line mass.

We can also see that, while the magnetic runs may be more efficient at producing star-forming gas as indicated in Table \ref{tab:gascuts}, these projections do not show near-critical or supercritical line masses in as large a fraction of the total gas. Overall, the line masses of the filaments would suggest that our non-magnetic models could have higher star formation rates due to the higher concentration of near-critical filaments. Instead, we have shown throughout our previous results that our magnetic models in fact have the higher star formation and produce more cluster sinks than our non-magnetic models. In fact, what is apparent from our line mass projections is that the magnetic fields have acted to speed up the formation of filamentary structure. In this way, filaments take less time to form, being able to start forming star clusters more rapidly than the hydrodynamical models, despite unmagnetized models starting to form clusters sooner than the magnetized. 

We can conclude that cluster formation  begins after filaments have reached supercritical line masses, whereupon they begin to fragment into clumps that ultimately host star clusters, lending support to beads on a string formation scenarios of star clusters in molecular filaments.

\subsection{Magnetic Field-Density Relation}

\begin{figure}
    \centering
    \includegraphics[width=0.9\linewidth]{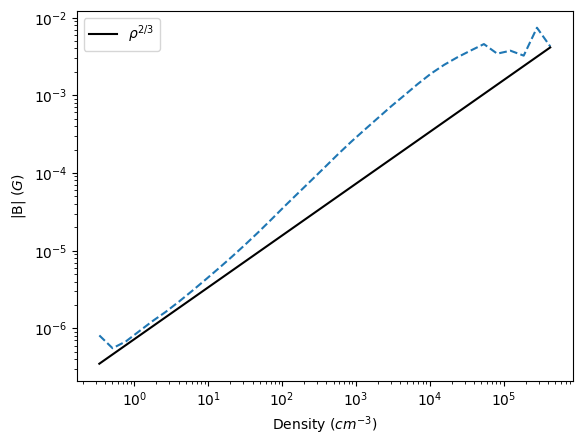}
    \caption{Relation between magnetic field magnitude and volume density of the disp8MHD model. Blue dashed line represents the simulation, and the solid black line shows a $B \propto \rho^{2/3}$ relation, matching observations from \citet{crutcher_brho_2010}.}
    \label{fig:bvrho}
\end{figure}

It is important to track the evolution of the magnetic field in our simulation as it is likely to evolve over time, changing it's strength and the role it plays on the gas. Magnetic fields may increase in strength simply due to the compression of the gas which contains them. In this event, $B \propto \rho^{2/3}$ discussed in \citet{crutcher_brho_2010}. On the other hand, some cases may see magnetic field strength increase disproportionately to the density increase. These instances are evidence of dynamo effects, where turbulent motions in the dense gas may feed magnetic fields without needing to increase density. In a dynamo case, one would see a deviation to steeper B vs. $\rho$ relation from the well-accepted trend of \citet{crutcher_brho_2010}.

In Figure \ref{fig:bvrho} we plot the relation between magnetic field magnitude and density, and compare it to  the scaling with $ \rho^{2/3}$. At densities below $10^2~cm^{-3}$, we have good agreement with the $2/3$ power-law relation, but notice a significant deviation from the scaling relation at densities above this value. As we deviate above the scaling from \citet{crutcher_brho_2010}, we see a higher magnetic field strength. This may indicate the operation of a turbulent dynamo effect that increases the field strength beyond values that can be reached by compression alone. At the highest densities, the trend flattens, with spiking present from the formation of sink particles artificially decreasing the amount of gas at $10^4-10^5~cm^{-3}$ from our formation criteria. Given the resolution limit and the formation of sink particles, it is not possible to conclude definitively that this flattening is a real physical result in our simulations, or a consequence of our resolution limit. 

\section{Conclusions}\label{sec:conclusion}

In this paper we have presented simulations of the effects of the CNM environment on molecular cloud and star cluster formation. We used initial conditions for typical CNM densities and temperatures, although at higher column densities than the local region - more in accord with typical massive star-forming regions, such as the CMZ and luminous starburst galaxies. We used the RAMSES code - that includes, MHD, cloud chemistry, turbulence, and self gravity, but without stellar feedback - to set up MHD simulations for a range of three different initial 3D velocity dispersions imposed on the gas but allowed to decay. Since we chose the decay times to be much longer than our simulation time scales (5 Myr), the turbulent amplitudes remain nearly constant over our runs. We then analyzed the structure formed in the various cases and found an optimal condition for structure and cluster formation. We list our conclusions below. 

Arguably our most important conclusion is that velocity dispersion in the CNM plays a significant role in structure formation. A dispersion of 5 km/s in a 0.5 kpc cube is dominated by its self-gravity. Its weak turbulence, even combined with thermal pressure, can barely support it. Dispersions of 8 and 10 km/s (initial energy ratios of 0.01 and 0.03 respectively) both allow for turbulence to create filamentary structure, though 8 km/s is able to reproduce some features that resemble observations of the Orion A GMC. Additionally, in all three models we note that structure formation begins and is set well before one crossing time has passed on the scale of individual clouds. The formation of dense filaments therefore does not solely rely on the turbulent crossing time in the gas, and we here stress the importance of gravity in creating stable, dense structure as well.

\medskip

Several important conclusions about the effects of magnetic fields are:

\begin{itemize} 

    \item Although both 8 km/s models contain $\sim$50\% of the total gas mass in molecular gas of densities 100 $cm^{-3}$ or higher by the onset of sink particle formation, the dense gas fraction between hydro and magnetic models is very different. Just 0.2\% of the total gas is converted to star-forming densities in unmagnetized runs, whereas the inclusion of magnetic fields doubles the fraction to 0.5\%. The relative rarity of reaching star-forming densities is mirrored in the sink particle formation, in which only a handful of cluster sinks form before the end of our simulations, exhibiting a low star formation rate in the early formation of clusters in the CNM.

    \item Magnetic field orientations are primarily perpendicular to the gas above densities of $100~\rm{cm}^{-3}$ once clusters have begun to form. Dispersions of 5 and 8 km/s see magnetic field orientations return to be parallel to the gas again above densities of $\simeq 10^3~\rm{cm}^{-3}$, and dispersions of 10 km/s are trending towards parallel alignment at higher densities. Our models see evidence for the effects of high density gas flows in the centres of filaments dragging magnetic field lines into parallel alignment, though higher dispersions may contain higher levels of turbulent support, needing higher density flows to pull the field into parallel alignment. 

    \item Magnetic fields also significantly contribute to the pressure of the ISM, with average pressures an order of magnitude higher than in our non-magnetic models. This results in an order of magnitude increase in temperature due to the lower compressibility of the gas arising through magnetic support. Both temperature and pressure plots showcase these compression differences in the high-density ranges, where both are limited to a very small spread in unmagnetized models. Additionally, Fig. \ref{fig:densityPDF} shows that the gas in the pure hydro simulation overcools by an order of magnitude compared to the MHD case - a clear indication that more of the shocked gas in our MHD models has lower cooling rates and will be warmer. While neither hydro nor MHD models contain a hot, ionized medium, the models do both recreate a multiphase ISM, containing both cold and warm neutral gas, with significantly more cold gas in our hydro models. In future work, we intend on introducing radiation effects in order to add the hot, ionized medium and recreate the three-phase ISM.

    \item Filament line masses are largely transcritical in unmagnetized simulations. Including magnetic fields leaves the line masses at subcritical values in all but the high-density gas. Additionally, magnetic fields can suppress the amount of gas that fragments to form clusters, such that even though the presence of fields correlates with more gas in high density regimes, the star formation rate is suppressed from the stabilizing effects of the fields on the line mass.
\end{itemize}

Other significant results are:

\begin{itemize}

    \item Temperatures of filaments are cold, sitting around 10-30 K, but they are surrounded by hot gas at temperatures $\geq 600$ K, heated by the shock waves that form the filaments. We note the intricate filamentary structure formed, specifically the braided features that dense filaments display in density projections, temperature slices and in magnetic field structure. These braids are crucial for the interaction of filaments to create dense structures. 
    
    \item Density PDFs for models with dispersions of 8 km/s produce a standard log-normal shape. The inclusion of magnetic fields supports the beginning of a secondary peak, formed by the cold gas in the simulation. High densities show evidence of power-law tails produced via self-gravity, regardless of presence of magnetic fields.

    \item Magnetic field strengths increase with density, where field strengths of up to $100 \mu G$ correspond to star-forming densities of $10^4 cm^{-3}$, showing a clear link between gas density and magnetic field density. The filamentary structure of the ISM can be seen in the magnetic field strength, displaying ridges of high magnetic field along the filaments, and the structure shows perpendicular field lines funnelling these filaments to their high densities. We also see, in our 8 km/s models, the majority of cluster formation happening along a high density filament and, thus, in highly magnetized environments. We link this to the delay we see in sink particle formation, as the magnetic field strengths in the high density gas prove very dynamically important.

    \item Turbulence produces no net, global rotation throughout the box, such that rotation and shear in the ISM are likely created mostly via galactic dynamics. The angular momentum magnitude per radius bin shows very low values, indicating a low and constant momentum flux which supports our conclusions towards low/no net rotation. Turbulence does produce local fluctuations in angular momentum, and these are likely to be important for the formation of disks on much smaller (and unresolved) scales. 

    \item We compare our models to \citet{crutcher_brho_2010} through the B vs. $\rho$ relation. We note that low densities follow a standard $2/3$ power-law relation, but high densities experience more compression, deviating above the $2/3$ line. This may indicate some turbulent dynamo action in the denser gas, though more work must be done to investigate thoroughly.

\end{itemize}

Finally, there are several interesting conclusions regarding the very earliest phases of star cluster formation: 

\begin{itemize}
    
    \item The differences in the sink particle formation onset times highlight the delay of formation that occurs in our MHD models. Between all three dispersions, we see a consistently later time to sink formation with the inclusion of magnetic fields. The inclusion of magnetic fields in our simulations delays the onset of sink particle formation by $\sim 0.4$ Myr in our 8 km/s dispersion models. The delay does not seem to affect the amount of molecular gas formed, but rather affects the high density star-forming gas, indicating a support role from the magnetic fields in preventing very high density clumps from forming as quickly. This result is further mirrored in the density PDFs, wherein our magnetic runs exhibit a different shape at high densities than our hydro runs. 

    \item Cluster mass sink particles form quickly, and we surpass a total mass in sinks of $10^3 M_{\odot}$ within 0.4 Myr, indicating rapid growth. As in previous works \citep[see, for instance,][]{howard}, mass growth is fast. After initial rapid growth, our models see average accretion rates of $10^{-5}~M_{\odot}~yr^{-1}$. While we only look at early evolution here, there are already signs of a turnover in total cluster mass, signifying cluster formation happens early on and finishes rapidly, as found in \citet{huili_paper1}.
    
\end{itemize}

\section*{Acknowledgements}
The authors thank an anonymous referee for a report that helped to clarify the manuscript considerably.  We would also like to acknowledge and thank the following people for helpful and interesting discussions during the course of this work: No{\'e} Brucy, Juan Soler, Marta Reina-Campos, Hector Robinson, Claude Cournoyer-Cloutier, Jeremy Karam, Davide Decataldo, Ralf Klessen, Alison Sills, James Wadsley, Henrik Beuther. We would like to especially thank Romain Teyssier for his help and support in implementing the RAMSES code. We thank Juan Soler and Patrick Hennebelle for allowing us the use of Magnetar to analyze magnetic field orientations. Computation was enabled through the Graham and Cedar clusters of the Digital Research Alliance of Canada, as well as support from their team. RP is grateful to the funding support of the Ontario Graduate Scholarship and the Joanne and Joseph Lee Scholarship at McMaster University. REP is supported by a Discovery Grant from NSERC Canada. 

\section*{Data Availability} 
The data underlying this article will be shared on reasonable request to the corresponding author.


\bibliographystyle{mnras}
\bibliography{Bibliography} % if your bibtex file is called example.bib

\begin{thebibliography}{}
\makeatletter
\relax
\def\mn@urlcharsother{\let\do\@makeother \do\$\do\&\do\#\do\^\do\_\do\%\do\~}
\def\mn@doi{\begingroup\mn@urlcharsother \@ifnextchar [ {\mn@doi@} {\mn@doi@[]}}
\def\mn@doi@[#1]#2{\def\@tempa{#1}\ifx\@tempa\@empty \href {http://dx.doi.org/#2} {doi:#2}\else \href {http://dx.doi.org/#2} {#1}\fi \endgroup}
\def\mn@eprint#1#2{\mn@eprint@#1:#2::\@nil}
\def\mn@eprint@arXiv#1{\href {http://arxiv.org/abs/#1} {{\tt arXiv:#1}}}
\def\mn@eprint@dblp#1{\href {http://dblp.uni-trier.de/rec/bibtex/#1.xml} {dblp:#1}}
\def\mn@eprint@#1:#2:#3:#4\@nil{\def\@tempa {#1}\def\@tempb {#2}\def\@tempc {#3}\ifx \@tempc \@empty \let \@tempc \@tempb \let \@tempb \@tempa \fi \ifx \@tempb \@empty \def\@tempb {arXiv}\fi \@ifundefined {mn@eprint@\@tempb}{\@tempb:\@tempc}{\expandafter \expandafter \csname mn@eprint@\@tempb\endcsname \expandafter{\@tempc}}}

\bibitem[\protect\citeauthoryear{{Abgrall}, {Le Bourlot}, {Pineau Des Forets}, {Roueff}, {Flower}  \& {Heck}}{{Abgrall} et~al.}{1992}]{AbgrallLeBourlot1992}
{Abgrall} H.,  {Le Bourlot} J.,  {Pineau Des Forets} G.,  {Roueff} E.,  {Flower} D.~R.,   {Heck} L.,  1992, Astronomy and Astrophysics, \href {https://ui.adsabs.harvard.edu/abs/1992A&A...253..525A} {253, 525}

\bibitem[\protect\citeauthoryear{{Alves de Oliveira} et~al.,}{{Alves de Oliveira} et~al.}{2014}]{herschel_filaments}
{Alves de Oliveira} C.,  et~al., 2014, \mn@doi [Astronomy \& Astrophysics] {10.1051/0004-6361/201423504}, \href {https://ui.adsabs.harvard.edu/abs/2014A&A...568A..98A} {568, A98}

\bibitem[\protect\citeauthoryear{{Andr{\'e}} et~al.,}{{Andr{\'e}} et~al.}{2010}]{andre2010}
{Andr{\'e}} P.,  et~al., 2010, \mn@doi [Astronomy \& Astrophysics] {10.1051/0004-6361/201014666}, \href {https://ui.adsabs.harvard.edu/abs/2010A&A...518L.102A} {518, L102}

\bibitem[\protect\citeauthoryear{{Andr{\'e}}, {Di Francesco}, {Ward-Thompson}, {Inutsuka}, {Pudritz}  \& {Pineda}}{{Andr{\'e}} et~al.}{2014}]{andre_paradigm}
{Andr{\'e}} P.,  {Di Francesco} J.,  {Ward-Thompson} D.,  {Inutsuka} S.~I.,  {Pudritz} R.~E.,   {Pineda} J.~E.,  2014, in {Beuther} H.,  {Klessen} R.~S.,  {Dullemond} C.~P.,   {Henning} T.,  eds, Protostars and Planets VI. p.~27 (\mn@eprint {arXiv} {1312.6232}), \mn@doi{10.2458/azu\_uapress\_9780816531240-ch002}

\bibitem[\protect\citeauthoryear{{Arzoumanian} et~al.,}{{Arzoumanian} et~al.}{2011}]{arzoumanian}
{Arzoumanian} D.,  et~al., 2011, \mn@doi [Astronomy \& Astrophysics] {10.1051/0004-6361/201116596}, \href {https://ui.adsabs.harvard.edu/abs/2011A&A...529L...6A} {529, L6}

\bibitem[\protect\citeauthoryear{{Arzoumanian}, {Andr{\'e}}, {Peretto}  \& {K{\"o}nyves}}{{Arzoumanian} et~al.}{2013}]{arzoumanian_2013}
{Arzoumanian} D.,  {Andr{\'e}} P.,  {Peretto} N.,   {K{\"o}nyves} V.,  2013, \mn@doi [Astronomy \& Astrophysics] {10.1051/0004-6361/201220822}, \href {https://ui.adsabs.harvard.edu/abs/2013A&A...553A.119A} {553, A119}

\bibitem[\protect\citeauthoryear{{Banerjee}, {V{\'a}zquez-Semadeni}, {Hennebelle}  \& {Klessen}}{{Banerjee} et~al.}{2009}]{banerjee_clumps}
{Banerjee} R.,  {V{\'a}zquez-Semadeni} E.,  {Hennebelle} P.,   {Klessen} R.~S.,  2009, \mn@doi [Monthly Notices of the Royal Astronomical Society] {10.1111/j.1365-2966.2009.15115.x}, \href {https://ui.adsabs.harvard.edu/abs/2009MNRAS.398.1082B} {398, 1082}

\bibitem[\protect\citeauthoryear{{Bellomi}, {Godard}, {Hennebelle}, {Valdivia}, {Pineau des For{\^e}ts}, {Lesaffre}  \& {P{\'e}rault}}{{Bellomi} et~al.}{2020}]{BellomiEcoGal}
{Bellomi} E.,  {Godard} B.,  {Hennebelle} P.,  {Valdivia} V.,  {Pineau des For{\^e}ts} G.,  {Lesaffre} P.,   {P{\'e}rault} M.,  2020, \mn@doi [Astronomy and Astrophysics] {10.1051/0004-6361/202038593}, \href {https://ui.adsabs.harvard.edu/abs/2020A&A...643A..36B} {643, A36}

\bibitem[\protect\citeauthoryear{Beuther et~al.,}{Beuther et~al.}{2018}]{beuthercore}
Beuther H.,  et~al., 2018, Astronomy \& Astrophysics, 617, A100

\bibitem[\protect\citeauthoryear{{Bieri}, {Naab}, {Geen}, {Coles}, {Pakmor}  \& {Walch}}{{Bieri} et~al.}{2022}]{satin}
{Bieri} R.,  {Naab} T.,  {Geen} S.,  {Coles} J.~P.,  {Pakmor} R.,   {Walch} S.,  2022, arXiv e-prints, \href {https://ui.adsabs.harvard.edu/abs/2022arXiv220906842B} {p. arXiv:2209.06842}

\bibitem[\protect\citeauthoryear{Bleuler \& Teyssier}{Bleuler \& Teyssier}{2014}]{ramses_sinks}
Bleuler A.,  Teyssier R.,  2014, \mn@doi [Monthly Notices of the Royal Astronomical Society] {10.1093/mnras/stu2005}, 445, 4015–4036

\bibitem[\protect\citeauthoryear{{Bleuler}, {Teyssier}, {Carassou}  \& {Martizzi}}{{Bleuler} et~al.}{2015}]{phew}
{Bleuler} A.,  {Teyssier} R.,  {Carassou} S.,   {Martizzi} D.,  2015, \mn@doi [Computational Astrophysics and Cosmology] {10.1186/s40668-015-0009-7}, \href {https://ui.adsabs.harvard.edu/abs/2015ComAC...2....5B} {2, 5}

\bibitem[\protect\citeauthoryear{{Bournaud}, {Elmegreen}  \& {Martig}}{{Bournaud} et~al.}{2009}]{BournaudElmegreen2009}
{Bournaud} F.,  {Elmegreen} B.~G.,   {Martig} M.,  2009, \mn@doi [The Astrophysical Journal] {10.1088/0004-637X/707/1/L1}, \href {https://ui.adsabs.harvard.edu/abs/2009ApJ...707L...1B} {707, L1}

\bibitem[\protect\citeauthoryear{{Brown} \& {Gnedin}}{{Brown} \& {Gnedin}}{2022}]{brown}
{Brown} G.,  {Gnedin} O.~Y.,  2022, \mn@doi [Monthly Notices of the Royal Astronomical Society] {10.1093/mnras/stac1164}, \href {https://ui.adsabs.harvard.edu/abs/2022MNRAS.514..280B} {514, 280}

\bibitem[\protect\citeauthoryear{{Brucy} \& {Hennebelle}}{{Brucy} \& {Hennebelle}}{2021}]{BrucyHennebelle2021}
{Brucy} N.,  {Hennebelle} P.,  2021, \mn@doi [Monthly Notices of the Royal Astronomical Society] {10.1093/mnras/stab738}, \href {https://ui.adsabs.harvard.edu/abs/2021MNRAS.503.4192B} {503, 4192}

\bibitem[\protect\citeauthoryear{{Brucy}, {Hennebelle}, {Bournaud}  \& {Colling}}{{Brucy} et~al.}{2020}]{BrucyHennebelle2020}
{Brucy} N.,  {Hennebelle} P.,  {Bournaud} F.,   {Colling} C.,  2020, \mn@doi [The Astrophysical Journal] {10.3847/2041-8213/ab9830}, \href {https://ui.adsabs.harvard.edu/abs/2020ApJ...896L..34B} {896, L34}

\bibitem[\protect\citeauthoryear{{Brucy}, {Hennebelle}, {Colman}  \& {Iteanu}}{{Brucy} et~al.}{2023}]{BrucyHennebelle2023}
{Brucy} N.,  {Hennebelle} P.,  {Colman} T.,   {Iteanu} S.,  2023, \mn@doi [arXiv e-prints] {10.48550/arXiv.2305.18012}, \href {https://ui.adsabs.harvard.edu/abs/2023arXiv230518012B} {p. arXiv:2305.18012}

\bibitem[\protect\citeauthoryear{{Brunetti} \& {Wilson}}{{Brunetti} \& {Wilson}}{2022}]{2022brunetti}
{Brunetti} N.,  {Wilson} C.~D.,  2022, \mn@doi [Monthly Notices of the Royal Astronomical Society] {10.1093/mnras/stac1975}, \href {https://ui.adsabs.harvard.edu/abs/2022MNRAS.515.2928B} {515, 2928}

\bibitem[\protect\citeauthoryear{{Calura} et~al.,}{{Calura} et~al.}{2022}]{ramsescosmo1}
{Calura} F.,  et~al., 2022, \mn@doi [Monthly Notices of the Royal Astronomical Society] {10.1093/mnras/stac2387}, \href {https://ui.adsabs.harvard.edu/abs/2022MNRAS.516.5914C} {516, 5914}

\bibitem[\protect\citeauthoryear{{Chevance}, {Krumholz}, {McLeod}, {Ostriker}, {Rosolowsky}  \& {Sternberg}}{{Chevance} et~al.}{2022}]{chevance}
{Chevance} M.,  {Krumholz} M.~R.,  {McLeod} A.~F.,  {Ostriker} E.~C.,  {Rosolowsky} E.~W.,   {Sternberg} A.,  2022, arXiv e-prints, \href {https://ui.adsabs.harvard.edu/abs/2022arXiv220309570C} {p. arXiv:2203.09570}

\bibitem[\protect\citeauthoryear{{Colman} et~al.,}{{Colman} et~al.}{2022}]{colman_ECOGal}
{Colman} T.,  et~al., 2022, \mn@doi [Monthly Notices of the Royal Astronomical Society] {10.1093/mnras/stac1543}, \href {https://ui.adsabs.harvard.edu/abs/2022MNRAS.514.3670C} {514, 3670}

\bibitem[\protect\citeauthoryear{{Coppola}, {Diomede}, {Longo}  \& {Capitelli}}{{Coppola} et~al.}{2011}]{CoppolaDiomede2011}
{Coppola} C.~M.,  {Diomede} P.,  {Longo} S.,   {Capitelli} M.,  2011, \mn@doi [The Astrophysical Journal] {10.1088/0004-637X/727/1/37}, \href {https://ui.adsabs.harvard.edu/abs/2011ApJ...727...37C} {727, 37}

\bibitem[\protect\citeauthoryear{{Crutcher}, {Wandelt}, {Heiles}, {Falgarone}  \& {Troland}}{{Crutcher} et~al.}{2010}]{crutcher_brho_2010}
{Crutcher} R.~M.,  {Wandelt} B.,  {Heiles} C.,  {Falgarone} E.,   {Troland} T.~H.,  2010, \mn@doi [The Astrophysical Journal] {10.1088/0004-637X/725/1/466}, \href {https://ui.adsabs.harvard.edu/abs/2010ApJ...725..466C} {725, 466}

\bibitem[\protect\citeauthoryear{{Decataldo}, {Ferrara}, {Pallottini}, {Gallerani}  \& {Vallini}}{{Decataldo} et~al.}{2017}]{decataldo2017}
{Decataldo} D.,  {Ferrara} A.,  {Pallottini} A.,  {Gallerani} S.,   {Vallini} L.,  2017, \mn@doi [Monthly Notices of the Royal Astronomical Society] {10.1093/mnras/stx1879}, \href {https://ui.adsabs.harvard.edu/abs/2017MNRAS.471.4476D} {471, 4476}

\bibitem[\protect\citeauthoryear{{Draine}}{{Draine}}{2011}]{draine}
{Draine} B.~T.,  2011, {Physics of the Interstellar and Intergalactic Medium}

\bibitem[\protect\citeauthoryear{{Falgarone}, {Pety}  \& {Phillips}}{{Falgarone} et~al.}{2001}]{falgarone2001}
{Falgarone} E.,  {Pety} J.,   {Phillips} T.~G.,  2001, \mn@doi [The Astrophysical Journal] {10.1086/321483}, \href {https://ui.adsabs.harvard.edu/abs/2001ApJ...555..178F} {555, 178}

\bibitem[\protect\citeauthoryear{Federrath, Roman-Duval, Klessen, Schmidt  \& Mac~Low}{Federrath et~al.}{2010a}]{federrathturb}
Federrath C.,  Roman-Duval J.,  Klessen R.~S.,  Schmidt W.,   Mac~Low M.~M.,  2010a, Astronomy \& Astrophysics, 512

\bibitem[\protect\citeauthoryear{Federrath, Banerjee, Clark  \& Klessen}{Federrath et~al.}{2010b}]{federrathsink}
Federrath C.,  Banerjee R.,  Clark P.~C.,   Klessen R.~S.,  2010b, The Astrophysical Journal, 713, 269

\bibitem[\protect\citeauthoryear{{Fiege} \& {Pudritz}}{{Fiege} \& {Pudritz}}{2000}]{fiege_pudritz}
{Fiege} J.~D.,  {Pudritz} R.~E.,  2000, \mn@doi [Monthly Notices of the Royal Astronomical Society] {10.1046/j.1365-8711.2000.03066.x}, \href {https://ui.adsabs.harvard.edu/abs/2000MNRAS.311...85F} {311, 85}

\bibitem[\protect\citeauthoryear{{Fisher}, {Bolatto}, {White}, {Glazebrook}, {Abraham}  \& {Obreschkow}}{{Fisher} et~al.}{2019}]{FisherBolatto2019}
{Fisher} D.~B.,  {Bolatto} A.~D.,  {White} H.,  {Glazebrook} K.,  {Abraham} R.~G.,   {Obreschkow} D.,  2019, \mn@doi [The Astrophysical Journal] {10.3847/1538-4357/aaee8b}, \href {https://ui.adsabs.harvard.edu/abs/2019ApJ...870...46F} {870, 46}

\bibitem[\protect\citeauthoryear{{Galv{\'a}n-Madrid} et~al.,}{{Galv{\'a}n-Madrid} et~al.}{2013}]{sma_observemassivecluster}
{Galv{\'a}n-Madrid} R.,  et~al., 2013, \mn@doi [The Astrophysical Journal] {10.1088/0004-637X/779/2/121}, \href {https://ui.adsabs.harvard.edu/abs/2013ApJ...779..121G} {779, 121}

\bibitem[\protect\citeauthoryear{{Girichidis} et~al.,}{{Girichidis} et~al.}{2016}]{silcc2}
{Girichidis} P.,  et~al., 2016, \mn@doi [Monthly Notices of the Royal Astronomical Society] {10.1093/mnras/stv2742}, \href {https://ui.adsabs.harvard.edu/abs/2016MNRAS.456.3432G} {456, 3432}

\bibitem[\protect\citeauthoryear{{Girichidis}, {Seifried}, {Naab}, {Peters}, {Walch}, {W{\"u}nsch}, {Glover}  \& {Klessen}}{{Girichidis} et~al.}{2018}]{silcc5}
{Girichidis} P.,  {Seifried} D.,  {Naab} T.,  {Peters} T.,  {Walch} S.,  {W{\"u}nsch} R.,  {Glover} S. C.~O.,   {Klessen} R.~S.,  2018, \mn@doi [Monthly Notices of the Royal Astronomical Society] {10.1093/mnras/sty2016}, \href {https://ui.adsabs.harvard.edu/abs/2018MNRAS.480.3511G} {480, 3511}

\bibitem[\protect\citeauthoryear{{G{\'o}mez} \& {V{\'a}zquez-Semadeni}}{{G{\'o}mez} \& {V{\'a}zquez-Semadeni}}{2014}]{gomez2014}
{G{\'o}mez} G.~C.,  {V{\'a}zquez-Semadeni} E.,  2014, \mn@doi [The Astrophysical Journal] {10.1088/0004-637X/791/2/124}, \href {https://ui.adsabs.harvard.edu/abs/2014ApJ...791..124G} {791, 124}

\bibitem[\protect\citeauthoryear{{Grisdale}}{{Grisdale}}{2021}]{grisdale}
{Grisdale} K.,  2021, \mn@doi [Monthly Notices of the Royal Astronomical Society] {10.1093/mnras/staa3524}, \href {https://ui.adsabs.harvard.edu/abs/2021MNRAS.500.3552G} {500, 3552}

\bibitem[\protect\citeauthoryear{{Gro{\ss}schedl} et~al.,}{{Gro{\ss}schedl} et~al.}{2018}]{3dshape_orion_gaia}
{Gro{\ss}schedl} J.~E.,  et~al., 2018, \mn@doi [Astronomy and Astrophysics] {10.1051/0004-6361/201833901}, \href {https://ui.adsabs.harvard.edu/abs/2018A&A...619A.106G} {619, A106}

\bibitem[\protect\citeauthoryear{{Grudi{\'c}}, {Kruijssen}, {Faucher-Gigu{\`e}re}, {Hopkins}, {Ma}, {Quataert}  \& {Boylan-Kolchin}}{{Grudi{\'c}} et~al.}{2021}]{GrudicKruijssen2021}
{Grudi{\'c}} M.~Y.,  {Kruijssen} J.~M.~D.,  {Faucher-Gigu{\`e}re} C.-A.,  {Hopkins} P.~F.,  {Ma} X.,  {Quataert} E.,   {Boylan-Kolchin} M.,  2021, \mn@doi [Monthly Notices of the Royal Astronomical Society] {10.1093/mnras/stab1894}, \href {https://ui.adsabs.harvard.edu/abs/2021MNRAS.506.3239G} {506, 3239}

\bibitem[\protect\citeauthoryear{{Grudi{\'{c}}}, {Guszejnov}, {Offner}, {Rosen}, {Raju}, {Faucher-Gigu{\`{e}}re}  \& {Hopkins}}{{Grudi{\'{c}}} et~al.}{2022}]{starforge}
{Grudi{\'{c}}} M.~Y.,  {Guszejnov} D.,  {Offner} S. S.~R.,  {Rosen} A.~L.,  {Raju} A.~N.,  {Faucher-Gigu{\`{e}}re} C.-A.,   {Hopkins} P.~F.,  2022, \mn@doi [Monthly Notices of the Royal Astronomical Society] {10.1093/mnras/stac526}, \href {https://ui.adsabs.harvard.edu/abs/2022MNRAS.512..216G} {512, 216}

\bibitem[\protect\citeauthoryear{{Guszejnov}, {Markey}, {Offner}, {Grudi{\'c}}, {Faucher-Gigu{\`e}re}, {Rosen}  \& {Hopkins}}{{Guszejnov} et~al.}{2022}]{starforge1}
{Guszejnov} D.,  {Markey} C.,  {Offner} S. S.~R.,  {Grudi{\'c}} M.~Y.,  {Faucher-Gigu{\`e}re} C.-A.,  {Rosen} A.~L.,   {Hopkins} P.~F.,  2022, \mn@doi [Monthly Notices of the Royal Astronomical Society] {10.1093/mnras/stac1737}, \href {https://ui.adsabs.harvard.edu/abs/2022MNRAS.515..167G} {515, 167}

\bibitem[\protect\citeauthoryear{{Hacar}, {Clark}, {Heitsch}, {Kainulainen}, {Panopoulou}, {Seifried}  \& {Smith}}{{Hacar} et~al.}{2022}]{hacar2022}
{Hacar} A.,  {Clark} S.,  {Heitsch} F.,  {Kainulainen} J.,  {Panopoulou} G.,  {Seifried} D.,   {Smith} R.,  2022, arXiv e-prints, \href {https://ui.adsabs.harvard.edu/abs/2022arXiv220309562H} {p. arXiv:2203.09562}

\bibitem[\protect\citeauthoryear{{Han}, {Kimm}, {Katz}, {Devriendt}  \& {Slyz}}{{Han} et~al.}{2022}]{han}
{Han} D.,  {Kimm} T.,  {Katz} H.,  {Devriendt} J.,   {Slyz} A.,  2022, arXiv e-prints, \href {https://ui.adsabs.harvard.edu/abs/2022arXiv220705745H} {p. arXiv:2207.05745}

\bibitem[\protect\citeauthoryear{{He}, {Wilson}, {Brunetti}, {Finn}, {Bemis}  \& {Johnson}}{{He} et~al.}{2022}]{he_wilson_observeYMC}
{He} H.,  {Wilson} C.,  {Brunetti} N.,  {Finn} M.,  {Bemis} A.,   {Johnson} K.,  2022, \mn@doi [The Astrophysical Journal] {10.3847/1538-4357/ac5628}, \href {https://ui.adsabs.harvard.edu/abs/2022ApJ...928...57H} {928, 57}

\bibitem[\protect\citeauthoryear{{Henning}, {Linz}, {Krause}, {Ragan}, {Beuther}, {Launhardt}, {Nielbock}  \& {Vasyunina}}{{Henning} et~al.}{2010}]{henning2010}
{Henning} T.,  {Linz} H.,  {Krause} O.,  {Ragan} S.,  {Beuther} H.,  {Launhardt} R.,  {Nielbock} M.,   {Vasyunina} T.,  2010, \mn@doi [Astronomy \& Astrophysics] {10.1051/0004-6361/201014635}, \href {https://ui.adsabs.harvard.edu/abs/2010A&A...518L..95H} {518, L95}

\bibitem[\protect\citeauthoryear{Howard, Pudritz  \& Harris}{Howard et~al.}{2018}]{howard}
Howard C.~S.,  Pudritz R.~E.,   Harris W.~E.,  2018, Nature Astronomy, 2, 725

\bibitem[\protect\citeauthoryear{{Jappsen} \& {Klessen}}{{Jappsen} \& {Klessen}}{2004}]{Jappsen2004}
{Jappsen} A.~K.,  {Klessen} R.~S.,  2004, \mn@doi [\aap] {10.1051/0004-6361:20040220}, \href {https://ui.adsabs.harvard.edu/abs/2004A&A...423....1J} {423, 1}

\bibitem[\protect\citeauthoryear{Jeffreson, Kruijssen, Keller, Chévance  \& C.O.}{Jeffreson et~al.}{2020}]{jeffreson}
Jeffreson S.,  Kruijssen J.,  Keller B.,  Chévance M.,   C.O. G.~S.,  2020, Monthly Notices of the Royal Astronomical Society, 498, 385*429

\bibitem[\protect\citeauthoryear{{Joung}, {Mac Low}  \& {Bryan}}{{Joung} et~al.}{2009}]{JoungMacLow2009}
{Joung} M.~R.,  {Mac Low} M.-M.,   {Bryan} G.~L.,  2009, \mn@doi [The Astrophysical Journal] {10.1088/0004-637X/704/1/137}, \href {https://ui.adsabs.harvard.edu/abs/2009ApJ...704..137J} {704, 137}

\bibitem[\protect\citeauthoryear{{Kainulainen}, {Federrath}  \& {Henning}}{{Kainulainen} et~al.}{2014}]{kainulainen}
{Kainulainen} J.,  {Federrath} C.,   {Henning} T.,  2014, \mn@doi [Science] {10.1126/science.1248724}, \href {https://ui.adsabs.harvard.edu/abs/2014Sci...344..183K} {344, 183}

\bibitem[\protect\citeauthoryear{{Kim} \& {Ostriker}}{{Kim} \& {Ostriker}}{2007}]{KimOstriker2007}
{Kim} W.-T.,  {Ostriker} E.~C.,  2007, \mn@doi [The Astrophysical Journal] {10.1086/513176}, \href {https://ui.adsabs.harvard.edu/abs/2007ApJ...660.1232K} {660, 1232}

\bibitem[\protect\citeauthoryear{{Kim} \& {Ostriker}}{{Kim} \& {Ostriker}}{2015}]{Kim_Ostriker2015}
{Kim} C.-G.,  {Ostriker} E.~C.,  2015, \mn@doi [\apj] {10.1088/0004-637X/815/1/67}, \href {https://ui.adsabs.harvard.edu/abs/2015ApJ...815...67K} {815, 67}

\bibitem[\protect\citeauthoryear{{Kim} \& {Ostriker}}{{Kim} \& {Ostriker}}{2017}]{tigress}
{Kim} C.-G.,  {Ostriker} E.~C.,  2017, \mn@doi [The Astrophysical Journal] {10.3847/1538-4357/aa8599}, \href {https://ui.adsabs.harvard.edu/abs/2017ApJ...846..133K} {846, 133}

\bibitem[\protect\citeauthoryear{{Kim}, {Kim}  \& {Ostriker}}{{Kim} et~al.}{2011}]{KimKim2011}
{Kim} C.-G.,  {Kim} W.-T.,   {Ostriker} E.~C.,  2011, \mn@doi [The Astrophysical Journal] {10.1088/0004-637X/743/1/25}, \href {https://ui.adsabs.harvard.edu/abs/2011ApJ...743...25K} {743, 25}

\bibitem[\protect\citeauthoryear{{Kim}, {Ostriker}  \& {Kim}}{{Kim} et~al.}{2013}]{KimOstriker2013}
{Kim} C.-G.,  {Ostriker} E.~C.,   {Kim} W.-T.,  2013, \mn@doi [The Astrophysical Journal] {10.1088/0004-637X/776/1/1}, \href {https://ui.adsabs.harvard.edu/abs/2013ApJ...776....1K} {776, 1}

\bibitem[\protect\citeauthoryear{{Kirk}, {Klassen}, {Pudritz}  \& {Pillsworth}}{{Kirk} et~al.}{2015}]{kirk}
{Kirk} H.,  {Klassen} M.,  {Pudritz} R.,   {Pillsworth} S.,  2015, \mn@doi [The Astrophysical Joural] {10.1088/0004-637X/802/2/75}, \href {https://ui.adsabs.harvard.edu/abs/2015ApJ...802...75K} {802, 75}

\bibitem[\protect\citeauthoryear{Kitsionas et~al.,}{Kitsionas et~al.}{2009}]{algorithmPDF}
Kitsionas S.,  et~al., 2009, Astronomy \& Astrophysics, 508, 541

\bibitem[\protect\citeauthoryear{Klassen, Pudritz  \& Peters}{Klassen et~al.}{2012}]{klassen}
Klassen M.,  Pudritz R.~E.,   Peters T.,  2012, Monthly Notices of the Royal Astronomical Society, 421, 2861

\bibitem[\protect\citeauthoryear{{Klassen}, {Pudritz}  \& {Kirk}}{{Klassen} et~al.}{2017}]{KlassenPudritz2017}
{Klassen} M.,  {Pudritz} R.~E.,   {Kirk} H.,  2017, \mn@doi [Monthly Notices of the Royal Astronomical Society] {10.1093/mnras/stw2889}, \href {https://ui.adsabs.harvard.edu/abs/2017MNRAS.465.2254K} {465, 2254}

\bibitem[\protect\citeauthoryear{{Klessen} \& {Glover}}{{Klessen} \& {Glover}}{2016}]{klessen2016}
{Klessen} R.~S.,  {Glover} S. C.~O.,  2016, \mn@doi [Saas-Fee Advanced Course] {10.1007/978-3-662-47890-5_2}, \href {https://ui.adsabs.harvard.edu/abs/2016SAAS...43...85K} {43, 85}

\bibitem[\protect\citeauthoryear{{K{\"o}rtgen}, {Banerjee}, {Pudritz}  \& {Schmidt}}{{K{\"o}rtgen} et~al.}{2019}]{kortgen2019}
{K{\"o}rtgen} B.,  {Banerjee} R.,  {Pudritz} R.~E.,   {Schmidt} W.,  2019, \mn@doi [Monthly Notices of the Royal Astronomical Society] {10.1093/mnras/stz2491}, \href {https://ui.adsabs.harvard.edu/abs/2019MNRAS.489.5004K} {489, 5004}

\bibitem[\protect\citeauthoryear{{Krumholz}, {McKee}  \& {Klein}}{{Krumholz} et~al.}{2005}]{krumholz_mckee}
{Krumholz} M.~R.,  {McKee} C.~F.,   {Klein} R.~I.,  2005, \mn@doi [The Astrophysical Journal Letters] {10.1086/427555}, \href {https://ui.adsabs.harvard.edu/abs/2005ApJ...618L..33K} {618, L33}

\bibitem[\protect\citeauthoryear{{Krumholz} et~al.,}{{Krumholz} et~al.}{2014}]{KrumholzBate2014}
{Krumholz} M.~R.,  et~al., 2014, in {Beuther} H.,  {Klessen} R.~S.,  {Dullemond} C.~P.,   {Henning} T.,  eds, Protostars and Planets VI. pp 243--266 (\mn@eprint {arXiv} {1401.2473}), \mn@doi{10.2458/azu_uapress_9780816531240-ch011}

\bibitem[\protect\citeauthoryear{{Kuffmeier}}{{Kuffmeier}}{2018}]{Kuffmeier2020}
{Kuffmeier} M.,  2018, in {Constraints from zoom-in simulations on the protostellar accretion process}. Cambridge University Press, pp 56--60, \mn@doi{10.1017/S1743921319001510}

\bibitem[\protect\citeauthoryear{{Kuiper} \& {Hosokawa}}{{Kuiper} \& {Hosokawa}}{2018}]{kuiper_hosokawa}
{Kuiper} R.,  {Hosokawa} T.,  2018, \mn@doi [Astronomy and Astrophysics] {10.1051/0004-6361/201832638}, \href {https://ui.adsabs.harvard.edu/abs/2018A&A...616A.101K} {616, A101}

\bibitem[\protect\citeauthoryear{{Kwon} et~al.,}{{Kwon} et~al.}{2022}]{bfield_filaments}
{Kwon} W.,  et~al., 2022, \mn@doi [The Astrophysical Journal] {10.3847/1538-4357/ac4bbe}, \href {https://ui.adsabs.harvard.edu/abs/2022ApJ...926..163K} {926, 163}

\bibitem[\protect\citeauthoryear{Lada \& Lada}{Lada \& Lada}{2003}]{ladalada2003}
Lada C.~J.,  Lada E.~A.,  2003, Annual Review of Astronomy \& Astrophysics, 41, 51

\bibitem[\protect\citeauthoryear{{Lane}, {Grudi{\'c}}, {Guszejnov}, {Offner}, {Faucher-Gigu{\`e}re}  \& {Rosen}}{{Lane} et~al.}{2022}]{lane}
{Lane} H.~B.,  {Grudi{\'c}} M.~Y.,  {Guszejnov} D.,  {Offner} S. S.~R.,  {Faucher-Gigu{\`e}re} C.-A.,   {Rosen} A.~L.,  2022, \mn@doi [Monthly Notices of the Royal Astronomical Society] {10.1093/mnras/stab3739}, \href {https://ui.adsabs.harvard.edu/abs/2022MNRAS.510.4767L} {510, 4767}

\bibitem[\protect\citeauthoryear{{Larson}}{{Larson}}{1981}]{larson}
{Larson} R.~B.,  1981, \mn@doi [Monthly Notices of the Royal Astronomical Society] {10.1093/mnras/194.4.809}, \href {https://ui.adsabs.harvard.edu/abs/1981MNRAS.194..809L} {194, 809}

\bibitem[\protect\citeauthoryear{{Lee} et~al.,}{{Lee} et~al.}{2022}]{phangs_OG}
{Lee} J.~C.,  et~al., 2022, \mn@doi [The Astrophysical Journal Supplement Series] {10.3847/1538-4365/ac1fe5}, \href {https://ui.adsabs.harvard.edu/abs/2022ApJS..258...10L} {258, 10}

\bibitem[\protect\citeauthoryear{{Li}, {Gnedin}, {Gnedin}, {Meng}, {Semenov}  \& {Kravtsov}}{{Li} et~al.}{2017}]{huili_paper1}
{Li} H.,  {Gnedin} O.~Y.,  {Gnedin} N.~Y.,  {Meng} X.,  {Semenov} V.~A.,   {Kravtsov} A.~V.,  2017, \mn@doi [The Astrophysical Journal] {10.3847/1538-4357/834/1/69}, \href {https://ui.adsabs.harvard.edu/abs/2017ApJ...834...69L} {834, 69}

\bibitem[\protect\citeauthoryear{{Lombardi}, {Alves}  \& {Lada}}{{Lombardi} et~al.}{2011}]{lombardi_2mass-orion}
{Lombardi} M.,  {Alves} J.,   {Lada} C.~J.,  2011, \mn@doi [Astronomy and Astrophysics] {10.1051/0004-6361/201116915}, \href {https://ui.adsabs.harvard.edu/abs/2011A&A...535A..16L} {535, A16}

\bibitem[\protect\citeauthoryear{{Lombardi}, {Bouy}, {Alves}  \& {Lada}}{{Lombardi} et~al.}{2014}]{lombardi_herschel}
{Lombardi} M.,  {Bouy} H.,  {Alves} J.,   {Lada} C.~J.,  2014, \mn@doi [Astronomy and Astrophysics] {10.1051/0004-6361/201323293}, \href {https://ui.adsabs.harvard.edu/abs/2014A&A...566A..45L} {566, A45}

\bibitem[\protect\citeauthoryear{{Lue}, {Guszejnov}, {Offner}  \& {Grudi{\'c}}}{{Lue} et~al.}{2021}]{lue}
{Lue} A.,  {Guszejnov} D.,  {Offner} S. S.~R.,   {Grudi{\'c}} M.~Y.,  2021, \mn@doi [Research Notes of the American Astronomical Society] {10.3847/2515-5172/ac2d37}, \href {https://ui.adsabs.harvard.edu/abs/2021RNAAS...5..225L} {5, 225}

\bibitem[\protect\citeauthoryear{{Mac Low} \& {Klessen}}{{Mac Low} \& {Klessen}}{2004}]{maclow_klessen}
{Mac Low} M.-M.,  {Klessen} R.~S.,  2004, \mn@doi [Reviews of Modern Physics] {10.1103/RevModPhys.76.125}, \href {https://ui.adsabs.harvard.edu/abs/2004RvMP...76..125M} {76, 125}

\bibitem[\protect\citeauthoryear{{Massey} \& {Hunter}}{{Massey} \& {Hunter}}{1998}]{hubble_observeR136}
{Massey} P.,  {Hunter} D.~A.,  1998, \mn@doi [The Astrophysical Journal] {10.1086/305126}, \href {https://ui.adsabs.harvard.edu/abs/1998ApJ...493..180M} {493, 180}

\bibitem[\protect\citeauthoryear{{McClure-Griffiths}, {Dickey}, {Gaensler}, {Green}  \& {Haverkorn}}{{McClure-Griffiths} et~al.}{2006}]{mcclure-griffiths}
{McClure-Griffiths} N.~M.,  {Dickey} J.~M.,  {Gaensler} B.~M.,  {Green} A.~J.,   {Haverkorn} M.,  2006, \mn@doi [The Astrophysical Journal] {10.1086/508706}, \href {https://ui.adsabs.harvard.edu/abs/2006ApJ...652.1339M} {652, 1339}

\bibitem[\protect\citeauthoryear{{Men'shchikov} et~al.,}{{Men'shchikov} et~al.}{2010}]{menshchikov2010}
{Men'shchikov} A.,  et~al., 2010, \mn@doi [Astronomy \& Astrophysics] {10.1051/0004-6361/201014668}, \href {https://ui.adsabs.harvard.edu/abs/2010A&A...518L.103M} {518, L103}

\bibitem[\protect\citeauthoryear{{Molinari} et~al.,}{{Molinari} et~al.}{2010}]{higal}
{Molinari} S.,  et~al., 2010, \mn@doi [Astronomy \& Astrophysics] {10.1051/0004-6361/201014659}, \href {https://ui.adsabs.harvard.edu/abs/2010A&A...518L.100M} {518, L100}

\bibitem[\protect\citeauthoryear{{Motte} et~al.,}{{Motte} et~al.}{2010}]{motte2010}
{Motte} F.,  et~al., 2010, \mn@doi [Astronomy \& Astrophysics] {10.1051/0004-6361/201014690}, \href {https://ui.adsabs.harvard.edu/abs/2010A&A...518L..77M} {518, L77}

\bibitem[\protect\citeauthoryear{{Myers}}{{Myers}}{2009}]{myers2009}
{Myers} P.~C.,  2009, \mn@doi [The Astrophysical Journal] {10.1088/0004-637X/700/2/1609}, \href {https://ui.adsabs.harvard.edu/abs/2009ApJ...700.1609M} {700, 1609}

\bibitem[\protect\citeauthoryear{{Ntormousi} \& {Hennebelle}}{{Ntormousi} \& {Hennebelle}}{2019}]{ntormousi}
{Ntormousi} E.,  {Hennebelle} P.,  2019, \mn@doi [Astronomy and Astrophysics] {10.1051/0004-6361/201834094}, \href {https://ui.adsabs.harvard.edu/abs/2019A&A...625A..82N} {625, A82}

\bibitem[\protect\citeauthoryear{{Pattle}, {Fissel}, {Tahani}, {Liu}  \& {Ntormousi}}{{Pattle} et~al.}{2022}]{pattle2022}
{Pattle} K.,  {Fissel} L.,  {Tahani} M.,  {Liu} T.,   {Ntormousi} E.,  2022, arXiv e-prints, \href {https://ui.adsabs.harvard.edu/abs/2022arXiv220311179P} {p. arXiv:2203.11179}

\bibitem[\protect\citeauthoryear{{Peretto} et~al.,}{{Peretto} et~al.}{2013}]{peretto2013}
{Peretto} N.,  et~al., 2013, \mn@doi [Astronomy \& Astrophysics] {10.1051/0004-6361/201321318}, \href {https://ui.adsabs.harvard.edu/abs/2013A&A...555A.112P} {555, A112}

\bibitem[\protect\citeauthoryear{{Pillai} et~al.,}{{Pillai} et~al.}{2020}]{pillai2020}
{Pillai} T. G.~S.,  et~al., 2020, \mn@doi [Nature Astronomy] {10.1038/s41550-020-1172-6}, \href {https://ui.adsabs.harvard.edu/abs/2020NatAs...4.1195P} {4, 1195}

\bibitem[\protect\citeauthoryear{{Pineda}, {Caselli}  \& {Goodman}}{{Pineda} et~al.}{2008}]{pineda2008}
{Pineda} J.~E.,  {Caselli} P.,   {Goodman} A.~A.,  2008, \mn@doi [The Astrophysical Journal] {10.1086/586883}, \href {https://ui.adsabs.harvard.edu/abs/2008ApJ...679..481P} {679, 481}

\bibitem[\protect\citeauthoryear{{Pineda}, {Goodman}, {Arce}, {Caselli}, {Foster}, {Myers}  \& {Rosolowsky}}{{Pineda} et~al.}{2010}]{2010pineda}
{Pineda} J.~E.,  {Goodman} A.~A.,  {Arce} H.~G.,  {Caselli} P.,  {Foster} J.~B.,  {Myers} P.~C.,   {Rosolowsky} E.~W.,  2010, \mn@doi [The Astrophysical Journal] {10.1088/2041-8205/712/1/L116}, \href {https://ui.adsabs.harvard.edu/abs/2010ApJ...712L.116P} {712, L116}

\bibitem[\protect\citeauthoryear{{Pineda}, {Goodman}, {Arce}, {Caselli}, {Longmore}  \& {Corder}}{{Pineda} et~al.}{2011}]{2011pineda}
{Pineda} J.~E.,  {Goodman} A.~A.,  {Arce} H.~G.,  {Caselli} P.,  {Longmore} S.,   {Corder} S.,  2011, \mn@doi [The Astrophysical Journal] {10.1088/2041-8205/739/1/L2}, \href {https://ui.adsabs.harvard.edu/abs/2011ApJ...739L...2P} {739, L2}

\bibitem[\protect\citeauthoryear{{Polychroni} et~al.,}{{Polychroni} et~al.}{2013}]{polychroni}
{Polychroni} D.,  et~al., 2013, \mn@doi [The Astrophysical Journal Letters] {10.1088/2041-8205/777/2/L33}, \href {https://ui.adsabs.harvard.edu/abs/2013ApJ...777L..33P} {777, L33}

\bibitem[\protect\citeauthoryear{{Ram{{\'i}}rez-Galeano}, {Ballesteros-Paredes}, {Smith}, {Camacho}  \& {Zamora-Avil{\'e}s}}{{Ram{{\'i}}rez-Galeano} et~al.}{2022}]{ramirez-galeano}
{Ram{{\'i}}rez-Galeano} L.,  {Ballesteros-Paredes} J.,  {Smith} R.~J.,  {Camacho} V.,   {Zamora-Avil{\'e}s} M.,  2022, \mn@doi [Monthly Notices of the Royal Astronomical Society] {10.1093/mnras/stac1848}, \href {https://ui.adsabs.harvard.edu/abs/2022MNRAS.515.2822R} {515, 2822}

\bibitem[\protect\citeauthoryear{{Rathjen} et~al.,}{{Rathjen} et~al.}{2021}]{silcc6}
{Rathjen} T.-E.,  et~al., 2021, \mn@doi [Monthly Notices of the Royal Astronomical Society] {10.1093/mnras/stab900}, \href {https://ui.adsabs.harvard.edu/abs/2021MNRAS.504.1039R} {504, 1039}

\bibitem[\protect\citeauthoryear{{Reina-Campos}, {Keller}, {Kruijssen}, {Gensior}, {Trujillo-Gomez}, {Jeffreson}, {Pfeffer}  \& {Sills}}{{Reina-Campos} et~al.}{2022}]{martapaper}
{Reina-Campos} M.,  {Keller} B.~W.,  {Kruijssen} J.~M.~D.,  {Gensior} J.,  {Trujillo-Gomez} S.,  {Jeffreson} S. M.~R.,  {Pfeffer} J.~L.,   {Sills} A.,  2022, \mn@doi [Monthly Notices of the Royal Astronomical Society] {10.1093/mnras/stac1934}, \href {https://ui.adsabs.harvard.edu/abs/2022MNRAS.tmp.1857R} {}

\bibitem[\protect\citeauthoryear{{Rezaei Kh.}, {Bailer-Jones}, {Soler}  \& {Zari}}{{Rezaei Kh.} et~al.}{2020}]{rezaei2020}
{Rezaei Kh.} S.,  {Bailer-Jones} C. A.~L.,  {Soler} J.~D.,   {Zari} E.,  2020, \mn@doi [Astronomy \& Astrophysics] {10.1051/0004-6361/202038708}, \href {https://ui.adsabs.harvard.edu/abs/2020A&A...643A.151R} {643, A151}

\bibitem[\protect\citeauthoryear{{Rice}, {Goodman}, {Bergin}, {Beaumont}  \& {Dame}}{{Rice} et~al.}{2016}]{RiceGoodman2016}
{Rice} T.~S.,  {Goodman} A.~A.,  {Bergin} E.~A.,  {Beaumont} C.,   {Dame} T.~M.,  2016, \mn@doi [The Astrophysical Journal] {10.3847/0004-637X/822/1/52}, \href {https://ui.adsabs.harvard.edu/abs/2016ApJ...822...52R} {822, 52}

\bibitem[\protect\citeauthoryear{{Rieder}, {Dobbs}, {Bending}, {Liow}  \& {Wurster}}{{Rieder} et~al.}{2022}]{rieder}
{Rieder} S.,  {Dobbs} C.,  {Bending} T.,  {Liow} K.~Y.,   {Wurster} J.,  2022, \mn@doi [Monthly Notices of the Royal Astronomical Society] {10.1093/mnras/stab3425}, \href {https://ui.adsabs.harvard.edu/abs/2022MNRAS.509.6155R} {509, 6155}

\bibitem[\protect\citeauthoryear{Schleicher, Schober, Federrath, Bovino  \& Schmidt}{Schleicher et~al.}{2013}]{Schleicher2013}
Schleicher D.,  Schober J.,  Federrath C.,  Bovino S.,   Schmidt W.,  2013, \mn@doi [New Journal of Physics] {10.1088/1367-2630/15/2/023017}, 15, 023017

\bibitem[\protect\citeauthoryear{{Schmidt}, {Federrath}, {Hupp}, {Kern}  \& {Niemeyer}}{{Schmidt} et~al.}{2009}]{schmidt2009}
{Schmidt} W.,  {Federrath} C.,  {Hupp} M.,  {Kern} S.,   {Niemeyer} J.~C.,  2009, \mn@doi [Astronomy and Astrophysics] {10.1051/0004-6361:200809967}, \href {https://ui.adsabs.harvard.edu/abs/2009A&A...494..127S} {494, 127}

\bibitem[\protect\citeauthoryear{{Seifried} et~al.,}{{Seifried} et~al.}{2017}]{silcc_zoom}
{Seifried} D.,  et~al., 2017, \mn@doi [Monthly Notices of the Royal Astronomical Society] {10.1093/mnras/stx2343}, \href {https://ui.adsabs.harvard.edu/abs/2017MNRAS.472.4797S} {472, 4797}

\bibitem[\protect\citeauthoryear{{Smith}, {Shetty}, {Stutz}  \& {Klessen}}{{Smith} et~al.}{2012}]{smith2012}
{Smith} R.~J.,  {Shetty} R.,  {Stutz} A.~M.,   {Klessen} R.~S.,  2012, \mn@doi [The Astrophysical Journal] {10.1088/0004-637X/750/1/64}, \href {https://ui.adsabs.harvard.edu/abs/2012ApJ...750...64S} {750, 64}

\bibitem[\protect\citeauthoryear{{Smith} et~al.,}{{Smith} et~al.}{2017}]{grackle}
{Smith} B.~D.,  et~al., 2017, \mn@doi [Monthly Notices of the Royal Astronomical Society] {10.1093/mnras/stw3291}, \href {https://ui.adsabs.harvard.edu/abs/2017MNRAS.466.2217S} {466, 2217}

\bibitem[\protect\citeauthoryear{{Smith}, {Bryan}, {Somerville}, {Hu}, {Teyssier}, {Burkhart}  \& {Hernquist}}{{Smith} et~al.}{2021}]{smith2021}
{Smith} M.~C.,  {Bryan} G.~L.,  {Somerville} R.~S.,  {Hu} C.-Y.,  {Teyssier} R.,  {Burkhart} B.,   {Hernquist} L.,  2021, \mn@doi [Monthly Notices of the Royal Astronomical Society] {10.1093/mnras/stab1896}, \href {https://ui.adsabs.harvard.edu/abs/2021MNRAS.506.3882S} {506, 3882}

\bibitem[\protect\citeauthoryear{{Soler}}{{Soler}}{2019}]{soler_planck}
{Soler} J.~D.,  2019, \mn@doi [Astronomy \& Astrophysics] {10.1051/0004-6361/201935779}, \href {https://ui.adsabs.harvard.edu/abs/2019A&A...629A..96S} {629, A96}

\bibitem[\protect\citeauthoryear{{Soler}, {Hennebelle}, {Martin}, {Miville-Desch{\^e}nes}, {Netterfield}  \& {Fissel}}{{Soler} et~al.}{2013}]{SolerHennebelle2013}
{Soler} J.~D.,  {Hennebelle} P.,  {Martin} P.~G.,  {Miville-Desch{\^e}nes} M.~A.,  {Netterfield} C.~B.,   {Fissel} L.~M.,  2013, \mn@doi [The Astrophysical Journal] {10.1088/0004-637X/774/2/128}, \href {https://ui.adsabs.harvard.edu/abs/2013ApJ...774..128S} {774, 128}

\bibitem[\protect\citeauthoryear{{Soler} et~al.,}{{Soler} et~al.}{2022}]{soler2022}
{Soler} J.~D.,  et~al., 2022, \mn@doi [Astronomy and Astrophysics] {10.1051/0004-6361/202243334}, \href {https://ui.adsabs.harvard.edu/abs/2022A&A...662A..96S} {662, A96}

\bibitem[\protect\citeauthoryear{{Solomon}}{{Solomon}}{2001}]{Solomon2001}
{Solomon} P.~M.,  2001, in {Tacconi} L.,  {Lutz} D.,  eds, Starburst Galaxies: Near and Far. p.~173 (\mn@eprint {arXiv} {astro-ph/0101482}), \mn@doi{10.48550/arXiv.astro-ph/0101482}

\bibitem[\protect\citeauthoryear{{Sun} et~al.,}{{Sun} et~al.}{2022}]{sun}
{Sun} J.,  et~al., 2022, arXiv e-prints, \href {https://ui.adsabs.harvard.edu/abs/2022arXiv220607055S} {p. arXiv:2206.07055}

\bibitem[\protect\citeauthoryear{{Tahani} et~al.,}{{Tahani} et~al.}{2022}]{tahani2022}
{Tahani} M.,  et~al., 2022, \mn@doi [Astronomy \& Astrophysics] {10.1051/0004-6361/202243322}, \href {https://ui.adsabs.harvard.edu/abs/2022A&A...660L...7T} {660, L7}

\bibitem[\protect\citeauthoryear{Teyssier}{Teyssier}{2002}]{ramses}
Teyssier R.,  2002, Astronomy and Astrophysics, 385, 337

\bibitem[\protect\citeauthoryear{{Theissen}, {Konopacky}, {Lu}, {Kim}, {Zhang}, {Hsu}, {Chu}  \& {Wei}}{{Theissen} et~al.}{2022}]{orion_kinematics}
{Theissen} C.~A.,  {Konopacky} Q.~M.,  {Lu} J.~R.,  {Kim} D.,  {Zhang} S.~Y.,  {Hsu} C.-C.,  {Chu} L.,   {Wei} L.,  2022, \mn@doi [The Astrophysical Journal] {10.3847/1538-4357/ac3252}, \href {https://ui.adsabs.harvard.edu/abs/2022ApJ...926..141T} {926, 141}

\bibitem[\protect\citeauthoryear{{Thilker} et~al.,}{{Thilker} et~al.}{2021}]{rosolowsky_phangs}
{Thilker} D.~A.,  et~al., 2021, arXiv e-prints, \href {https://ui.adsabs.harvard.edu/abs/2021arXiv210613366T} {1, arXiv:2106.13366}

\bibitem[\protect\citeauthoryear{{Townsley}, {Broos}, {Garmire}, {Bouwman}, {Povich}, {Feigelson}, {Getman}  \& {Kuhn}}{{Townsley} et~al.}{2014}]{chandra_observestarform}
{Townsley} L.~K.,  {Broos} P.~S.,  {Garmire} G.~P.,  {Bouwman} J.,  {Povich} M.~S.,  {Feigelson} E.~D.,  {Getman} K.~V.,   {Kuhn} M.~A.,  2014, \mn@doi [The Astrophysical Journal Supplement] {10.1088/0067-0049/213/1/1}, \href {https://ui.adsabs.harvard.edu/abs/2014ApJS..213....1T} {213, 1}

\bibitem[\protect\citeauthoryear{{Tress}, {Sormani}, {Glover}, {Klessen}, {Battersby}, {Clark}, {Hatchfield}  \& {Smith}}{{Tress} et~al.}{2020}]{TressSormani2020}
{Tress} R.~G.,  {Sormani} M.~C.,  {Glover} S. C.~O.,  {Klessen} R.~S.,  {Battersby} C.~D.,  {Clark} P.~C.,  {Hatchfield} H.~P.,   {Smith} R.~J.,  2020, \mn@doi [Monthly Notices of the Royal Astronomical Society] {10.1093/mnras/staa3120}, \href {https://ui.adsabs.harvard.edu/abs/2020MNRAS.499.4455T} {499, 4455}

\bibitem[\protect\citeauthoryear{{Turner} et~al.,}{{Turner} et~al.}{2022}]{2022turner}
{Turner} J.~A.,  et~al., 2022, \mn@doi [Monthly Notices of the Royal Astronomical Society] {10.1093/mnras/stac2559}, \href {https://ui.adsabs.harvard.edu/abs/2022MNRAS.516.4612T} {516, 4612}

\bibitem[\protect\citeauthoryear{{Wang} et~al.,}{{Wang} et~al.}{2020}]{WangBihr2020}
{Wang} Y.,  et~al., 2020, \mn@doi [Astronomy and Astrophysics] {10.1051/0004-6361/201935866}, \href {https://ui.adsabs.harvard.edu/abs/2020A&A...634A.139W} {634, A139}

\bibitem[\protect\citeauthoryear{{Watkins} et~al.,}{{Watkins} et~al.}{2023a}]{WatkinsKreckel2023}
{Watkins} E.~J.,  et~al., 2023a, \mn@doi [arXiv e-prints] {10.48550/arXiv.2302.03699}, \href {https://ui.adsabs.harvard.edu/abs/2023arXiv230203699W} {p. arXiv:2302.03699}

\bibitem[\protect\citeauthoryear{{Watkins} et~al.,}{{Watkins} et~al.}{2023b}]{WatkinsBarnes2023}
{Watkins} E.~J.,  et~al., 2023b, \mn@doi [The Astrophysical Journal] {10.3847/2041-8213/aca6e4}, \href {https://ui.adsabs.harvard.edu/abs/2023ApJ...944L..24W} {944, L24}

\bibitem[\protect\citeauthoryear{{Wilson}, {Jefferts}  \& {Penzias}}{{Wilson} et~al.}{1970}]{wilson1970}
{Wilson} R.~W.,  {Jefferts} K.~B.,   {Penzias} A.~A.,  1970, \mn@doi [The Astrophysical Journal Letters] {10.1086/180567}, \href {https://ui.adsabs.harvard.edu/abs/1970ApJ...161L..43W} {161, L43}

\bibitem[\protect\citeauthoryear{{Wilson}, {Elmegreen}, {Bemis}  \& {Brunetti}}{{Wilson} et~al.}{2019}]{WilsonElmegreen2019}
{Wilson} C.~D.,  {Elmegreen} B.~G.,  {Bemis} A.,   {Brunetti} N.,  2019, \mn@doi [The Astrophysical Journal] {10.3847/1538-4357/ab31f3}, \href {https://ui.adsabs.harvard.edu/abs/2019ApJ...882....5W} {882, 5}

\bibitem[\protect\citeauthoryear{{Wilson}, {Bemis}, {Ledger}  \& {Klimi}}{{Wilson} et~al.}{2023}]{WilsonBemis2023}
{Wilson} C.~D.,  {Bemis} A.,  {Ledger} B.,   {Klimi} O.,  2023, \mn@doi [Monthly Notices of the Royal Astronomical Society] {10.1093/mnras/stad560}, \href {https://ui.adsabs.harvard.edu/abs/2023MNRAS.521..717W} {521, 717}

\bibitem[\protect\citeauthoryear{{Zhao}, {Pudritz}, {Pillsworth}, {Robinson}  \& {Wadsley}}{{Zhao} et~al.}{2023}]{bopaper}
{Zhao} B.,  {Pudritz} R.~E.,  {Pillsworth} R.,  {Robinson} H.,   {Wadsley} J.,  2023, {Filamentary Hierarchies and Supperbubbles: Galactic Multiscale MHD Simulations of GMC to Star Cluster Formation}, submitted

\bibitem[\protect\citeauthoryear{Zucker et~al.,}{Zucker et~al.}{2021}]{Zucker_2021}
Zucker C.,  et~al., 2021, \mn@doi [The Astrophysical Journal] {10.3847/1538-4357/ac1f96}, 919, 35

\makeatother
\end{thebibliography}

%%%%%%%%%%%%%%%%%%%%%%%%%%%%%%%%%%%%%%%%%%%%%%%%%%

%%%%%%%%%%%%%%%%% APPENDICES %%%%%%%%%%%%%%%%%%%%%
\appendix

\section{Density and temperature PDFs}

\begin{figure}
    \centering
    \includegraphics[width=0.9\linewidth]{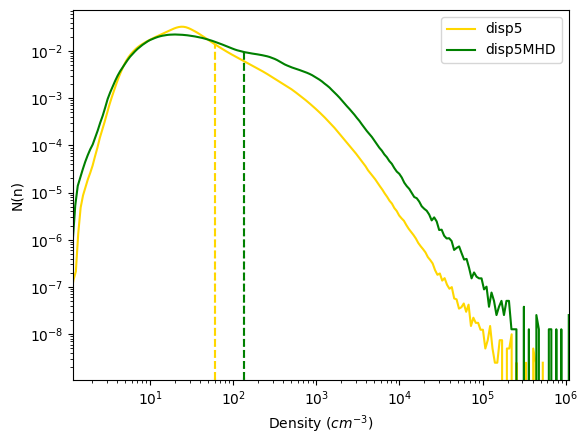}
    \includegraphics[width=0.9\linewidth]{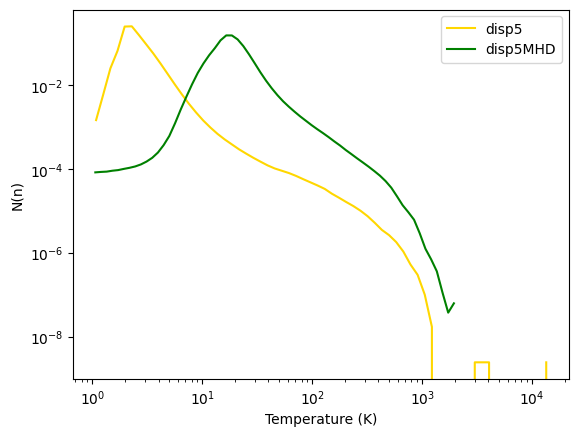}
    \caption{Density and temperature PDFs for our 5 km/s models. Dotted vertical lines on the density PDF (top) mark the average density of all gas. Dashed lines represent the average density of cold gas in the simulation. Drastically low temperatures in the disp5 hydro model showcase the overcooling of the gas brought on by significantly higher compression allowed due to the lack of magnetic support, as discussed in Sec. \ref{moleculargas}.}
    \label{fig:disp5pdfs}
\end{figure}

\begin{figure}
    \centering
    \includegraphics[width=0.9\linewidth]{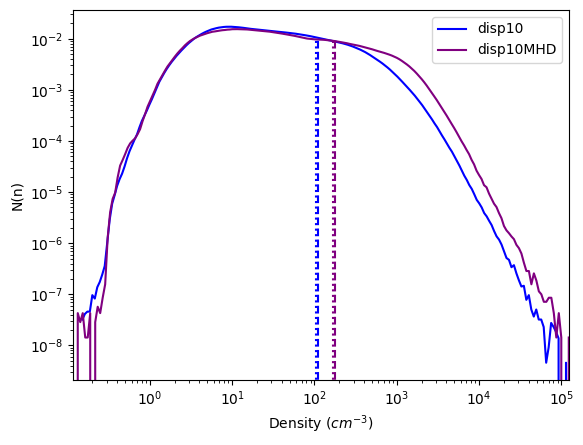}
    \includegraphics[width=0.9\linewidth]{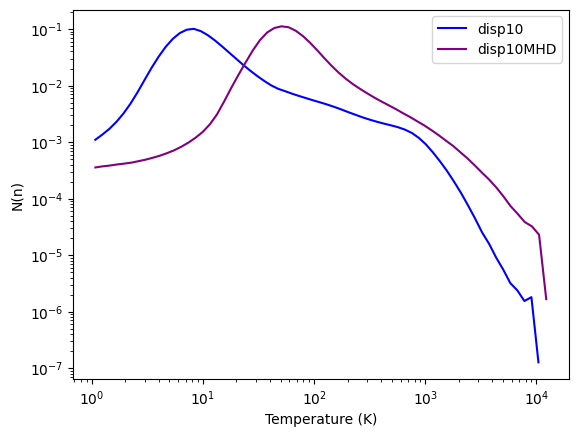}
    \caption{Density and temperature PDFs for our 10 km/s models. Dotted vertical lines on the density PDF (top) mark the average density of all gas. Dashed lines represent the average density of cold gas in the simulation.}
    \label{fig:disp10pdfs}
\end{figure}

Figures \ref{fig:disp5pdfs} and \ref{fig:disp10pdfs} show the isolated density and temperature PDFs for the 5 km/s and 10 km/s models, respectively. Dotted vertical lines pinpoint the average density for all gas in the simulation, while dashed vertical lines pinpoint the average density when considering only the gas contained in or below the peak temperature. We choose this temperature cutoff based on where the temperature curve has begun to flatten out.

For both the 5 and 10 km/s models, these averages are approximately the same, as evidenced by the overlapping vertical lines in the plots. Additionally, while these models show different average densities between their magnetized and unmagnetized runs, they have similar average densities to each other despite their different density ranges. Our 10 km/s models show more gas at low densities and reach maximum densities of $\simeq 10^5~\rm{cm}^{-3}$. The 5 km/s models, on the other hand, do not contain densities much lower than $\simeq 1~\rm{cm}^{-3}$, but reach high densities of $10^6~\rm{cm}^{-3}$.

Though the high density range above $10^5~\rm{cm}^{-3}$ is affected by the resolution limit of the simulation, we can still discern some general information. First, the higher density limit in the 5 km/s case shows that the lower total number of sink particles formed in the simulation is not due to high density gas not being present at all, but rather due to that high density gas never forming collapsing clumps. On the other hand, and to our second point, the 10 km/s models are not necessarily missing sink particles because of a lack of high density gas either, but more likely due to high turbulence creating shocks which are actively tearing high density filaments apart, preventing clumps from forming. Compared to our results for the 8 km/s model in Section \ref{sec:results}, these indicate that the balance between the gravity and turbulence plays a key role in all structure formation and directly correlates to cluster formation. We see no direct evidence of the magnetic field affecting these balances in this case, though we also not that a more detailed study of the magnetic field strength and its effects would be necessary to make strong conclusions.

\begin{figure}
    \centering
    \includegraphics[width=0.9\linewidth]{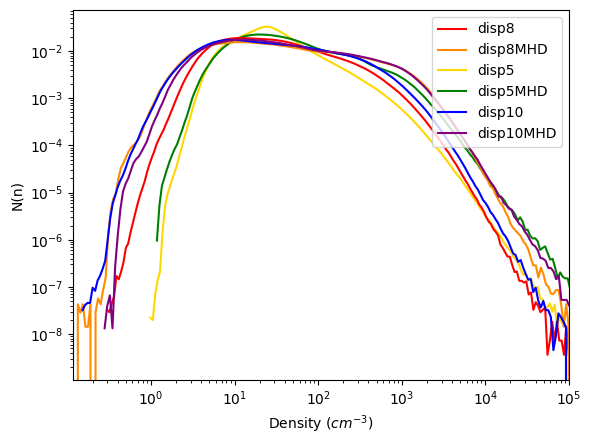}
    \includegraphics[width=0.9\linewidth]{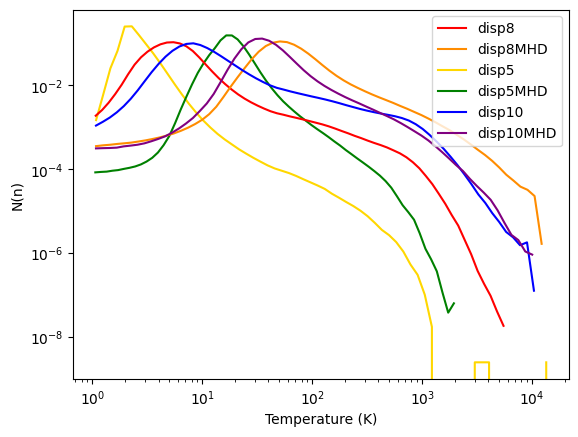}
    \caption{Density and temperature PDFs for all of our models.}
    \label{fig:pdfs_all}
\end{figure}

For completeness and direct comparison, we include in Figure \ref{fig:pdfs_all} the density and temperature PDFs of all six models together. For similar density PDFs, all of our models have a unique temperature PDF. Th unmagnetized hydro models follow an increasing peak temperature with increasing velocity dispersion, as one may expect from increased shocks contributing to high overall heating rates in the simulation. However, the inclusion of magnetic fields changes this trend. Instead we see the highest temperature peak in our 8 km/s MHD model, as well as more gas in the higher temperature range ($\geq 100 \rm{K}$) than the other models. 

\section{Line mass stability analysis}

We introduce in this paper a simple and rudimentary approach to analyzing the line mass of filaments in our simulations, and here describe the method followed. We define the line mass as the thermal line mass, that is where the critical line mass accounts for sound speed but not the velocity dispersion if the gas. We then have a critical line mass of the form

\begin{equation}
    M_{crit} = \frac{2c_s^2}{G}
\end{equation}

\begin{figure}
    \centering
    \includegraphics[width=1.0\linewidth]{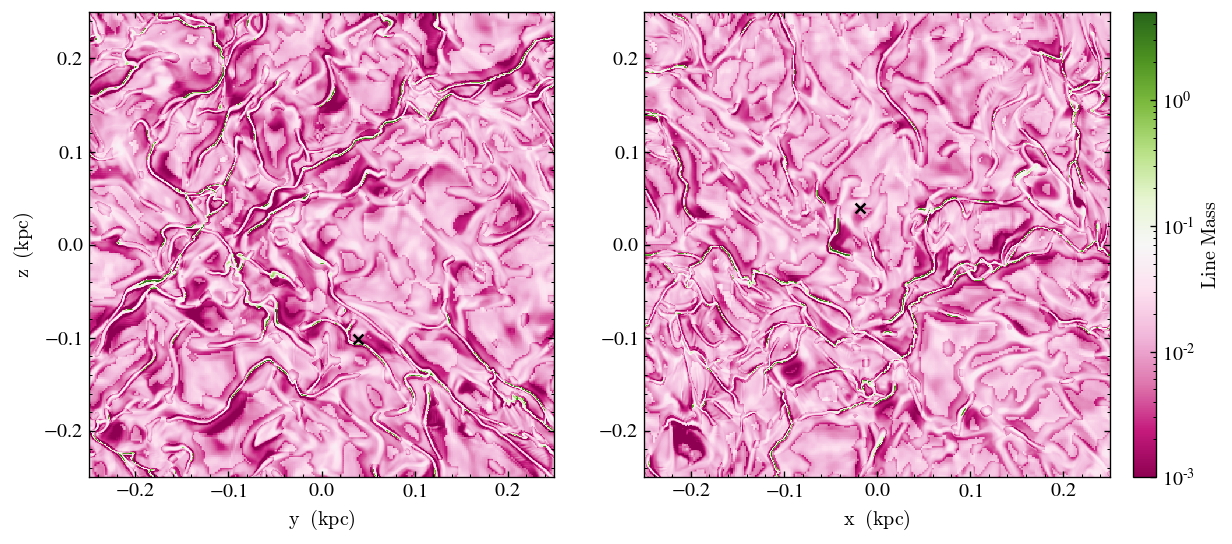}
    \includegraphics[width=1.0\linewidth]{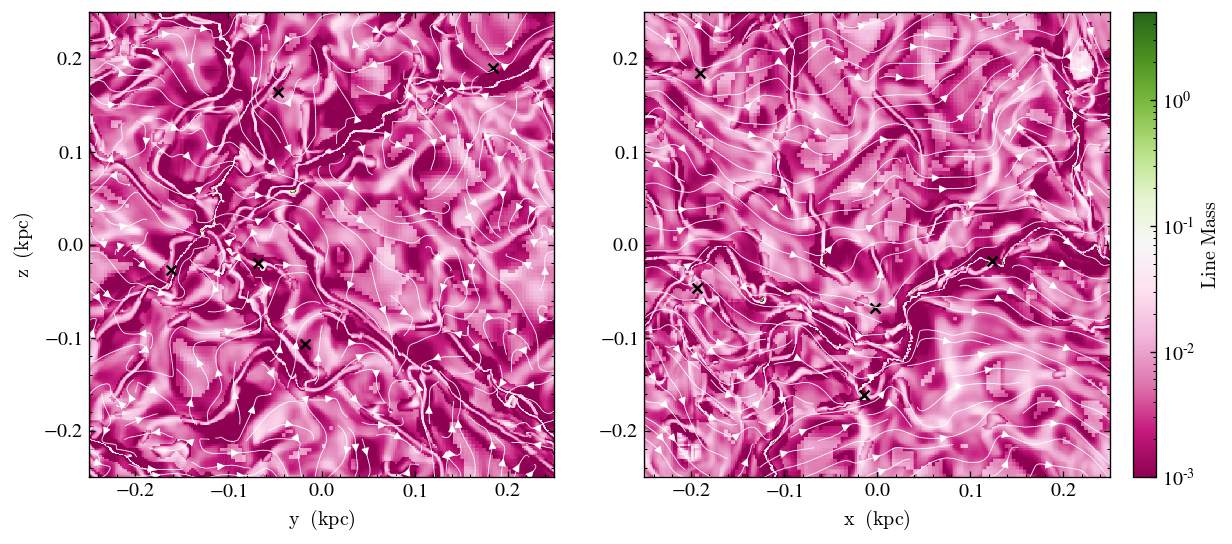}
    \caption{Examples of slices through the cube of our disp8 and disp8MHD models showing the line mass in evaluated in the slice cell by cell.}
    \label{fig:linemass_slices}
\end{figure}

\noindent The actual line mass we calculate as total mass of a cell divided by its length, a physical line mass. The ratio of this value to the critical line mass gives an indication of criticality, as discussed in Sec. \ref{sec:bridge}. This cell by cell construction allows us to visualize the line mass in multiple ways. In Figure \ref{fig:linemass_slices}, we show the line mass ratios for a slice through our disp8 and disp8MHD models. As all the cells have an associated line mass ratio, one could also create a 3D map of the line masses in order to show a map of the criticality throughout the filaments.

In Sec. \ref{sec:linemass}, we show the projected line masses of both of our 8 km/s models. In that case we have taken all the cells stacked on top of each other in the direction of the projection, and integrated the line mass field over the path length. Due to this integration, we then correct for the extra term of total path length by dividing the value in each cell by the length of the box. We note here that this is relatively easy to do given that our box is symmetrical, but non-symmetric boxes would of course require the definition of as many fields as projection directions one wishes to take. From this correction, we thus have both 3D and 2D planar information about the critical line masses of filaments, and we note the similarities in criticality between the slices above and the projections in the body of this paper. 

\section{Turbulent Energy Spectra}
We show here our turbulent energy spectra for our disp8MHD model at times of 1.90 Myr and 3.79 Myr. These show that our turbulent energy is not fully saturated to the Burger's relation until sink formation begins, therefore no appreciable decay happens in the timescales we study.

\begin{figure}
    \centering
    \includegraphics[width=0.99\linewidth]{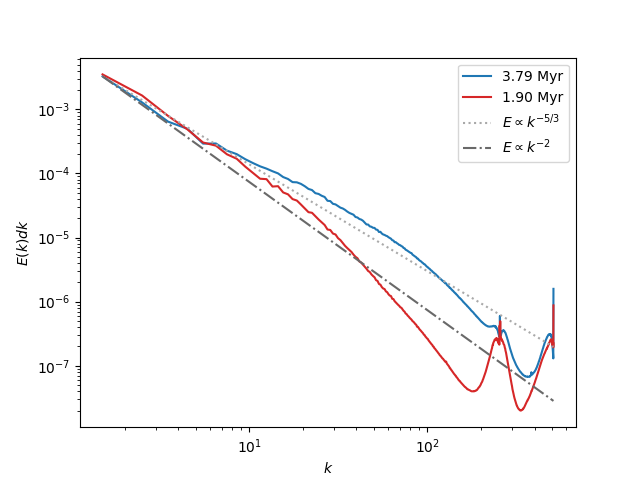}
    \caption{Turbulent kinetic energy spectra for our disp8MHD model at a time of 1.90 Myr (\textit{top}) and 3.79 Myr (\textit{bottom}). The dotted line represents the Kolmogorov spectrum $E(k) \propto k^{-5/3}$. The dash dotted line shows the Burgers spectrum $E(k) \propto k^{-2}$. The maximum wavenumber (k) corresponds to a cell of 0.48 pc width (our highest resolution), but we note that some resolution effects still arise in large k values. These spikes in energy are likely due to small numbers of cells of the specific corresponding resolution.}
    \label{fig:turbspectra}
\end{figure}
%%%%%%%%%%%%%%%%%%%%%%%%%%%%%%%%%%%%%%%%%%%%%%%%%%

% Don't change these lines
\bsp	% typesetting comment
\label{lastpage}
\end{document}